# Magnetic Chiral Light

Zaid Haddadin,[1] Lukas Novotny,[2] and Lisa V. Poulikakos[3,4,*]

[1]Department of Electrical & Computer Engineering, UC San Diego, La Jolla, CA 92093-0021, USA
[2]Photonics Laboratory, ETH Zürich, CH-8093 Zürich, Switzerland
[3]Department of Mechanical & Aerospace Engineering, UC San Diego, La Jolla, CA 92093-0021, USA
[4]Program of Materials Science & Engineering, UC San Diego, La Jolla, CA 92093-0021, USA
[*]Contact author: lpoulikakos@ucsd.edu

**POPULAR SUMMARY**. Chirality, or "handedness", is a fundamental property of nature, where objects cannot be superimposed onto their mirror image. In optics, chirality appears in circularly polarized light – where light's electric field rotates in a corkscrew shape, either clockwise or counterclockwise as light travels. Materials that are themselves chiral in shape can interact differently with each handedness of light, a phenomenon that underpins technologies from biosensing and drug development to quantum devices. Until now, chiral signals have relied on light's electric field.

In this work, we demonstrate a chiral optical signal can arise from interactions with light's magnetic field alone. We introduce a new theoretical framework, derived from of a fundamental equation in optics – the optical chirality continuity equation – to show that chiral signals can be generated through magnetic resonances. To test this idea, we designed planar arrays of dielectric nanoparticles. These structures acted like nanoscale antennas that transmit magnetic fields by creating an in-plane rotation of the electric field. This optical magnetism response created a highly selective chiral signal.

Our results open a new design principle for chiral photonics, showing that magnetic fields can play a central role in controlling chiral signal generation. This insight could enable new strategies for sensitive molecular detection, nanoscale imaging, and photonic technologies.

**ABSTRACT**. We present a result derived from the optical chirality continuity equation that shows the existence of a source term describing optical chirality generation through the interactions of the magnetic induction field with a source current, without contributions from the electric field. This framework is validated through a metasurface-based methodology. Using a lossless, all-dielectric nanoparticle array, we engineer an in-plane rotating electric field that generates out-of-plane optical magnetism. This array exhibits a far-field circular dichroism of 0.47 and a high-quality factor on the order of $10^6$. The presented findings demonstrate the feasibility of inducing chiral optical responses from a magnetic source. This work establishes a new paradigm for structured photonic media, offering insights into the design of nanophotonic devices that exploit optical magnetism for chiral light-matter interactions.

## I. INTRODUCTION.

The interaction between light and chiral matter has been a cornerstone of scientific inquiry for over two centuries [1–3] that shaped our understanding of molecular biology [4,5], chemistry [6–8], and optics [9]. Chirality – the geometrical property that prevents an object from being superimposed onto its mirror image [10–12] – appears at all length scales from fundamental particles [13,14] to astrophysical features [15]. A prominent example of chirality in optics is circularly polarized light (CPL) – light for which three-dimensional motion follows a corkscrew-like trajectory [16]. The CPL corkscrew can traverse through space in a circular clockwise or circular counterclockwise manner, which is determined by the rotation direction of a propagating electric field vector [3,16]. The two corkscrew rotations are mirror images that cannot be transformed into one another through any combination of rotation and translation [3,16], which underpins the essence of chiral asymmetry [10–12].

Chiral materials uniquely interact with each orientation of CPL [16,17], giving rise to phenomena such as optical activity [18–21] and its manifestations of circular dichroism (CD; the spin selective absorption of light) [22–24] and circular birefringence (the differential refraction of CPL) [24,25]. These effects have been widely exploited in applications ranging from biomolecular detection [26–28] to the design of quantum materials [29,30] and advanced photonic devices [31–35].

Advancements in nanophotonics enabled the engineering of tailored chiral optical responses through structured materials [36–41], particularly chiral metasurfaces [42–47] – structured arrays of elements with sub-wavelength transverse spacing and dimensions that control the propagation and polarization of chiral light. One research goal has been to enhance the strength of CD signals in these systems [48–60].In parallel, studies investigating the optical chirality continuity equation in lossy, dispersive media demonstrated that optical chirality dissipation can enable the generation of optical chirality flux, proportional to the degree of circular polarization in the far field [61–63].

Initial research showed that it's possible to create a magnetic source of chiral light in a single nanoparticle resonator [64]; however, the underlying physics is underexplored. This work investigates this phenomenon as a consequence of the optical chirality continuity equation [1,65]. We utilize this approach to achieve strong CD by transforming a lossless, all-dielectric metasurface into a source of optical chirality flux at its high-quality-factor magnetic dipolar resonance [1,65–68].

Analogous to Poynting's theorem for electromagnetic energy conservation – where a source or sink of electromagnetic energy describes the interaction between a source current and electric field [61–63,69,70] – we show that the interaction of a source current with the magnetic induction field can serve as a source or sink of optical chirality flux. This framework suggests that asymmetric optical chirality flux can emerge when the source term describing optical chirality generation is nonzero and differs between clockwise and counterclockwise CPL. While previous studies explored chiral light-matter interactions from magnetoelectric resonances [71,72] or the simultaneous excitation of electric and magnetic dipoles [73,74], our result demonstrates how optical chirality flux generation can also occur through optical magnetism alone [64,75–77].

To test this framework, we analyze metasurfaces composed of planar arrays of all-dielectric nanoparticles. In our previous work [78], we hypothesized that dichroic signals can be enhanced by modulating electric field lines through nanoparticle geometry. Building on this, we performed numerical simulations in COMSOL Multiphysics v6.0 [79] to investigate how geometric asymmetry influences chiral field generation. Our results reveal that metasurfaces with near two-dimensional asymmetry – specifically, L-shaped nanoparticles formed by an asymmetric corner cut in square-shaped nanoparticles – exhibit a highly localized and spectrally precise chiral response. This response originates from an in-plane electric field rotation, which, via the Ampère-Maxwell Law [67], induces a perpendicular out-of-plane magnetic field. This magnetic response drives the generation of a far-field chiral signal via optical magnetism [75] (Fig. 1). The resulting chiral resonance exhibits a linewidth below 1 nm, characteristic of a high-quality (high-Q) factor resonance [80]. Notably, this high-Q magnetic response must be confined between two electric resonances and is sensitive to incident CPL handedness: reversing the incident CPL switches the far-field signal amplitude at the magnetic resonance wavelength, while non-magnetic and off-resonant wavelengths remain largely unaffected or negligibly affected.

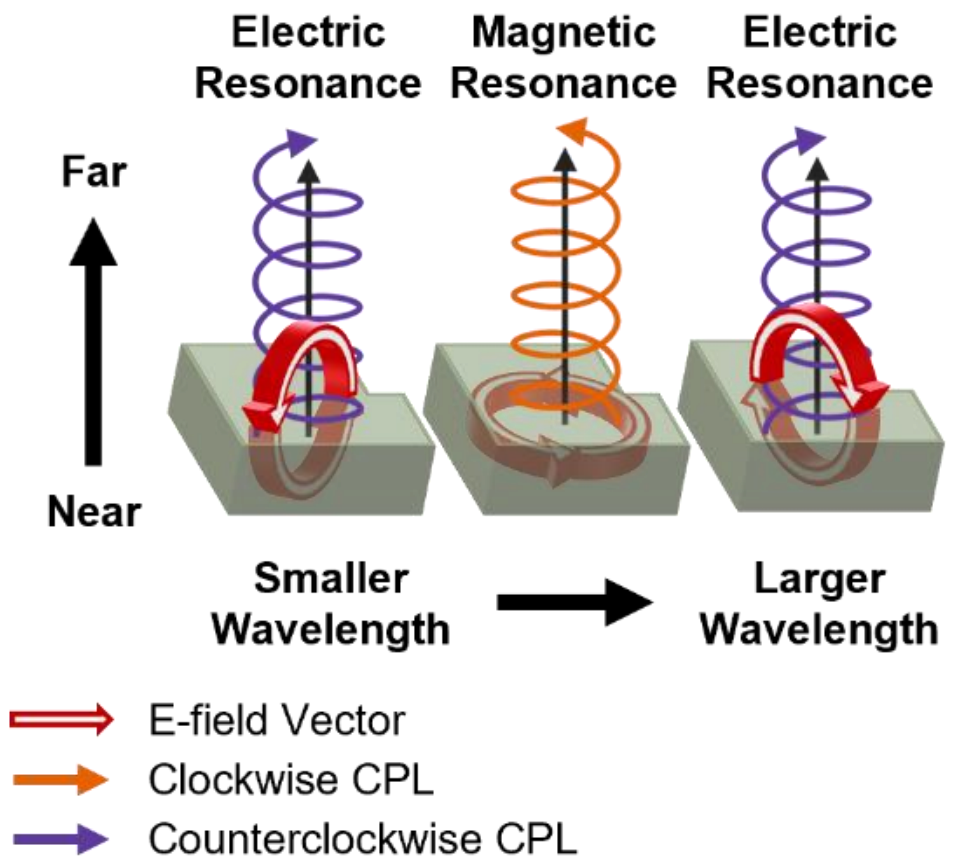

FIG 1. Conceptual schematic of this work's primary finding: a rotating electric field (shown as white arrows with red outlines) induces a perpendicular, far-field magnetic resonance. This response is confined between two far-field electric resonances. Notably, we find the electric and magnetic resonances generate far-field chiral signals of opposite handedness, denoted by clockwise (orange) or counterclockwise (purple) CPL arrows.

The significance of this result is twofold. First, it establishes our framework – developed as a consequence of the optical chirality continuity equation – as a quantitative tool for understanding chiral signal emergence from magnetic sources in structured photonic media. Second, it introduces a new design principle for engineering high-Q chiral resonances in metasurfaces, with potential applications in optical sensing, chiral quantum optics, and next-generation photonic technologies.

## II. METHODS

### A. Numerical simulations for generating magnetic chirality

Lattice arrays consisting of silicon nitride particles on a silicon dioxide substrate were modelled in COMSOL Multiphysics v6.0 [81], a finite-element method simulation software, using the Wave Optics Module [82] (see Supplementary Material Section S1, which includes Ref. [79,83–88]). Each particle had a fixed height of 270 nm and the gap among particles was held at 100 nm (Fig. 2(a)). The two-dimensional shape of the nanoparticles varied across arrays but was retained within an array. The tested shapes include squares, squares with a symmetric corner cut (achiral L-shaped), and squares with an asymmetric corner cut (chiral L-shaped) (Fig. 2(b)). CPL was used to excite the nanostructure arrays.

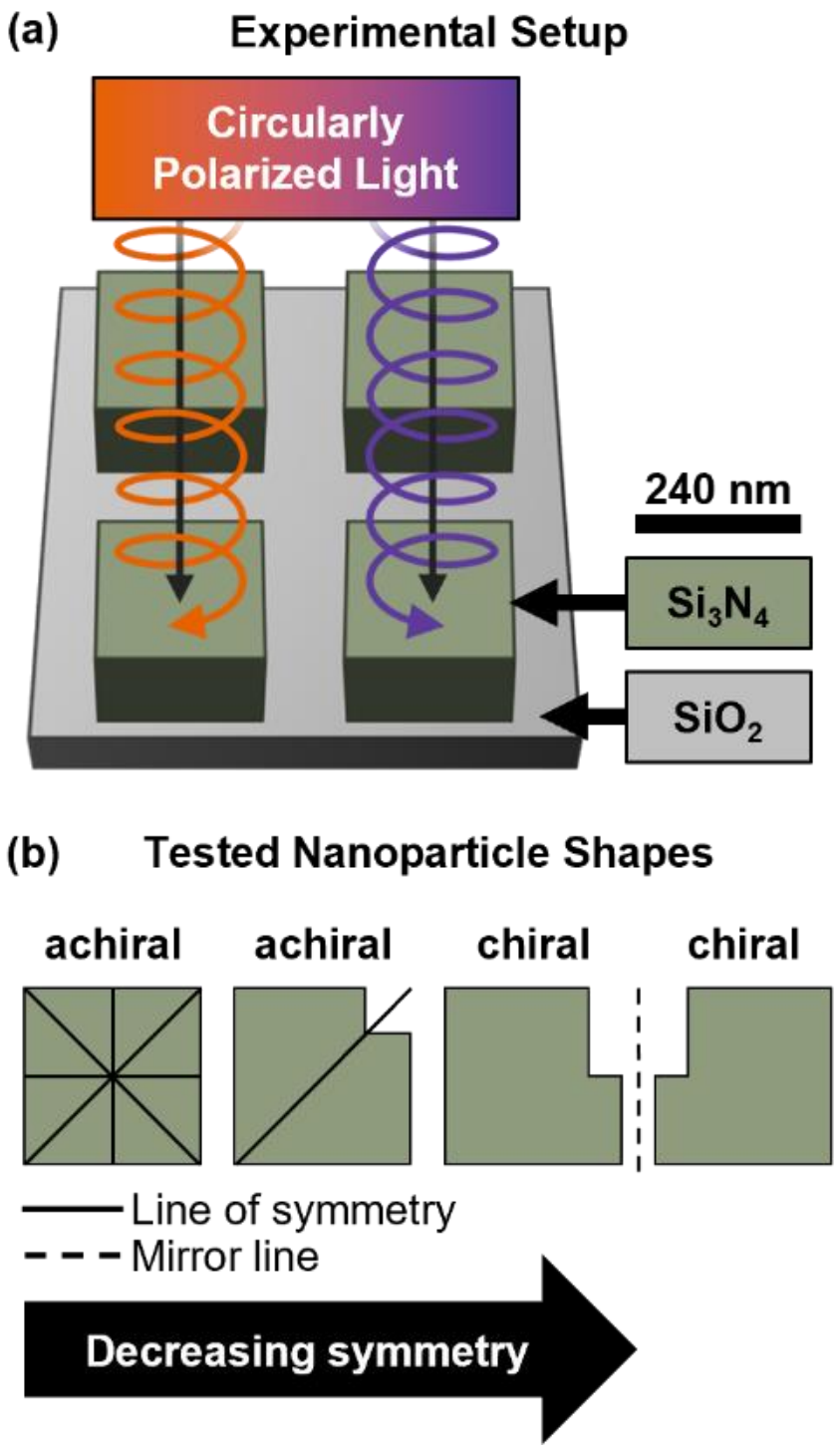

FIG 2. (a) Schematic of the metasurface, consisting of periodically arranged square-shaped silicon nitride nanoparticles on a silicon dioxide substrate. Though only four unit cells are shown, the numerical calculations approximate an infinite array using periodic boundary conditions. (b) Top-down view of the tested nanoparticle shapes: a square, a square with a symmetric corner cut, and two mirror images of a square with an asymmetric corner cut. Overlaid symmetry lines (black, solid) highlight that chiral shapes lack reflection symmetry. Created in BioRender. Poulikakos, L. (2026) https://BioRender.com/m06q553.

COMSOL Multiphysics v6.0 [81] functionalities were used to obtain far-field data and near-field data. Far-field data include (i) reflectance spectra and (ii) the Jones vectors [87] – from which the $S_3$ Stokes parameter was calculated [87,89] as it was shown to be proportional to the optical chirality flux [61,66]. Near-field data include (i) magnetic and electric field vectors and enhancement, (ii) the net magnetic and electric fluxes, (iii) the optical chirality density, and (iv) the optical chirality source term (see Supplementary Material Section S1 for full derivation). The far-field responses were used to characterize the circular dichroism in reflectance and polarization state of the outgoing light. The near-field responses were used to understand the underlying nanoparticle-level mechanisms that led to the far-field observations, and to show under what conditions a chiral signal can be generated via a magnetic resonance. This setup builds on previous works detailed by our team and others [78,90,91].

The nanoparticles in this work are categorized by whether they're two-dimensionally symmetric or asymmetric along their orthogonal axes, whether they're two-dimensionally chiral or achiral, or a combination of both. These categories arise from the type of corner cut induced upon a square-shaped particle: no corner cut (symmetric and achiral), a symmetrical corner cut (asymmetric and achiral), or an asymmetrical corner cut (asymmetric and chiral). We must note that the uncut and symmetrical corner cut structures within this work have undergone an analysis in our previous work [78]; however, they act as suitable controls to understand the behavior of the experimental group: the asymmetrical corner cut.

## III. RESULTS & DISCUSSION

### A. The optical chirality continuity equation

To investigate the physical origin of optical chirality generation, we begin by identifying the optical chirality continuity equation [1,65]:

$$\frac{1}{\mu_0}\nabla \cdot \boldsymbol{F} + \frac{\partial C}{\partial t} = -\frac{1}{2}[\boldsymbol{j_{tot}} \cdot (\nabla \times \boldsymbol{E}) + \boldsymbol{E} \cdot (\nabla \times \boldsymbol{j_{tot}})] \tag{3.1}$$

where $C \equiv \frac{\epsilon_0}{2}\boldsymbol{E} \cdot (\nabla \times \boldsymbol{E}) + \frac{1}{2\mu_0}\boldsymbol{B} \cdot (\nabla \times \boldsymbol{B})$ is the optical chirality density, $\boldsymbol{F} \equiv \frac{1}{2}[\boldsymbol{E} \times (\nabla \times \boldsymbol{B}) - \boldsymbol{B} \times (\nabla \times \boldsymbol{E})]$ is the optical chirality flux, and $\boldsymbol{j_{tot}}$ is the total current density containing all primary and secondary sources. $\boldsymbol{E}$ & $\boldsymbol{B}$ are the time-dependent electric and magnetic induction fields, respectively, and $\mu_0$ & $\epsilon_0$ are the vacuum magnetic permeability and electric permittivity, respectively. The right-hand side of Eqn. (3.1) corresponds to the generation or dissipation of optical chirality.

Eqn. (3.1) parallels Poynting's theorem, which governs energy conservation in electromagnetism [1,92]:

$$\frac{1}{\mu_0}\nabla \cdot \boldsymbol{S} + \frac{\delta u}{\delta t} = -\boldsymbol{j_{tot}} \cdot \boldsymbol{E} \tag{3.2}$$

where $u = \frac{\epsilon_0}{2}\boldsymbol{E} \cdot \boldsymbol{E} + \frac{1}{2\mu_0}\boldsymbol{B} \cdot \boldsymbol{B}$ is the energy density, $\boldsymbol{S} = \boldsymbol{E} \times \boldsymbol{B}$ is the Poynting vector representing energy flow, and $\boldsymbol{j_{tot}} \cdot \boldsymbol{E}$ describes the power supplied by the current $\boldsymbol{j_{tot}}$.

The structural similarities between Eqns. (3.1) and (3.2) led previous studies to draw parallels between energy conservation and chiral conservation [61,62,66]. Building on this structural analogy, we show that a reformulation of Eqn. (3.1) can elucidate the physical origin of optical chirality sources. A full derivation is provided in the supplementary material (see Supplementary Material Section S2, which includes Ref. [1,61,62,65–67,92–101]), and the key steps are outlined below.

We begin by taking the divergence of the optical chirality flux:

$$\nabla \cdot \boldsymbol{F} = \nabla \cdot \frac{1}{2}[\boldsymbol{E} \times (\nabla \times \boldsymbol{B}) - \boldsymbol{B} \times (\nabla \times \boldsymbol{E})] \tag{3.3}$$

Applying the vector identity $\nabla \cdot (\boldsymbol{a} \times \boldsymbol{b}) = \boldsymbol{b} \cdot (\nabla \times \boldsymbol{a}) - \boldsymbol{a} \cdot (\nabla \times \boldsymbol{b})$ [93], where $\boldsymbol{a}$ and $\boldsymbol{b}$ are arbitrary vector fields, and simplifying; we obtain:

$$\nabla \cdot \boldsymbol{F} = \frac{1}{2}\left[\boldsymbol{B} \cdot \big(\nabla \times (\nabla \times \boldsymbol{E})\big) - \boldsymbol{E} \cdot \big(\nabla \times (\nabla \times \boldsymbol{B})\big)\right] \tag{3.4}$$

Substituting in the wave equations for $\boldsymbol{E}$ and $\boldsymbol{B}$ [67,94] allows us to obtain:

$$\frac{1}{\mu_0}\nabla \cdot \boldsymbol{F} - \frac{\partial}{\partial t}\left[\frac{1}{2\mu_0 c^2}\left(-\boldsymbol{B} \cdot \frac{\partial \boldsymbol{E}}{\partial t} + \boldsymbol{E} \cdot \frac{\partial \boldsymbol{B}}{\partial t}\right)\right] = -\frac{1}{2}\left[\boldsymbol{B} \cdot \frac{\partial \boldsymbol{j_{tot}}}{\partial t} + \boldsymbol{E} \cdot (\nabla \times \boldsymbol{j_{tot}})\right] \tag{3.5}$$

Recognizing the bracketed term on the left-hand side corresponds to the optical chirality density $C$ and the bracketed term on the right-hand side corresponds to the source term describing optical chirality generation or dissipation, then Eqn. (3.4) maintains the form of the optical chirality continuity equation [1] presented in Eqn. (3.1).

By applying the vector identity introduced when deriving Eqn. (3.4) [93] and Faraday's Law [67], the bracketed term on the right-hand side can be written as:

$$\frac{1}{\mu_0}\nabla\cdot\boldsymbol{F}-\frac{\partial}{\partial t}\left[\frac{1}{2\mu_0 c^2}\left(-\boldsymbol{B}\cdot\frac{\partial\boldsymbol{E}}{\partial t}+\boldsymbol{E}\cdot\frac{\partial\boldsymbol{B}}{\partial t}\right)\right]=-\frac{1}{2}\left[2\boldsymbol{B}\cdot\frac{\partial\boldsymbol{j_{tot}}}{\partial t}-\frac{\partial}{\partial t}(\boldsymbol{j_{tot}}\cdot\boldsymbol{B})+\nabla\cdot(\boldsymbol{j_{tot}}\times\boldsymbol{E})\right] \tag{3.6}$$

With the Ampère-Maxwell Law and Faraday's Law [67], the bracketed term on the left-hand side can be simplified:

$$\frac{1}{\mu_0}\nabla\cdot\boldsymbol{F}-\frac{\partial}{\partial t}\left(-\frac{1}{2}\,\boldsymbol{B}\cdot\boldsymbol{j_{tot}}\right)=-\frac{1}{2}\left[2\boldsymbol{B}\cdot\frac{\partial\boldsymbol{j_{tot}}}{\partial t}-\frac{\partial}{\partial t}(\boldsymbol{j_{tot}}\cdot\boldsymbol{B})+\nabla\cdot(\boldsymbol{j_{tot}}\times\boldsymbol{E})\right] \tag{3.7}$$

Further simplification allows us to obtain:

$$\frac{1}{\mu_0}\nabla\cdot\boldsymbol{F}=-\boldsymbol{B}\cdot\frac{\partial\boldsymbol{j_{tot}}}{\partial t}-\frac{1}{2}\nabla\cdot(\boldsymbol{j_{tot}}\times\boldsymbol{E}) \tag{3.8}$$

Seeking a physically measurable form, we consider the time-averaged version of Eqn. (3.8) over a period $T$ (denoted by $<\cdot>_T$) under steady-state harmonic fields. $\boldsymbol{\mathcal{F}}$, $\boldsymbol{\mathcal{B}}$, $\boldsymbol{\mathcal{J}_{tot}}$, and $\boldsymbol{\mathcal{E}}$ are the time-harmonic optical chirality flux, magnetic induction field, total current density, and electric field, respectively. For some real-valued, time-dependent field $\boldsymbol{X}$, it relates to its time-harmonic field $\boldsymbol{\mathcal{X}}$ by $\boldsymbol{X}=Re[\boldsymbol{\mathcal{X}}e^{-i\omega t}]$, where $\omega$ denotes the angular frequency, and that each field varies sinusoidally with a single frequency [96,98]. In such notation, we have:

$$\frac{1}{\mu_0}<\nabla\cdot\boldsymbol{\mathcal{F}}>_T=-<\boldsymbol{\mathcal{B}}\cdot\frac{\partial\boldsymbol{\mathcal{J}_{tot}}}{\partial t}>_T-\frac{1}{2}<\nabla\cdot(\boldsymbol{\mathcal{J}_{tot}}\times\boldsymbol{\mathcal{E}})>_T \tag{3.9}$$

Next, we integrate the time-averaged form in Eqn. (3.9) over a volume $V$, which encompasses all primary and secondary sources:

$$\frac{1}{\mu_0}\iiint_V<\nabla\cdot\boldsymbol{\mathcal{F}}>_T\,dV=-\iiint_V<\boldsymbol{\mathcal{B}}\cdot\frac{\partial\boldsymbol{\mathcal{J}_{tot}}}{\partial t}>_T\,dV-\frac{1}{2}\iiint_V<\nabla\cdot(\boldsymbol{\mathcal{J}_{tot}}\times\boldsymbol{\mathcal{E}})>_T\,dV \tag{3.10}$$

We turn our attention to the right-most term in Eqn. (3.10): $\iiint_V<\nabla\cdot(\boldsymbol{\mathcal{J}_{tot}}\times\boldsymbol{\mathcal{E}})>_T$. Applying Gauss' theorem to this term $\left(\iiint_V<\nabla\cdot(\boldsymbol{\mathcal{J}_{tot}}\times\boldsymbol{\mathcal{E}})>_T\,dV=\iint_{\partial V}<\boldsymbol{\mathcal{J}_{tot}}\times\boldsymbol{\mathcal{E}}>_T\cdot\,\boldsymbol{\hat{n}}\,da\right.$, where $\boldsymbol{\hat{n}}$ is the unit vector$\left.\right)$ [99] can show that the divergence of the current density for any closed surface defining a volume will equal to zero under steady-state conditions. This is because the surface integral $\iint_{\partial V}<\boldsymbol{\mathcal{J}_{tot}}\times\boldsymbol{\mathcal{E}}>_T\cdot\,\hat{n}\,da$ represents the net flux of the time-averaged vector field $\boldsymbol{\mathcal{J}_{tot}}\times\boldsymbol{\mathcal{E}}$ through the closed surface $\partial V$, which will equate to zero when there are no sources in the volume $V$ [100,101]. Thus, the volume integral $\iiint_V<\nabla\cdot(\boldsymbol{\mathcal{J}_{tot}}\times\boldsymbol{\mathcal{E}})>_T$ will also equal to zero. Therefore,

$$\frac{1}{\mu_0}\iiint_V<\nabla\cdot\boldsymbol{\mathcal{F}}>_T\,dV=-\iiint_V<\boldsymbol{\mathcal{B}}\cdot\frac{\partial\boldsymbol{\mathcal{J}_{tot}}}{\partial t}>_T\,dV \tag{3.11}$$

Evaluating using the relation $\frac{\partial\mathcal{X}}{\partial t}=i\omega\boldsymbol{\mathcal{X}}$ for some time-harmonic field $\boldsymbol{\mathcal{X}}$, along with the time-averaged identity $<\boldsymbol{\mathcal{X}}\cdot\boldsymbol{\mathcal{Y}}>_T=\frac{1}{2}Re(\boldsymbol{\mathcal{X}}\cdot\boldsymbol{\mathcal{Y}}^*)$ for time-harmonic fields $\boldsymbol{\mathcal{X}}$ and $\boldsymbol{\mathcal{Y}}$ [96], and applying Gauss' theorem [99] to the term on the left, we obtain:

$$\iiint_V\nabla\cdot\mathrm{Re}(\boldsymbol{\mathcal{F}})dV=-\mu_0\omega_0\iiint_V Im(\boldsymbol{\mathcal{B}}\cdot\boldsymbol{\mathcal{J}}_{\mathrm{tot}}^*)dV \tag{3.12}$$

This result shows that a flux of optical chirality requires a nonzero $Im(\boldsymbol{\mathcal{B}}\cdot\boldsymbol{\mathcal{J}}_{tot}^*)$. Notably, Eqn. (3.12) is a general result valid for any material properties as $\boldsymbol{\mathcal{J}_{tot}}$ contains all primary and secondary sources. Analogous to how electromagnetic energy dissipation is governed by the interaction of $\boldsymbol{E}$ and $\boldsymbol{j}$ via the source term $\iiint_V\boldsymbol{E}\cdot\boldsymbol{j}\;dV$ found in the time-averaged Poynting's theorem [70], here the source or sink of optical chirality flux describing chiral generation is governed by the interaction of the magnetic induction field and current density – via the source term $\iiint_V Im(\boldsymbol{\mathcal{B}} \cdot \boldsymbol{\mathcal{J}}_{tot}^{*})dV$.

Eqn. (3.12) forms the theoretical foundation of this work. As an illustrative example, we consider the lowest-order term in the multipole expansion of the current density with respect to an origin $r_0$ [102], where $\boldsymbol{\mathcal{J}}_{tot} \approx -i\omega\boldsymbol{p}\delta(\boldsymbol{r} - \boldsymbol{r}_0)$. For this case, we can express the optical chirality flux as:

$$\iint_{\partial V} \mathrm{Re}(\boldsymbol{\mathcal{F}}) \cdot \hat{\boldsymbol{n}}\, da = -\frac{\mu_0\omega^2}{2} Re[\boldsymbol{p}^* \cdot \boldsymbol{\mathcal{B}}(r_0)] \tag{3.13}$$

where $\boldsymbol{p}$ is the dipole moment.

Eqn. (3.13) reveals that optical chirality flux can be generated whenever the dipole moment $\boldsymbol{p}$ has a nonzero projection with the local magnetic induction field $\boldsymbol{\mathcal{B}}(r_0)$. While magnetoelectric dipolar sources of chiral optical fields have been extensively studied [73,74], Eqn. (3.13) reveals previously unexplored potential of generating chiral optical fields from magnetic dipolar sources – achievable, for example, via optical magnetism [75–77]. In the next section, we explore this concept numerically using the setup introduced in Fig. 2.

### B. Numerical simulations for generating magnetic chirality

The reflectance spectra of the achiral and chiral nanostructures reveal distinct resonance characteristics under CPL (Fig. 3(a)). The achiral structures exhibit perfect overlap between clockwise and counterclockwise CPL in the reflectance spectra, indicating no optical chirality – as expected [3,37,45,46]. The square-shaped structures exhibit a single resonance peak at 550 nm, while the achiral L-shaped structure displays two distinct resonance peaks at 536 nm and 550 nm. In contrast, the chiral L-shaped structures exhibit three distinct reflectance resonances at 527 nm, ~539 nm, and 548 nm. The ~539 nm resonance shows a measurable wavelength shift between clockwise and counterclockwise CPL. For the left-handed chiral structures, the resonances appear at 539.9980 nm (clockwise CPL) and 539.0025 nm (counterclockwise CPL); for the right-handed structures, the resonances appear at 539.0156 nm (clockwise CPL) and 539.0113 nm (counterclockwise CPL). This small (~0.0044 nm) but consistent spectral discrepancy may be a characteristic feature of the system, originating from an asymmetric optical transition between the outer resonances at 527 nm and 548 nm. These outer peaks correspond to those observed in the achiral L-shaped structures, suggesting shared underlying modes. In contrast, the ~539 nm resonance is unique to the chiral structures. The presence of these resonances accompanies a transition in the electric field behavior, as will be discussed below.

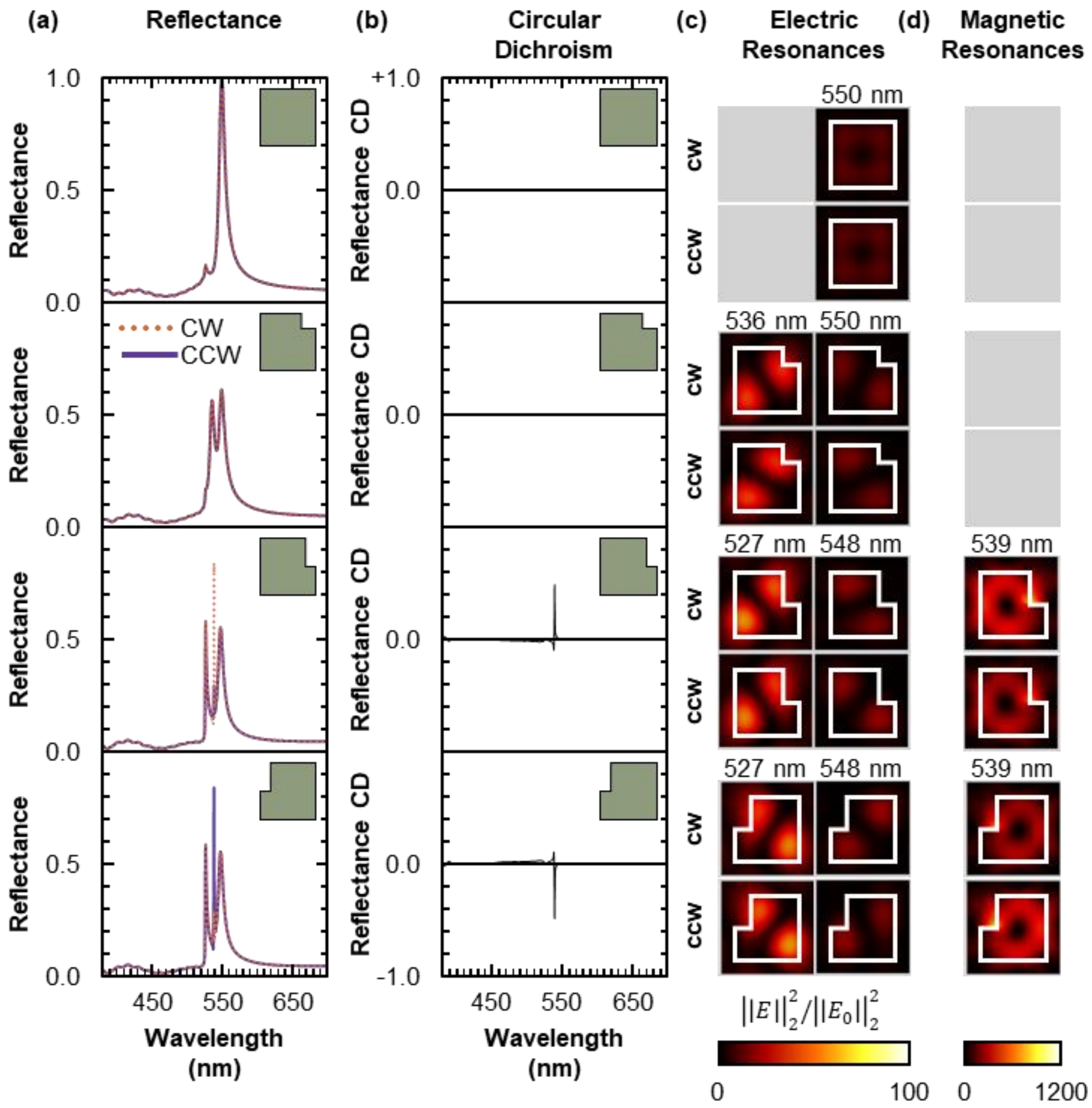

FIG 3. (a) Reflectance spectra for the square-shaped, achiral L-shaped, and both mirror images of the chiral L-shaped structures under broadband (380-700 nm) CPL illumination. Spectra for clockwise (CW; orange, dotted) and counterclockwise (CCW; purple, solid) CPL are shown. Reflectance spectra for the chiral L-shaped structures were taken at a step size of 0.0001 nm between 538.9 nm and 539.1 nm; outside of those ranges and for the achiral structures, the spectra were taken at steps of 1 nm. (b) Circular dichroism (CD) in reflectance was calculated from the reflectance spectra using the formula $(|CW| - |CCW|)/(|CW| + |CCW|)$. Only the chiral structures exhibit a nonzero CD signal. (c-d) Near-field electric field enhancement plots at resonance wavelengths, measured at the cross-section of the horizontal center plane through the nanoparticle, for the (c) electric and (d) magnetic resonances. White outlines delineate the nanoparticle boundaries under (Top Row) CW and (Bottom Row) CCW illumination. The color bar represents the value of the value of electric field enhancement. $E$ is the local electric field vector, $E_0$ is the incident electric field vector, and $||\cdot||_2$ is the Euclidean norm of a vector.

CD in the reflectance spectra further confirms the chiral nature of the middle resonance (Fig. 3(b)). The achiral structures exhibit zero CD, consistent with the perfect overlap of clockwise and counterclockwise plots. Conversely, the chiral structures show a differential reflectance of 0.47 at ~539 nm, along with smaller differences of 0.009 at 527 nm and 0.0045 at 548 nm. Additionally, we observe differentiable CD at the transition between these resonances, suggesting that the movement to and from the underlying modes is part of the overall chiral response. The left-handed chiral L-shaped structures exhibit a stronger middle peak under clockwise CPL compared to counterclockwise CPL, while the right-handed structures display the mirrored response. This confirms that the structural handedness directly influences the interaction of the system with CPL, supporting the existence of the ~539 nm resonances as that of a chiral response. Notably, the ~539 nm peaks have an approximate linewidth at the half-maximum of 0.22 nm, yielding a high-Q factor on the order of $10^6$.

Near-field examinations of the electric field enhancement provide insight into the physical origins of the observed reflectance and CD results. The square-shaped structure produces four antinodes (i.e., resonator mode peaks), while the achiral L-shaped structures exhibit two antinodes per resonance (Fig. 3(c)). This behavior stems from the fundamental composition of CPL, which consists of two orthogonal, quarter-wave phase-shifted linearly polarized components [16,78]. In the two-dimensionally-symmetric square-shaped structures, these components interfere constructively to form four antinodes at a single resonance. However, breaking the symmetry by introducing the L-shaped structures force each linear component to resonate at distinct wavelengths, which produces two separate antinodes per resonance. This behavior is typical of a standing wave that forms at the fundamental mode of a resonance when the nanoparticle dimensions are much larger than the wavelength of light [103,104]. However, because the wavelength approaches the size of our nanoparticles in this work, this phenomenon can be treated as a transient standing wave where the resonance leaks to the far field, i.e. a Mie resonance [103–107].

For the chiral L-shaped structures, the three resonances (527 nm, ~539 nm, and 548 nm) correspond to distinct electric field configurations. The outer resonances (527 nm and 548 nm) resemble those of the achiral structures, but the middle resonances (~539 nm) introduce a notable in-plane curling of the electric field (Fig. 3(d)). The transition from the 527 nm resonance to the ~539 nm resonances shift the electric field from an out-of-plane configuration to an in-plane rotating motion. This behavior reverses as the system transitions from the ~539 nm resonances to the 548 nm resonance, returning the electric field to an out-of-plane configuration that is oriented along the opposite diagonal from the original configuration at 527 nm. The asymmetric nature of this transition between the out-of-plane and in-plane field configurations under clockwise and counterclockwise CPL is attributed to the geometry of the presented chiral L-shaped structures.

The in-plane curling of the electric field at the ~539 nm resonance suggests that this far-field peak originates from an out-of-plane magnetic field component, as predicted by the Maxwell-Ampère Law [67]. Such behavior is similar to the first demonstrations of magnetic dipole resonances in the visible light using dielectric nanospheres [76,77] and to the phenomenon of optical magnetism [75].

To test this hypothesis, we qualitatively analyzed the near-field electric and magnetic induction field vectors at 539 nm in the achiral and ~539 nm in the chiral structures (Fig. 4(a)). The achiral structures exhibit mirror-symmetric or near-mirror-symmetric vector distributions between clockwise and counterclockwise CPL, with some out-of-plane electric field vectors but no out-of-plane magnetic induction components. In contrast, the chiral L-shaped structures display a rotating effect in the electric field vectors, with corresponding magnetic induction field vectors oriented distinctly under different CPL orientations. For example, in the left-handed chiral structures, clockwise CPL produces magnetic vectors pointing outward from the page ("toward the observer"; +z direction); whereas counterclockwise CPL reverses this direction to point inward ("away from the observer"; -z direction). The right-handed chiral structures show the opposite behavior. These observations indicate the generation of an out-of-plane magnetic field.

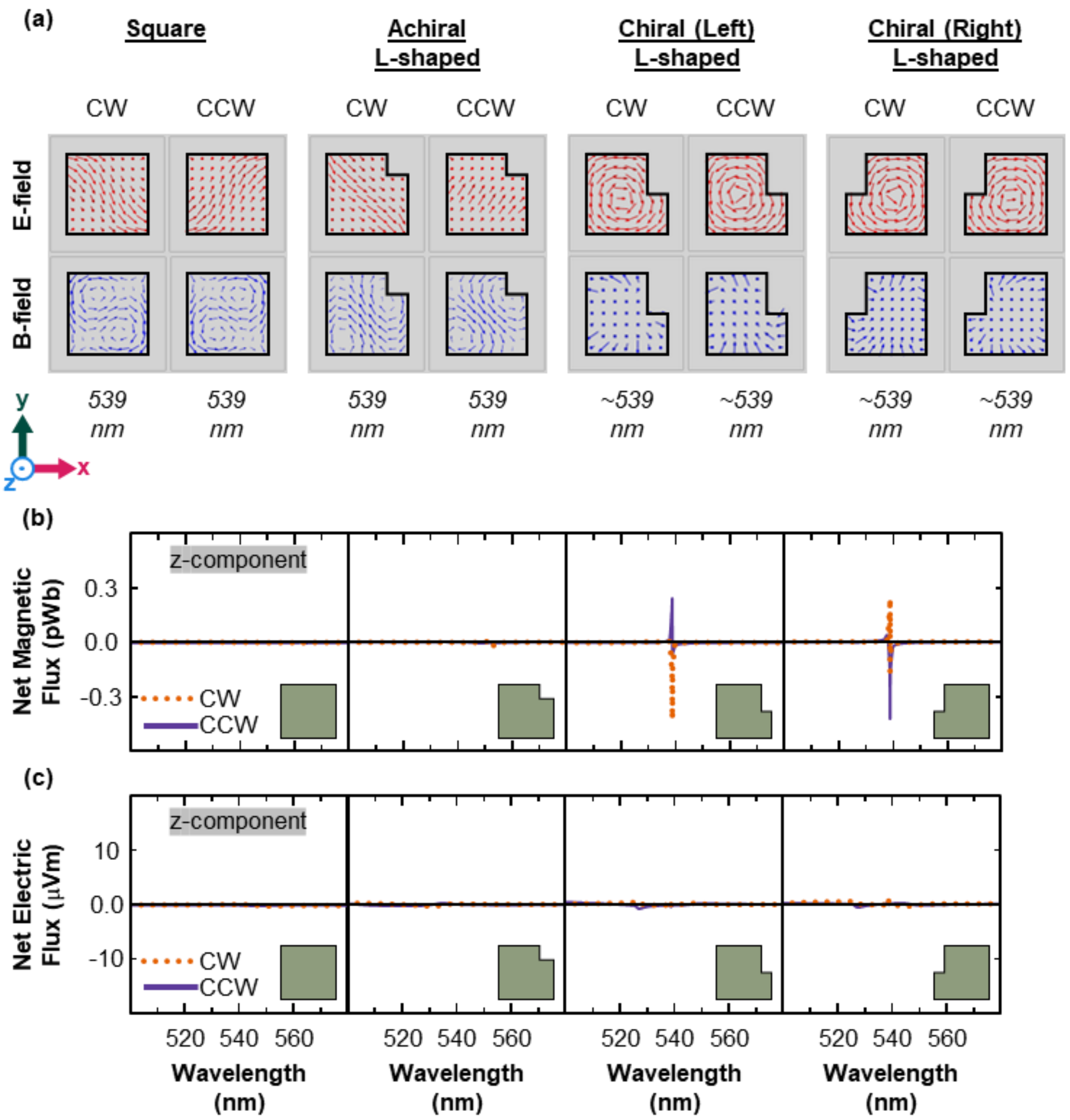

FIG 4. (a) (Top) Electric field (E-field, red arrows) and (Bottom) magnetic induction field (B-field, blue arrows) vectors at 539 nm for the achiral structures and ~539 nm for the chiral L-shaped structures, measured at the horizontal center plane through the nanoparticles, under clockwise (CW) and counterclockwise (CCW) illumination. Black outlines indicate nanoparticle boundaries. The ~539 nm resonances of the chiral (left) L-shaped structures were at 538.9980 under CW CPL and 539.0025 for CCW CPL. The ~539 nm resonances of the chiral (right) L-shaped structures were at 539.0156 nm for CW CPL and 539.0113 for CCW CPL. (b) Net magnetic flux and (c) net electric flux under clockwise (orange, dotted) and counterclockwise (purple, solid) CPL illumination for the (Left to Right) achiral square-shaped, achiral L-shaped, chiral (left) L-shaped, and chiral (right) L-shaped structures for z-component of the net flux. The x- and y-components of the net magnetic and net electric fluxes are shown in Supplementary Material Section S2.

To further validate this interpretation, we examined the out-of-plane (z-component) of the net magnetic flux (Fig. 4(b)). As expected, the achiral structures exhibit a near-zero net magnetic flux, indicating no preferential magnetic field orientation along the z-axis – consistent with their symmetry [3]. Conversely, the chiral structures exhibit a prominent peak at the ~539 nm resonance wavelengths, with oppositely oriented flux signals for clockwise and counterclockwise CPL. The result aligns with our previous observation that the magnetic induction field vectors are oriented in opposite directions between clockwise and counterclockwise CPL illuminations, which reinforces the suggestion that the ~539 nm peaks are signals generated by magnetic fields.

Observations of the out-of-plane (z-component) of the net electric flux further support this hypothesis (Fig. 4(c)). Similar to the net magnetic flux results, the achiral structures exhibit zero or near-zero net electric flux, as expected for non-chiral, symmetric systems interacting with CPL [3]. Unlike the net magnetic flux results, the chiral structures exhibit a near-zero net electric flux, implying the out-of-plane flux generated is not due to the electric field, supporting the idea that the chiral far-field field signals originate from magnetic fields.

Overall, the results suggest that the chiral L-shaped structures may induce a forced twisting of the electric field that generates an in-plane rotation within the nanoparticle.

To confirm, in Fig. 5. we analyzed the time-averaged optical chirality density ($\mathcal{C} = -\frac{\omega}{2} Im(\mathcal{D}^* \cdot \mathcal{B})$ [62,66]) – a measure of how strongly the structure twists light to generate a local chiral field [1,41]. We begin with a volume-integrated measurement for each nanoparticle design (Fig. 6(a)). In achiral structures, the optical chirality density plots for clockwise and counterclockwise CPL are mirrored, indicating no net chiral response. However, in chiral L-shaped structures, this symmetry is broken. A sharp peak emerges at the ~539 nm resonance in both clockwise and counterclockwise CPL plots, with both peaks skewed toward the same twisting direction. This suggests that, despite excitation with opposite input polarizations, the chiral structures enforce the same type of twisting on both CPL orientations. This directional enforcement correlates with the previously observed reflectance asymmetry, such that a higher reflectance is observed for the input polarization that aligns with the directionality of the enforced twisting. This twisting effect is most pronounced at ~539 nm, as shown by the difference graphs between the clockwise and counterclockwise optical chirality density plots (Fig. 6(b)). Between 527 nm and 548 nm, a nonzero signal indicates a lack of mirror symmetry, with the strongest amplitude difference occurring at ~539 nm. Outside this range, the signal returns to zero, reinforcing the idea that asymmetry in the chiral field of this system arises during the transition between electric and magnetic field resonances.

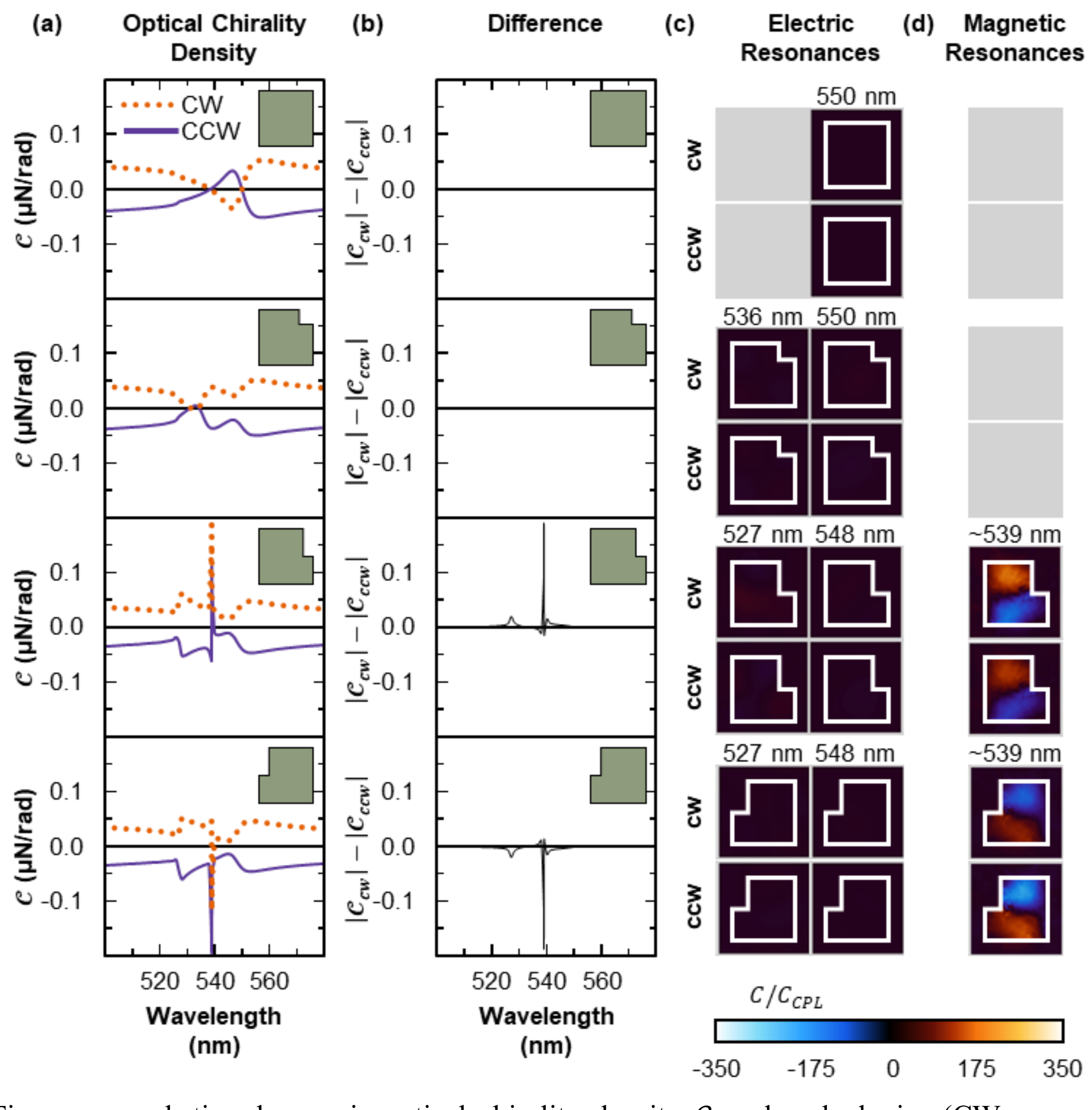

FIG 5. (a) Time-averaged, time-harmonic optical chirality density $\mathcal{C}$ under clockwise (CW; orange, dotted) and counterclockwise (CCW; purple, solid) CPL illumination for the achiral structures and both mirror images of the chiral L-shaped structures. $\mathcal{C} = -\frac{\omega}{2} Im(\mathcal{D}^* \cdot \mathcal{B})$, where $\mathcal{D}$ is the time-harmonic electric displacement and $\mathcal{B}$ is the time-

harmonic magnetic induction. (b) The difference between the absolute optical chiral density under clockwise CPL ($|\mathcal{C}_{cw}|$) and the absolute optical chiral density under counterclockwise CPL ($|\mathcal{C}_{ccw}|$). (c-d) Near-field time-averaged, time-harmonic optical chirality density enhancements at resonance wavelengths, measured at the cross-section of the horizontal center plane through the nanoparticle, for the (c) electric and (d) magnetic resonances. White outlines delineate the nanoparticle boundaries under (Top Row) CW and (Bottom Row) CCW CPL illumination. The color bar represents the normalized optical chiral density enhancement relative to the incident CPL illumination, using the time-averaged, time-harmonic formulation of the optical chirality density.

Next, we examined the spatial enhancement of the time-averaged optical chirality density at the resonances across all four structures (Fig. 5(c-d)). The optical chirality density enhancement of the chiral L-shaped structures is an order of magnitude greater at the ~539 nm resonance when compared to either of the outer peaks (527 nm or 548 nm) or any resonances from the achiral structures. Also, at ~539 nm: the enhancement plots reveal an almost-even split through the middle of the nanostructure between the directions of twisting, where the rotating electric field results in opposite twists – left-to-right then right-to-left.

Notably, although Fig. 4(a) shows the electric field vectors for the clockwise and counterclockwise CPL illuminations to rotate in opposite directions in the chiral L-shaped structure, this is not a contradiction to our findings in Fig. 5 that show an enforced twisting directionality. To resolve this, it is crucial to recognize that the time-averaged, time-harmonic optical chirality density calculation involves the material response via both electric displacement ($\mathcal{D}$) and magnetic induction ($\mathcal{B}$) fields [62,66]: $\mathcal{C} = -\frac{\omega}{2} Im(\mathcal{D}^* \cdot \mathcal{B})$. Thus, field vector movements in Fig. 4(a) cannot, in their isolation, be directly compared to the calculated optical chirality density. The twisting observed in Fig. 5 arises from the joint interaction of the $\mathcal{D}$ and $\mathcal{B}$ fields; from which it follows that the switching in electric field polarization and the switching in magnetic field directionality contribute to the observed matching signs of the optical chirality density for clockwise and counterclockwise CPL illumination.

The continuity equation introduced in Eqn. (3.12) establishes the relationship between the optical chirality flux and the source term that enables optical chirality generation or dissipation. Because there is a non-symmetric response between 527 nm and 548 nm in the chiral L-shape structures, then a similar asymmetric response should be seen in the optical chirality flux and optical chirality generation. Previous studies have shown that the $S_3$ Stokes parameter – when measured in the far-field – serves as a direct signature of optical chirality flux [61,62]. This is because the time-averaged, time-harmonic form of optical chirality flux $\boldsymbol{\mathcal{F}}$ can be expressed as a difference in the power flux of clockwise ($\boldsymbol{S}_{CW}$) and counterclockwise ($\boldsymbol{S}_{CCW}$) CPL: $\boldsymbol{\mathcal{F}} = \frac{\omega}{c}(|CW|^2 \boldsymbol{S}_{CW} - |CCW|^2 \boldsymbol{S}_{CCW})$, where $|CW|^2$ and $|CCW|^2$ are weighting factors of clockwise and counterclockwise CPL, respectively [61]. This expression directly corresponds to the definition of the $S_3$ parameter, which quantifies the intensity difference between clockwise and counterclockwise CPL [87,89].

As expected, the achiral structures exhibit perfectly mirror-symmetric $S_3$ responses for clockwise and counterclockwise CPL (Fig. 6(a)). However, unlike the square-shaped structures, the achiral L-shaped structures introduce switching in light handedness in between the resonances. Likewise, the chiral L-shaped structures also display switching in the light handedness. However, the chiral L-shaped structures have a more complex switching effect that is asymmetric between clockwise and counterclockwise CPL. Focusing on the left-handed chiral L-shape structures (while noting that the right-handed chiral L-shape structures will exhibit the mirrored behavior): (i) under clockwise CPL, the light handedness fully switches twice and partially once, with the transition at ~539 nm being incomplete (i.e. a "partial" switch); while (ii) under clockwise CPL, the light handedness fully switches three times, with a complete transition at ~539 nm. The first switch occurs at the shorter-wavelength outer peak (527 nm), corresponding to an electric field resonance, where the clockwise CPL converts to counterclockwise CPL, and vice versa. At the middle resonance (~539 nm), the clockwise CPL plot (now a counterclockwise far-field signal) exhibits a sharp peak but does not fully transition back to clockwise, resulting in the "partial-switch" effect. Conversely, the counterclockwise CPL plot (now a clockwise far-field signal) undergoes a full transition. The final switch happens at the longer-wavelength outer peak (548 nm), where both plots return to their original handedness. This behavior resembles that of electromagnetically induced transparency [108]; however, instead of transparency in terms of energy transmission, our system manifests this phenomenon as selective transparency for one handedness of chirality within a narrow bandwidth. Thus, the system functions as a chiral switch that only allows a specific CPL orientation to pass at a distinct resonance wavelength.

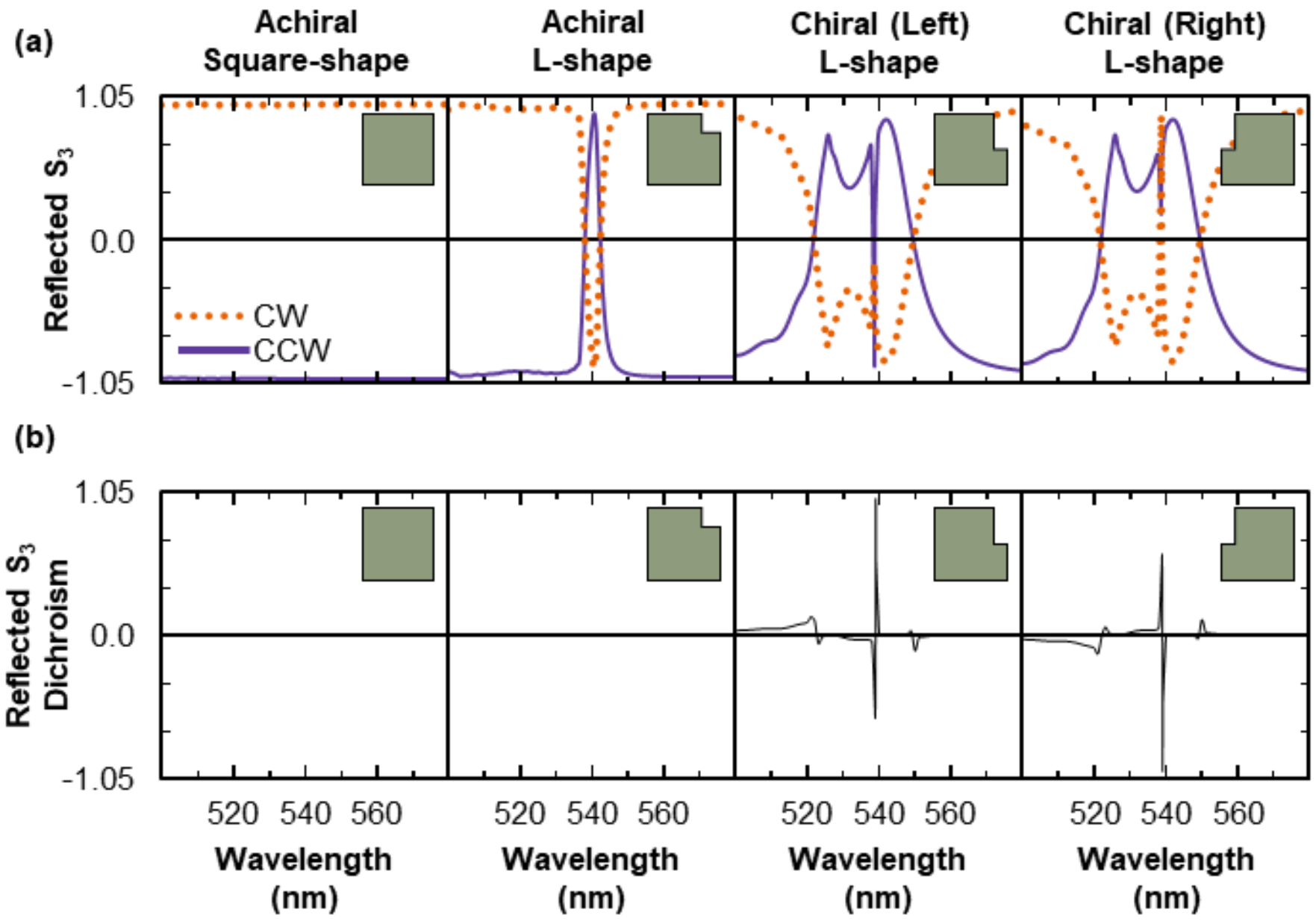

FIG 6. (a) Reflected $S_3$ Stokes parameter under clockwise (CW; orange, dotted) and counterclockwise (CCW; purple, solid) CPL illumination for the achiral structures and both mirror image of the chiral L-shaped structures. (b) The dichroic signal for the reflected $S_3$ Stokes parameter calculated as the normalized difference between clockwise and counterclockwise values: $\left(|S_{3_{CW}}| - |S_{3_{CCW}}|\right)/\left(|S_{3_{CW}}| + |S_{3_{CCW}}|\right)$.

The $S_3$ results reinforce the observations from the optical chirality density data: at the ~539 nm resonances, both clockwise and counterclockwise CPL experience the same handedness of twisting. Notably, it is the orientation of CPL that is forced to transition to the opposite handedness that correlates with the lower-amplitude reflectance signal observed in the far-field spectra of Fig. 3. To quantify this effect, we take the dichroic signal of the $S_3$ plots (Fig. 6(b)). The achiral structures show a zero signal, indicating perfectly mirrored behaviors. As for the chiral L-shaped structures, the result is a sharp, bisignate peak at ~539 nm; for the left-handed chiral structure, the signal first drops to -0.6 before jumping to +1, and vice versa for the right-handed chiral L-shaped structures. This behavior resembles the CD plot from Fig. 3(b), and the small peaks flanking the bisignate feature align with the differential reflectance signals observed at the outer resonances of 527 nm and 548 nm.

So far, we have shown that the chiral L-shaped structure exhibits a non-mirror-symmetric twisting of light at the ~539 nm magnetic field resonance; this is captured in the optical chiral density and the $S_3$ Stokes parameter. The time-averaged, time-harmonic optical chiral density, $\mathcal{C} = -\frac{\omega}{2} Im(\mathcal{D}^* \cdot \mathcal{B})$, is material-dependent and is not a conserved quantity [62,66]. In contrast, the optical chirality density that is explicitly found in the standard form of the optical chirality continuity equation (Eqn. (3.1)) does not account for material currents [1,65].

Likewise, the optical chirality flux is conserved in the standard form of the optical chirality continuity equation (Eqn. (3.1)). However, Fig. 6 represents a proportional measurement of the far-field, time-averaged, and time-harmonic optical chirality flux $\boldsymbol{\mathcal{F}} = \frac{\omega}{c}\left(|CW|^2 \boldsymbol{\mathcal{S}}_{CW} - |CCW|^2 \boldsymbol{\mathcal{S}}_{CCW}\right)$. This is an experimentally measurable quantity via the reflected $S_3$ Stokes parameter (in the +z direction) [61,62]. Note that the optical chirality flux is only a conserved quantity, fulfilling the optical chirality continuity equation, when integrated over a closed surface [1,65].

In contrast, our framework – rooted in the optical chirality continuity equation (Eqn. (3.12)) – introduces a source term that allows us to quantitatively track the generation of optical chirality. Fig. 7(a) compares this term under illumination with clockwise and counterclockwise CPL. Both achiral structures reveal mirror symmetric plots between the clockwise and counterclockwise CPL illuminations, implying an equal amount of conversion of chiral light into chiral material currents [61,109,110], which matches the lack of chiral differentiation observed so far. Unlike the graphs of the achiral structures, the graph of the chiral L-shaped structures shows oppositely oriented peaks at the

~539 nm resonance wavelengths between the clockwise and counterclockwise CPL illuminations. This is in accordance with the reflectance, CD, optical chirality density, and $S_3$ plots. Furthermore, the chiral L-shaped structures show the greatest differential response at the ~539 nm peaks (Fig. 7(b)), further showcasing that this magnetic field resonance has the greatest asymmetry in optical chirality generation, acting as an illustrative example of the mechanism behind the source of optical chirality shown in the chirality equation (Eqn. 3.6). Thus, with this result, we were able to show that the emergence of a chiral signal can emerge from an optical magnetism effect [75–77] that only requires the contributions of a source current and the magnetic induction field.

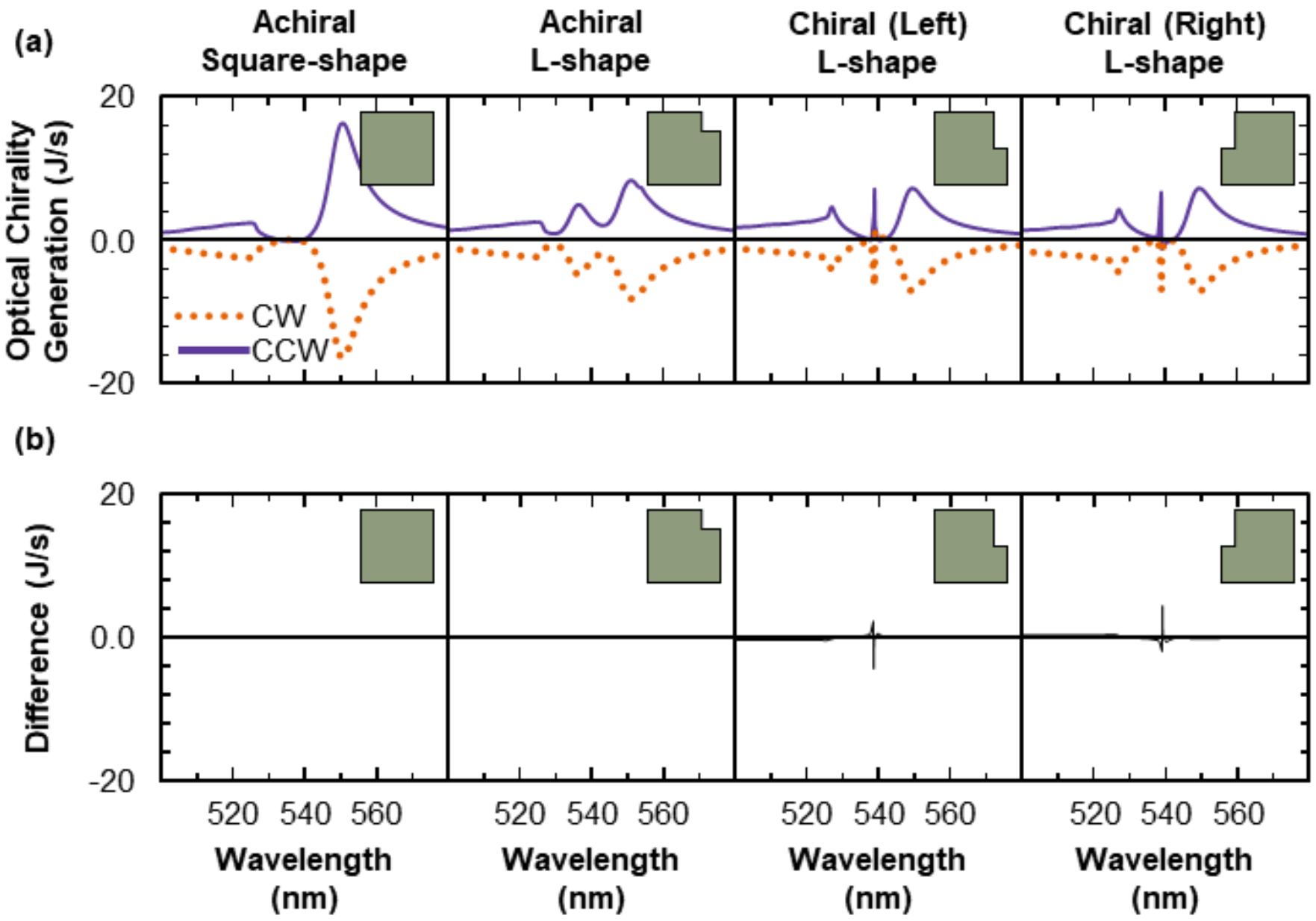

FIG 7. (a) Optical chirality generation under clockwise (CW; orange, dotted) and counterclockwise (CCW; purple, solid) CPL illumination for the achiral structures and both mirror image of the chiral L-shaped structures. The optical chirality generation is volume-integrated measurement of the nanoparticle. (b) The difference between the absolute values under CW CPL and the absolute values under CCW CPL for the optical chirality generation.

## V. CONCLUSIONS

In this work, we introduced a theoretical framework based on the optical chirality continuity equation [1], which demonstrates that an optical chirality flux signal can, in principle, solely arise from the contributions of source current and the magnetic induction. We illustrated these theoretical findings by designing an all-dielectric, lossless metasurface that generates an in-plane rotating electric field, inducing an out-of-plane magnetic field. Our numerical analysis confirmed this effect. Ultimately, we demonstrate the underlying physical mechanism from which optical chirality flux can be generated by optical magnetism [75–77].

However, our metasurface geometry was not fully optimized. The reflectance spectra revealed resonance signals for both clockwise and counterclockwise CPL at the magnetic field resonance, whereas an ideal structure would suppress the signal for one polarization while enhancing it for the other. Further evidence of this suboptimal design is seen in the asymmetries between the electric field vector distributions and the differing magnitudes of the out-of-plane magnetic flux components. Future work can focus on optimizing the metasurface geometry – either through computational algorithms or experimental refinement – to achieve maximal CD signals and an ideal chiral response. Nevertheless, these limitations do not diminish the primary finding: a chiral signal can be generated by optical magnetism [75–77].

Our numerical models did not incorporate smoothed corners to simulate nanofabrication artifacts. Prior studies suggest such artifacts can reduce near-field enhancements by having smoothed corners [111], lower peak amplitudes, and broadened resonance linewidths [90,112]. This effect is well reported for the electric field resonances, but we also expect it to appear for the magnetic field resonances. We do note that, in the creation of the magnetic field resonance, we realized that it is highly sensitive to structural perturbations in our system. This sensitivity implies that with sufficiently precise nanofabrication techniques, experimental realization of this effect is possible; though, large enough fabrication imperfections may pose challenges. However, the purpose of this work is not fabrication, but rather the theoretical introduction of a methodology for generating a magnetic source of chirality.

Looking ahead, exploring alternative structured photonic media to replicate this effect could establish a new class of chiral nanostructures. The metasurfaces we designed exhibit similarities to electromagnetically induced transparency [108] – but instead of observing narrow bandwidth transparency for energy transmission, we observe it for a specific CPL orientation. Unlike electromagnetically induced transparency, our theoretical framework does not inherently require the resonance to be confined between two others. Instead, if an in-plane rotating electric field can be made in a chiral system (as seen in optical magnetism [75–77]), the underlying introduced principles should remain true.

This study provides a new perspective on how to design chiral structures, reveals new avenues for optimizing chiral structures to achieve stronger dichroic signals, and supports a new path to generate chiral optical fields based on magnetic sources.

## ACKNOWLEDGMENTS

Z.H. and L.V.P. gratefully acknowledge funding from the Arnold and Mabel Beckman Foundation (Beckman Young Investigator Award, Project Number: 30155266). Z.H. thanks Saaj Chattopadhyay for providing feedback on the manuscript and Paula Kirya for helpful comments on designing figures.

**Magnetic Chiral Light: Supplementary Material**

Zaid Haddadin,[1] Lukas Novotny,[2] and Lisa V. Poulikakos[3,4,*]

[1]Department of Electrical & Computer Engineering, UC San Diego, La Jolla, CA 92093-0021, USA
[2]Photonics Laboratory, ETH Zürich, CH-8093 Zürich, Switzerland
[3]Department of Mechanical & Aerospace Engineering, UC San Diego, La Jolla, CA 92093-0021, USA
[4]Program of Materials Science & Engineering, UC San Diego, La Jolla, CA 92093-0021, USA
[*]Contact author: lpoulikakos@ucsd.edu

**CONTENTS**

## S1. COMSOL MULTIPHYSICS V6.0 IMPLEMENTATION

The COMSOL Multiphysics v6.0 base software [1] and corresponding Wave Optics Module [2] were used for this work. The COMSOL file was made using the Model Wizard by selecting a "3D" space dimension, adding an "Electromagnetic Waves, Frequency Domain" physics interface, and creating an "Empty Study".

The instructions in this section concern the design of a simulation file for the left-handed chiral L-shaped structures presented in the main text, which is provided as a supplemental material and can be downloaded from Ref. [3]. Similar methodologies were followed to make all the other structures; the only differences in methodology have to do with the geometry of the nanostructure shape. Data extracted from COMSOL for all structures may be found in Ref. [3].

This section is split into sub-sections reflecting COMSOL's default "Model Builder" groupings (Fig. S01): A. Global Definitions; B. Component; C. Study; and D. Results. The "B. Component" sub-section will detail additional, default sub-groupings: B.1. Definitions; B.2. Geometry; B.3. Materials; B.4. Electromagnetic Waves, Frequency Domain; and B.5. Mesh.

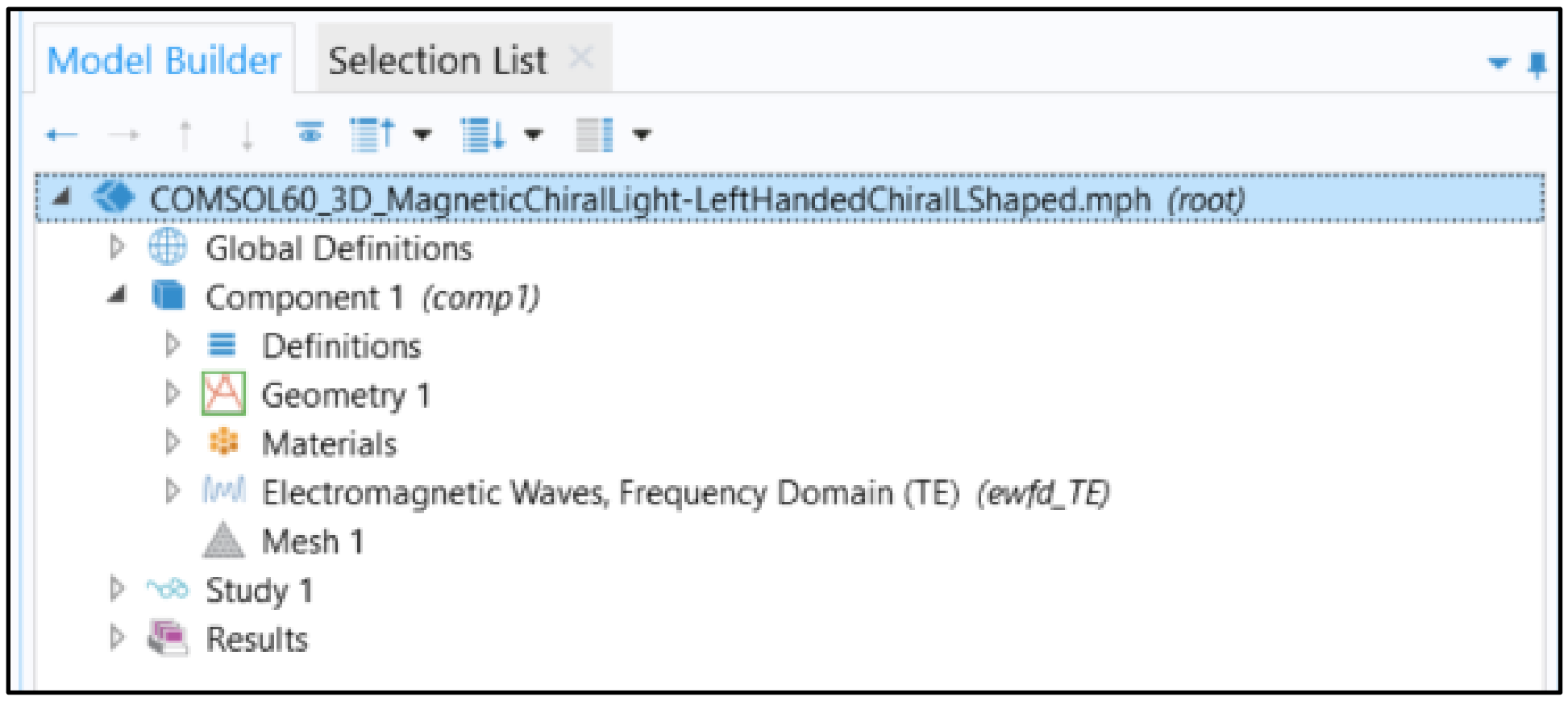

FIG S01. A screenshot of the default Model Builder groupings when creating a COMSOL Multiphysics v6.0 file using the Model Wizard by selecting a "3D" space dimension, adding an "Electromagnetic Waves, Frequency Domain" physics interface, and creating an "Empty Study". Within this screenshot, "COMSOL60_3D_MagneticChiralLight-LeftHandedChiralLshaped.mph" is the file name and "Electromagnetic Waves, Frequency Domain (TE)" is a renaming of the default "Electromagnetic Waves, Frequency Domain".

This section will move through the sub-sections in the sequence that they appear in COMSOL's Model Builder. However, we note that building the file is not necessarily linear like that. For example, we may define certain parameters early on whose purpose will not be realized until later.

### A. Global Definitions

In Global Definitions, users can add features that apply to the entire model. This is like how a global variable can be defined in many programming languages. We create three "Parameters" layers and appropriately label them (Fig. S02): "Model Parameters", for defining variables related to the construction of the entire 3D model; "Nanostructure Parameters", for defining variables related to the geometry of the nanostructure; and "Optics Parameters", for defining variables related to the incident illumination of light. The variables, expressions, values, and descriptions defined for each variable in the three parameters layers are provided in Tables S01-S03.

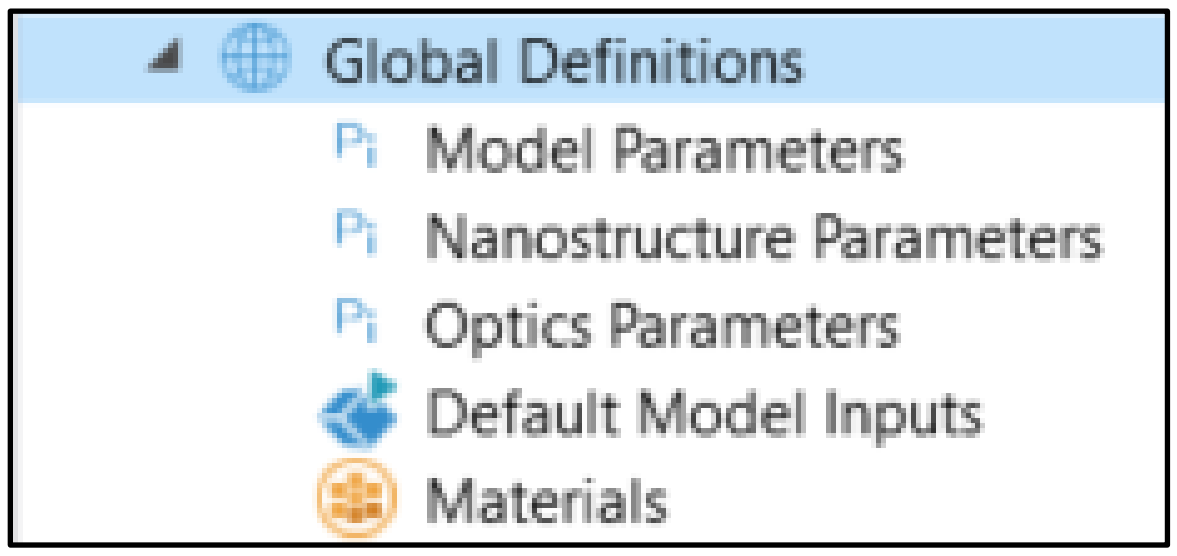

FIG S02. A screenshot of the Global Definitions groupings. "Default Model Inputs" and "Materials" are default created by the software and ignored for this work. The three parameters layers are named "Model Parameters", "Nanostructure Parameters" and "Optics Parameters".

TABLE S01. Model Parameters

| Name | Expression | Value | Description |
|---|---|---|---|
| domain_depth | lattice_depth_period | 3.4042E-7 m | The depth of the entire domain. |
| domain_height | incident_height + transmitted_height + nanostructure_thickness | 4.27E-6 m | The height of the entire domain. This is composed of all parts of the model. |
| domain_width | lattice_width_period | 3.4042E-7 m | The width of the entire domain. |
| incident_height | 2000 [nm] | 2E-6 m | The height of the domain through which light is incident towards the thin film. |
| transmitted_height | 2000 [nm] | 2E-6 m | The height of the domain through which light is transmitted after passing through the thin film. |
| lattice_width_period | nanostructure_pitch_width + interresonator_gap | 3.4042E-7 m | The interval of distance between successive repetitions of the nanoresonator along the width of the nanostructure. |
| lattice_depth_period | nanostructure_pitch_depth + interresonator_gap | 3.4042E-7 m | The interval of distance between successive repetitions of the nanoresonator along the depth of the nanostructure. |
| lattice_circumradius | sqrt((lattice_depth_period)^2 + (lattice_width_period)^2) / 2 | 2.4071E-7 m | Radius of the circle circumscribing the domain on the xy-plane. |

TABLE S02. Nanostructure Parameters

| Name | Expression | Value | Description |
|---|---|---|---|
| nanostructure_thickness | 270 [nm] | 2.7E-7 m | The thickness (or height) of the nanostructure. |
| nanostructure_pitch_width | nanostructure_pitch_depth * aspect_ratio_factor | 2.4042E-7 m | The width of the nanostructure. |
| nanostructure_pitch_depth | sqrt((170 [nm] * 340 [nm]) / aspect_ratio_factor) | 2.4042E-7 m | The depth of the nanostructure. The 170 nm and 340 nm numbers refer to the dimensions of a 1:2 aspect ratio shape, which will be the constrain to define all other aspect ratios. |
| interresonator_gap | 100 [nm] | 1E-7 m | The gap between two adjacent nanostructures. |
| aspect_ratio_factor | 1 | 1 | This factor is x in 1:x, assuming the shorter end of the structure will always be the "1". |
| nanostructure_cornercut_depth | nanostructure_pitch_depth / cornercut_ratio_factor_depth | 6.0104E-8 m | The depth of the cut that'll happen at the nanostructure's corner. |
| nanostructure_cornercut_width | nanostructure_pitch_width / cornercut_ratio_factor_width | 1.2021E-7 m | The width of the cut that'll happen at the nanostructure's corner. |
| cornercut_ratio_factor_depth | 4 | 4 | This is what the depth of the nanostructure should be divided by to get the corner cut depth. |
| cornercut_ratio_factor_width | 2 | 2 | This is what the length of the nanostructure should be divided by to get the corner cut depth. |

TABLE S03. Optics Parameters

| Name | Expression | Value | Description |
|---|---|---|---|
| wavelength_min* | 538.9 [nm] | 5.389E-7 m | The minimum value the wavelength of light can take for this study. |
| wavelength_max* | 539.1 [nm] | 5.391E-7 m | The maximum value the wavelength of light can take for this study. |
| wavelength_step* | 0.0001 [nm] | 1E-13 m | The steps taken in incrementing/decrementing the wavelength when running the simulation. |
| polarisation_rotation_angle** | 90 [deg] | 1.5708 rad | The trigger between CW and CCW CPL |

*The wavelength_min, wavelength_max, and wavelength_step can be altered depending on the needed resolution of the simulation. For example, most simulations were done between 380 nm and 700 nm with a step size of 1 nm. However, around the magnetic resonance, the resolution was increased to a step size of 0.0001 nm to better capture the behavior of this resonance. The 0.0001 nm step size was chosen as that was where the behavior converged and stopped changing, after testing increasing step sizes: 1 nm, 0.1 nm, 0.01 nm, 0.001 nm, 0.0001 nm.
**The polarization_rotation_angle can be set to -90 deg or 90 deg to switch between clockwise (CW) and counterclockwise (CCW) circularly polarized light (CPL), respectively.

## B. Component

### *B.1. Definitions*

Like Global Definitions, features can be added; however, unlike Global Definitions, the features in Definitions act more like local variables and are only applicable to the layers within the component. Fig. S03 shows the four variables, four selections, and two integration layers that were created under Definitions.

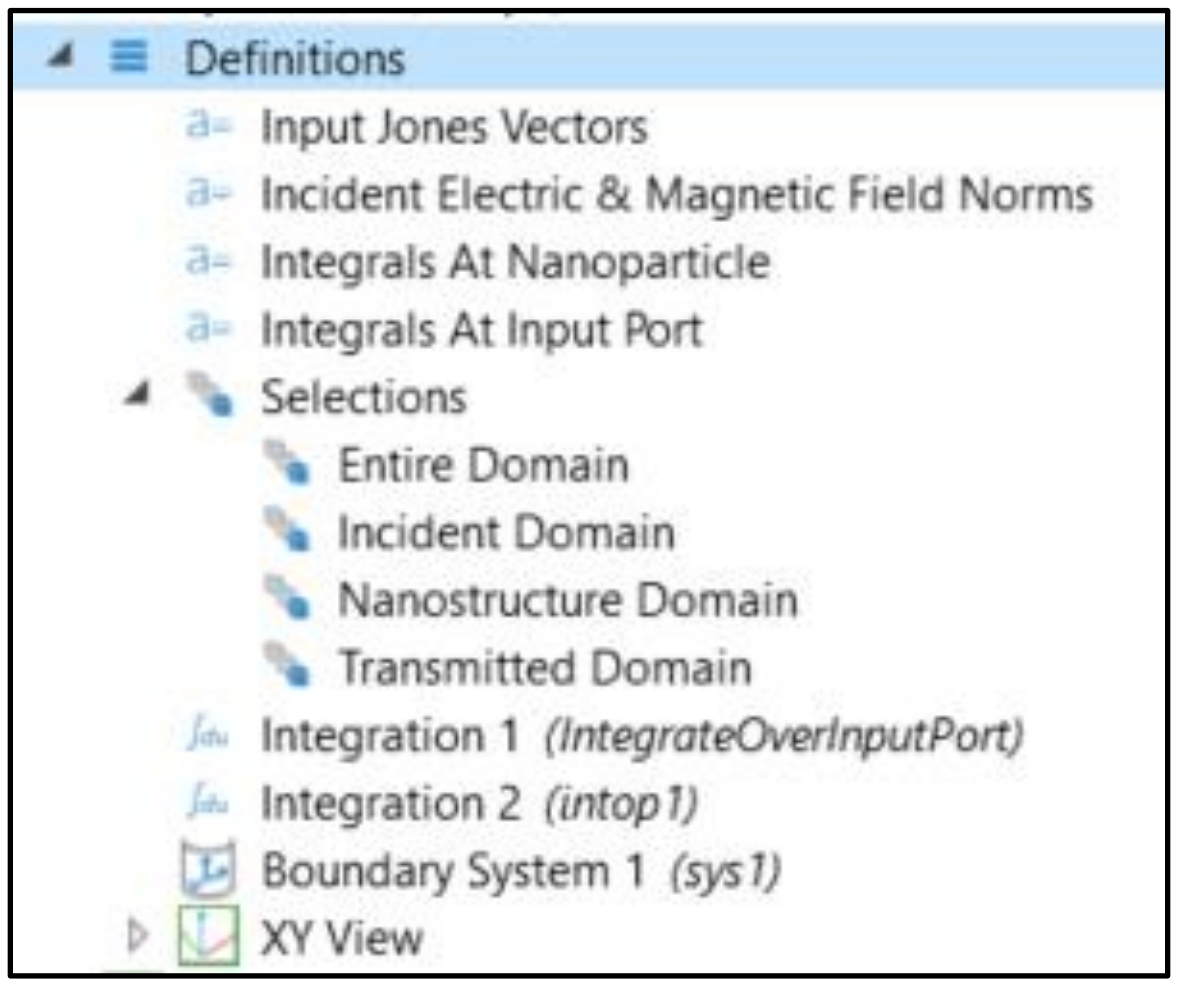

FIG S03. A screenshot of the Definitions groupings. “Boundary System 1” and “XY View” are default created by the software and ignored for this work. Four variables, four selections, and two integration layers were created as part of this work.

The “Input Jones Vectors” variables layer defined the incident polarization of light (Table S04); the “Incident Electric & Magnetic Field Norms” variables layer is used to retrieve the norms of the incident electric and magnetic fields by performing an integral over the input port (Table S05); and, the “Integrals At Nanoparticle” and “Integrals At Input Port” variables layers define a portion of the equations used to calculate the optical chirality generation and optical chirality density at the nanoparticle (Table S06) and in reflection at the input port (Table S07).

TABLE S04. Input Jones Vectors

| Name | Expression | Value | Description |
|---|---|---|---|
| E0x | 1 / sqrt(2) | | The x-component of the electric field amplitude. |
| E0y* | (1 / sqrt(2)) * i * sin(polarisation_rotation_angle) | | The y-component of the electric field amplitude. |

*By switching polarisation_rotation_angle between -90 deg and 90 deg in Global Definitions, it becomes possible to get the Jones Vector for clockwise and counterclockwise circularly polarized light, respectively.

TABLE S05. Incident Electric & Magnetic Field Norms

| Name | Expression | Value | Description |
|---|---|---|---|
| IncidentElectricFieldNorm | IntegrateOverInputPort(ewfd_TE.normE) / (domain_depth * domain_width) | V/m | The norm of the incident electric field. |
| IncidentMagneticFieldNorm | IntegrateOverInputPort(ewfd_TE.normH) / (domain_depth * domain_width) | A/m | The norm of the incident magnetic field. |

TABLE S06. Integrals At Nanoparticle

| Name | Expression | Value | Description |
|---|---|---|---|
| chiral_generation_NP | intop1(imag(conj(ewfd_TE.Jx) * ewfd_TE.Bx)) + intop1(imag(conj(ewfd_TE.Jy) * ewfd_TE.By)) + intop1(imag(conj(ewfd_TE.Jz) * ewfd_TE.Bz)) | N | Optical chirality generation integrated over the nanoparticle |
| chiral_density_NP | intop1(imag(conj(ewfd_TE.Dx) * ewfd_TE.Bx)) + intop1(imag(conj(ewfd_TE.Dy) * ewfd_TE.By)) + intop1(imag(conj(ewfd_TE.Dz) * ewfd_TE.Bz)) | N·s | Optical chirality density 4integrated over the nanoparticle |

TABLE S07. Integrals At Input Port

| Name | Expression | Value | Description |
|---|---|---|---|
| chiral_generation_InpP | IntegrateOverInputPort(imag(conj(ewfd_TE.Jx) * ewfd_TE.Bx)) + IntegrateOverInputPort(imag(conj(ewfd_TE.Jy) * ewfd_TE.By)) + IntegrateOverInputPort(imag(conj(ewfd_TE.Jz) * ewfd_TE.Bz)) | N/m | Optical chirality generation integrated over the input port. |
| chiral_density_InpP | IntegrateOverInputPort(imag(conj(ewfd_TE.Dx) * ewfd_TE.Bx)) + IntegrateOverInputPort(imag(conj(ewfd_TE.Dy) * ewfd_TE.By)) + IntegrateOverInputPort(imag(conj(ewfd_TE.Dz) * ewfd_TE.Bz)) | kg/s | Optical chirality density 4integrated over the input port. |

The “Selections” layers allow us to define specific areas of our model for easy reference (Fig. S04). The “Integration 1” and “Integration 2” layers allow us to select specific boundaries or volumes in our model that can be easily referenced to perform integral calculations at (Fig. S05).

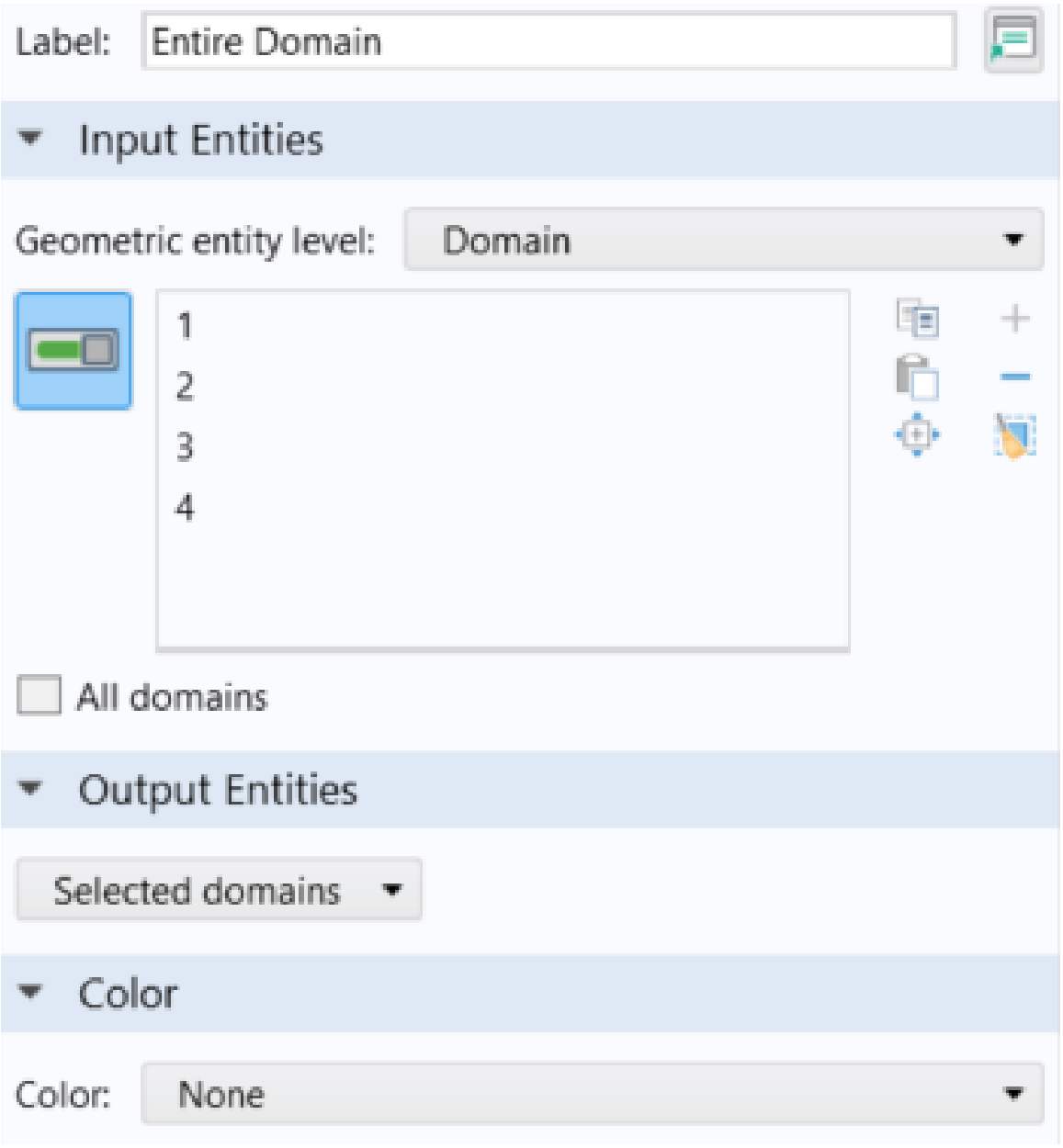

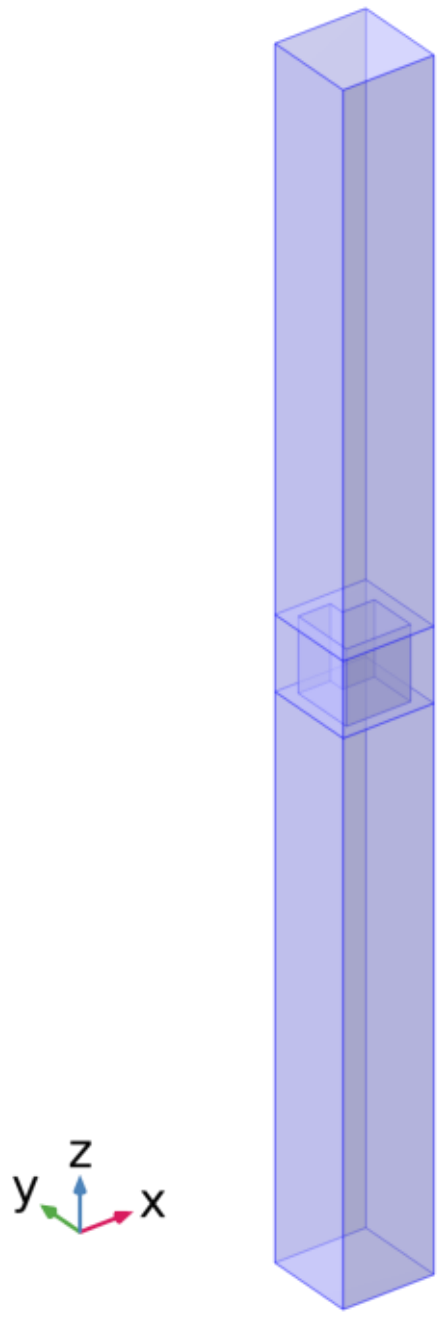
z
y
x

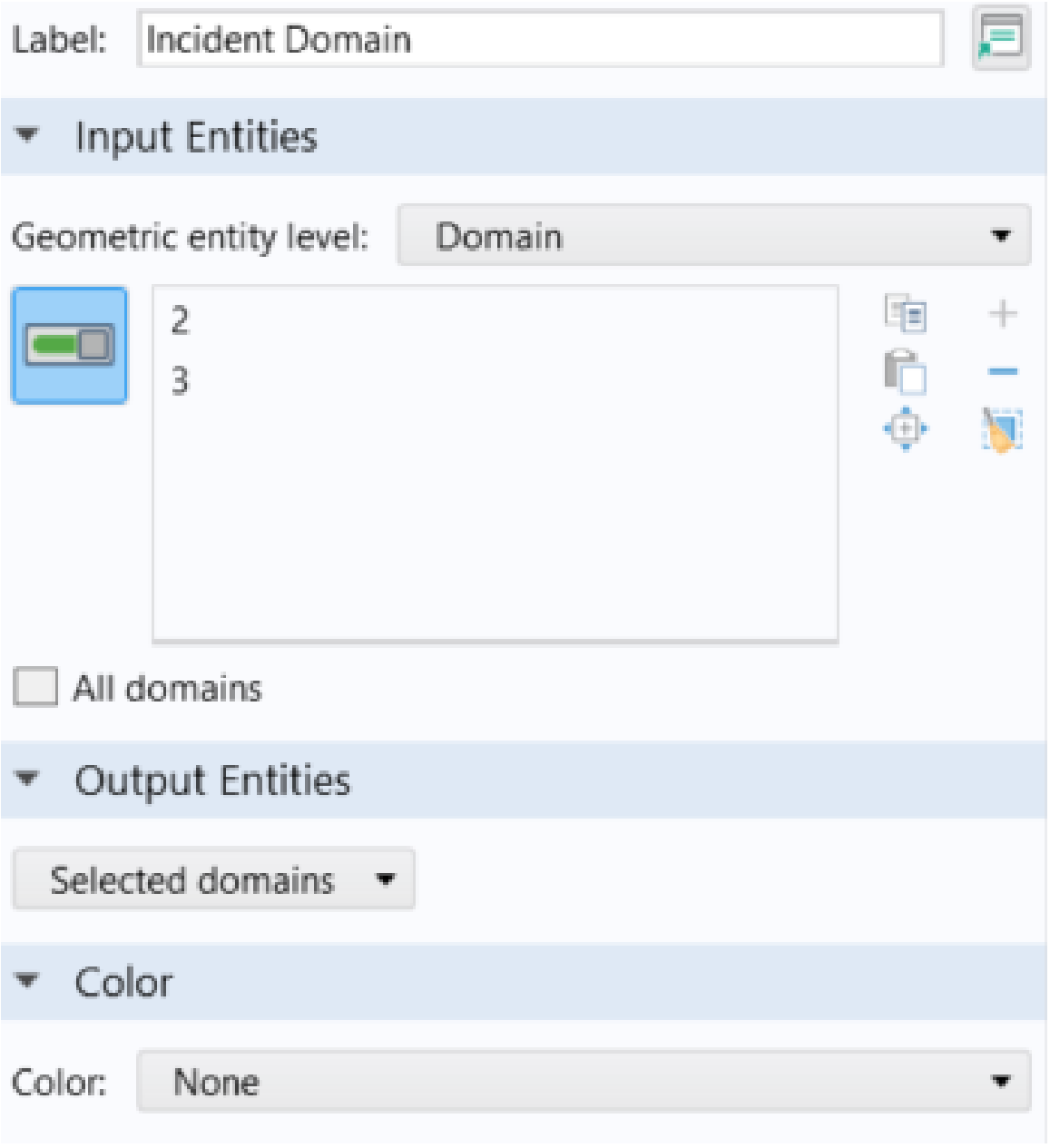

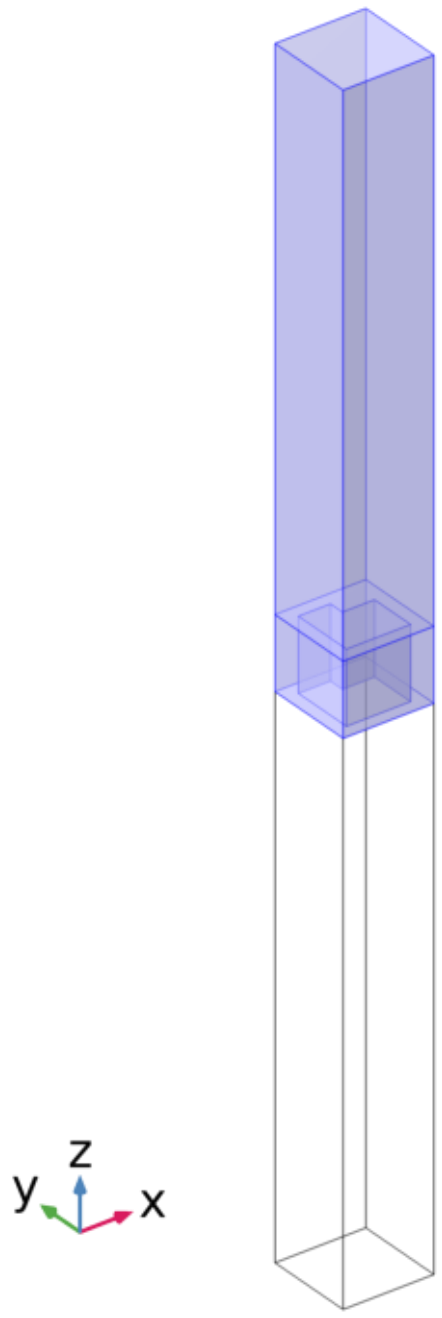
z
y
x

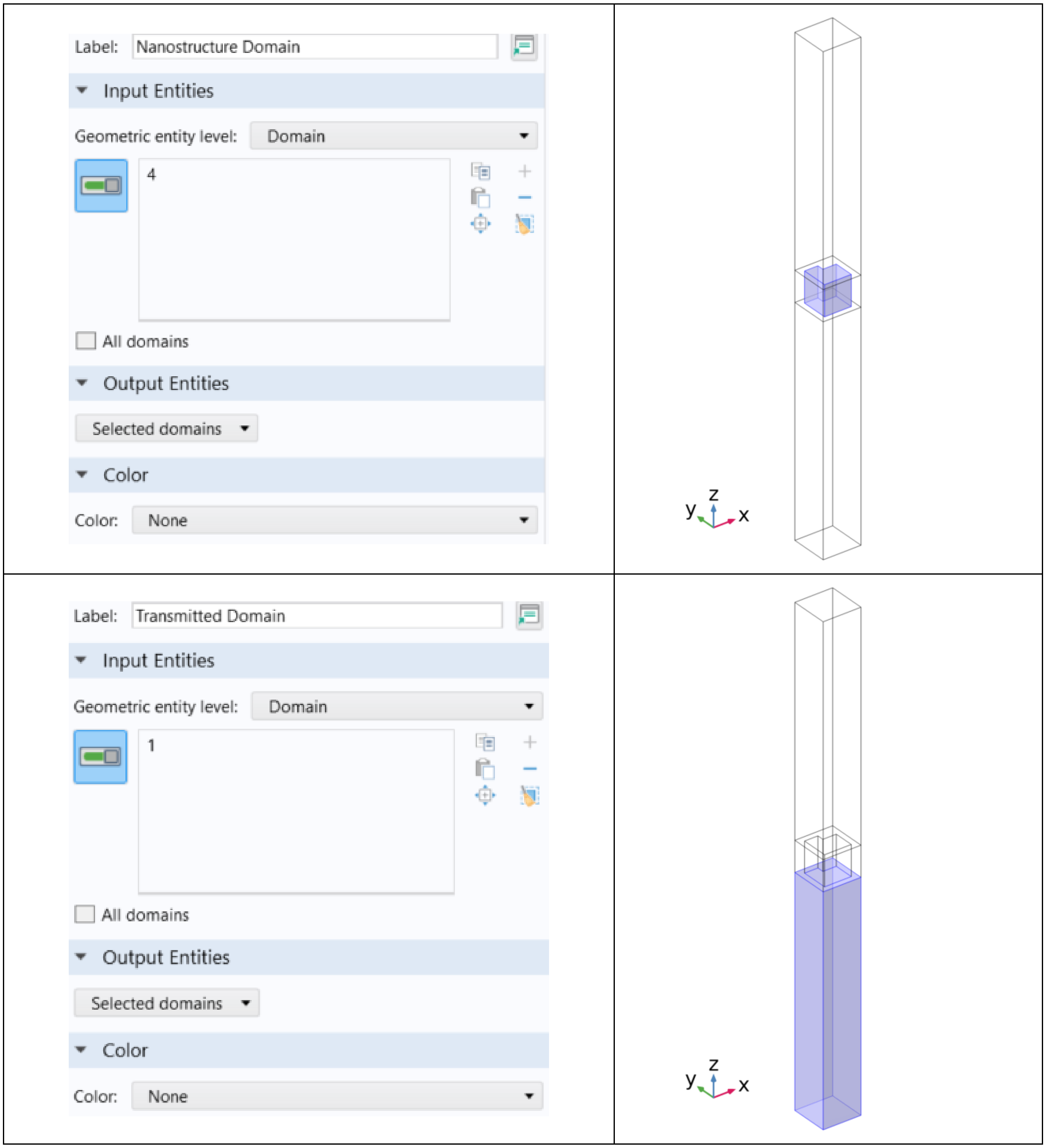

FIG S04. (Left Column) Screenshots of the Selections layers of (Top to Bottom) whole domain, incident domain, nanostructure domain, and transmitted domain; and their respective (Right Column) selected boundaries in the built model. The incident and transmitted domain refer to the portions of the model through which the incident light travels before or after interacting with the nanostructure, respectively.

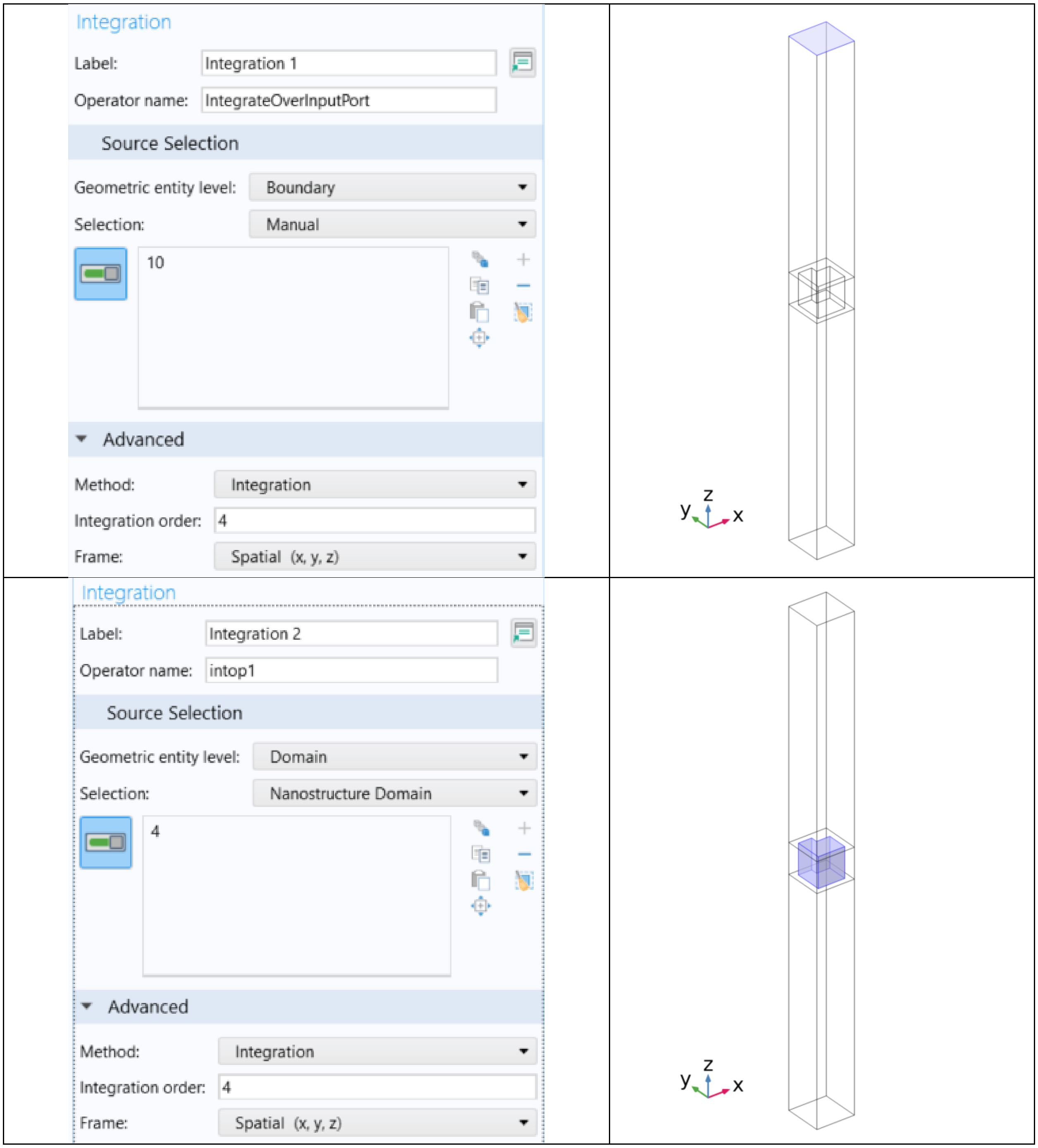

FIG S05. (Left Column) Screenshots of the Integration layers and their respective (Right Column) selected boundaries in the built model. (Top) “Integration 1” layer that provides a function for surface integrating over the input port. (Bottom) “Integration 2” layer that provides a function for volume integrating over the nanostructure.

### *B.2. Geometry*

In the Geometry section, we build up the entire domain of the model and the nanostructure within it. In this section, we create a block layer that we split up into the four selections defined in the previous sub-section; and then use a work plane layer to draw out our the 2D geometry of our nanostructure – that later becomes extruded to 3D. The parent of the geometry grouping of layers also defines the length unit to be used (Fig. S06).

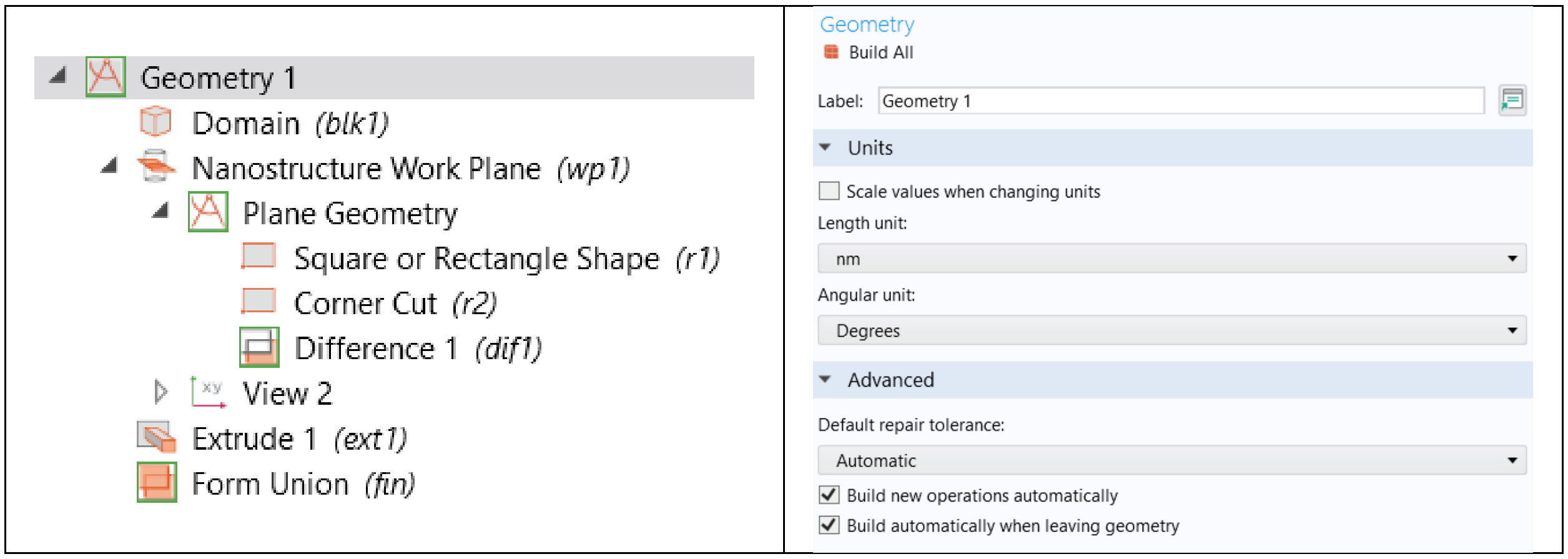

FIG S06. (Left) A screenshot of the Geometry groupings. (Right) The unit definitions used in the parent of the Geometry layers.

The domain layer uses the global definition variables defined to define the entire model's dimensions (Fig. S07) – in particular, the domain_width, domain_depth, and domain_height variables. This entire model is then split into three layers by using the incident_height and nanostructure_thickness variables to demarcate where to split the model.

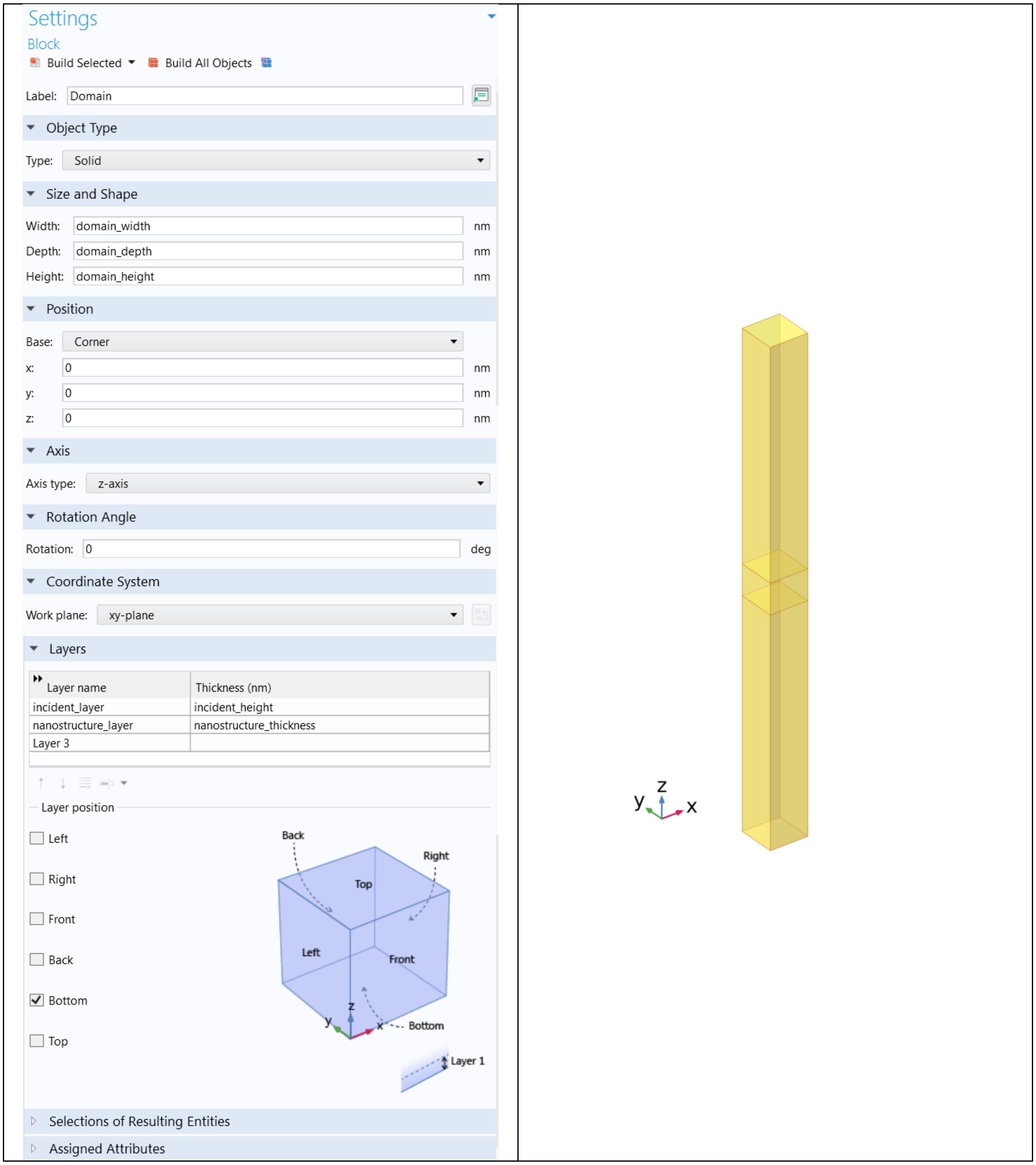

FIG S07. (Left) Screenshots of the parameters used in the Domain layer. (Right) The resulting model that is created when the Domain layer is built; the center of the model contains the demarcations that split it into three sections.

Once the overall model is built, we begin to define the 2D shape of our nanostructure. This is done using a Work Plane layer (Fig. S08). The Work Plane layer contains a Plane Geometry sub-layer that houses two Rectangle layers and a Difference layer (Fig. S09). One Rectangle layer is used to define the overall shape (square or oblong, and respective dimensions). The second Rectangle layer is used to define the dimensions and location of the corner to be cut. The cutting of the corner occurs using a Difference operation between the two Rectangle layers. Like the Domain layer, all these Work Plane layers will also use Global Definition variables that were defined for building the nanostructure and the model.

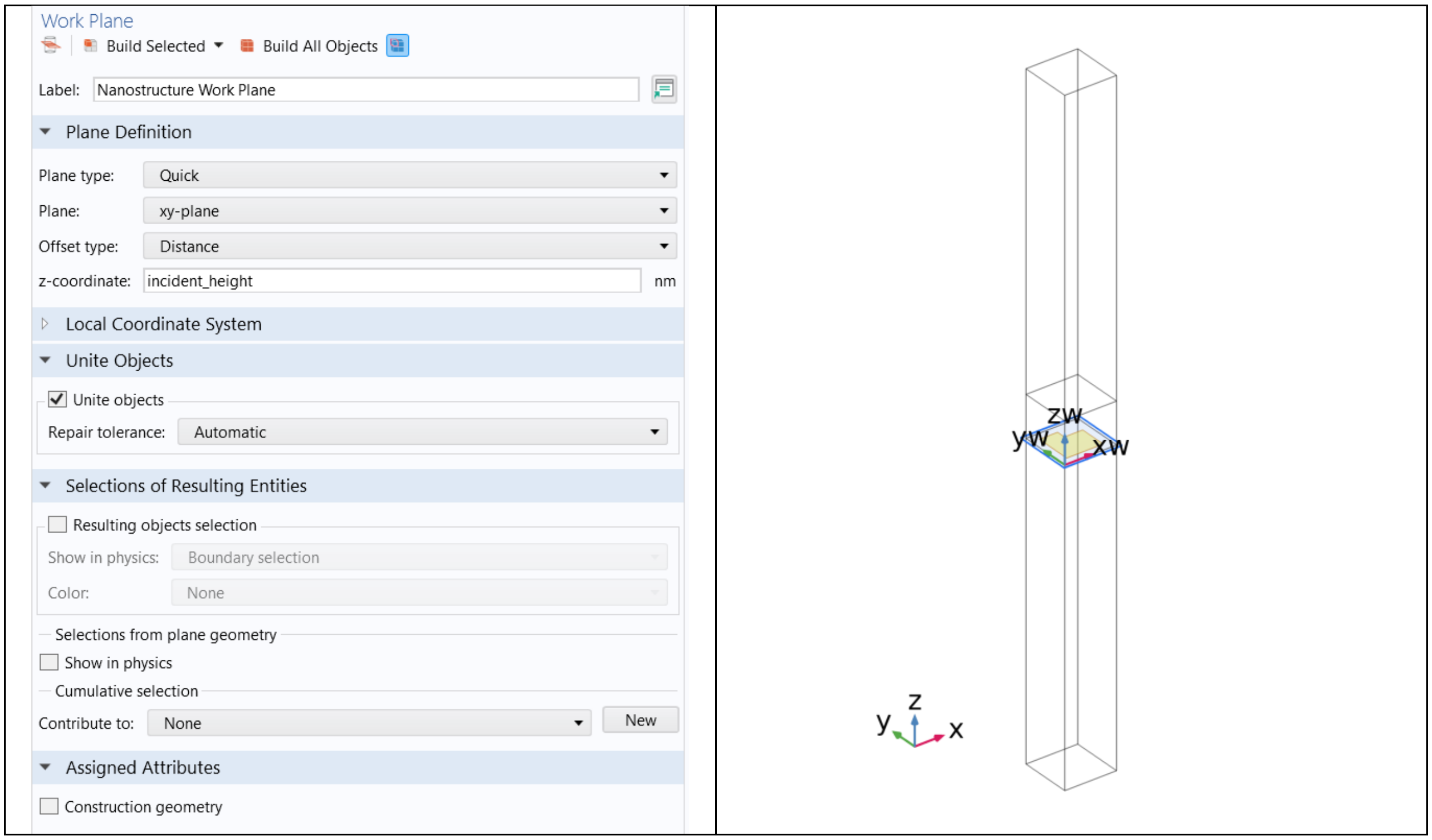

FIG S08. (Left) A screenshot of the Work Plane layer and the arguments & parameters used within. (Right) Highlighted in yellow is where the Work Plane layer is defined in the overall model, i.e. where the nanostructure will be created. "zw", "yw", and 'xw" are the cartesian coordinates for the Work Plane.

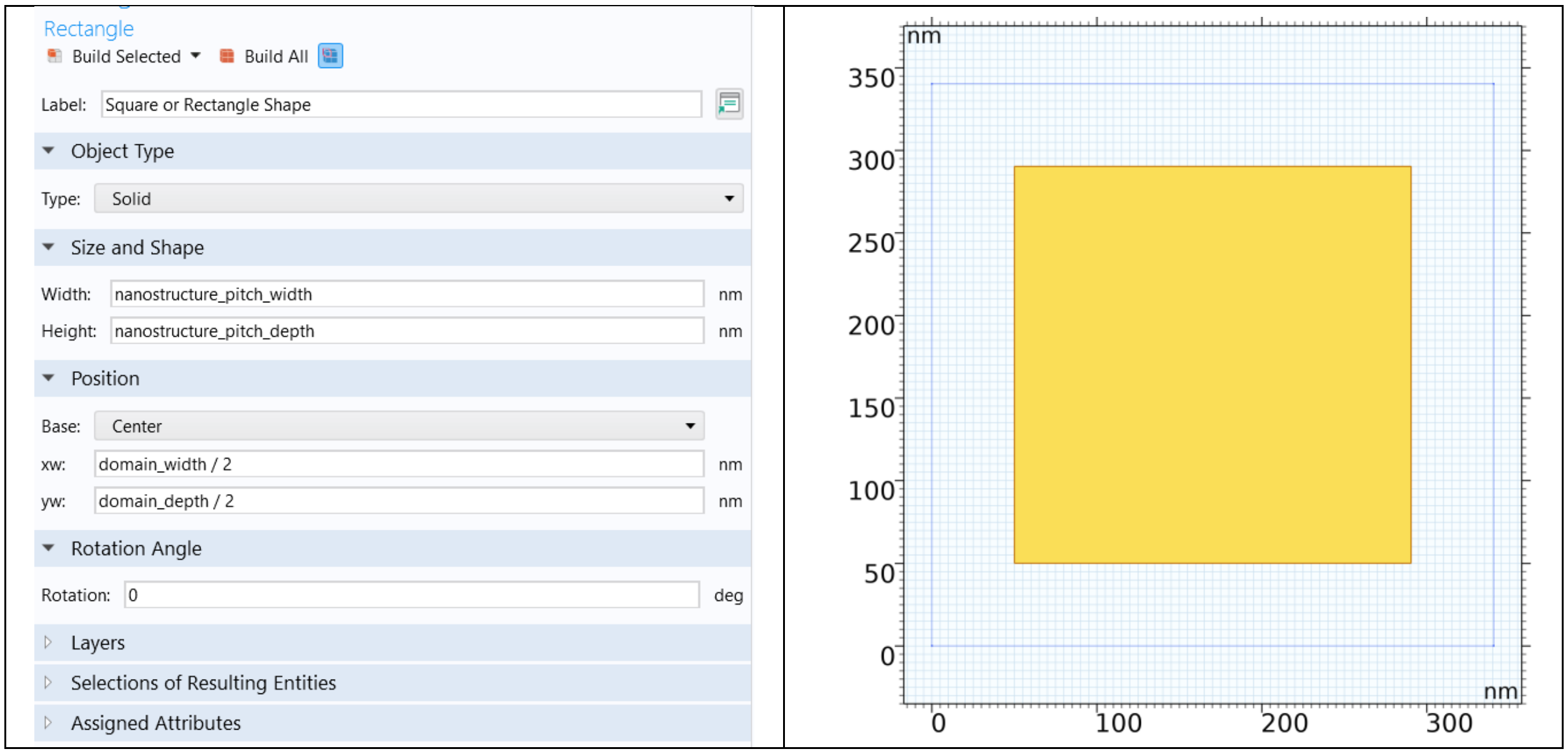

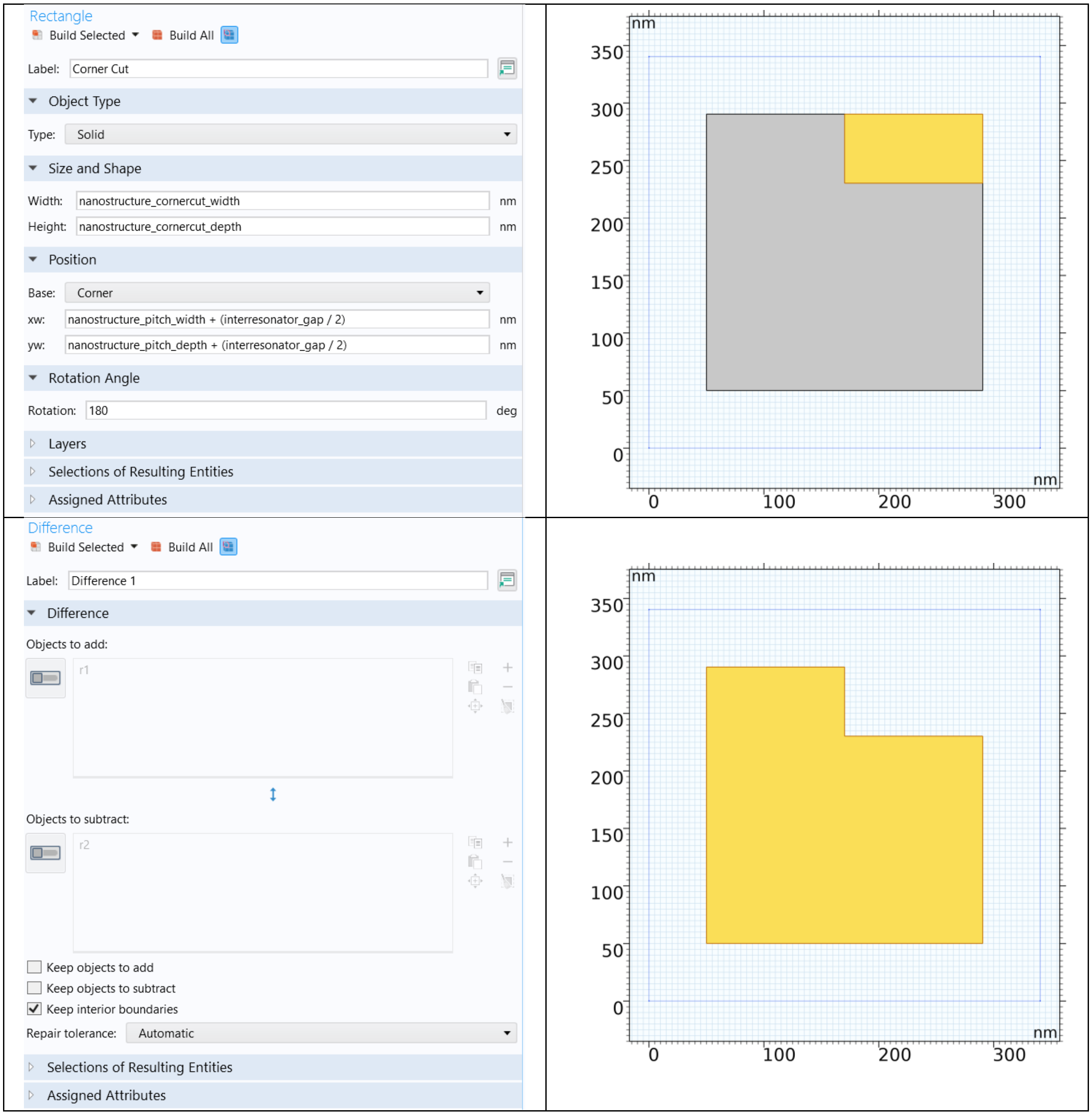

FIG S09. (Left) The arguments & parameters for both Rectangle layers and the Difference layer. (Right) Highlighted in yellow is the resultant shape from their respective layers.

With the 2D dimensions of the nanostructure defined, an Extrude layer is introduced in the parent Geometry layer to transform it into a 3D structure (Fig. S10).

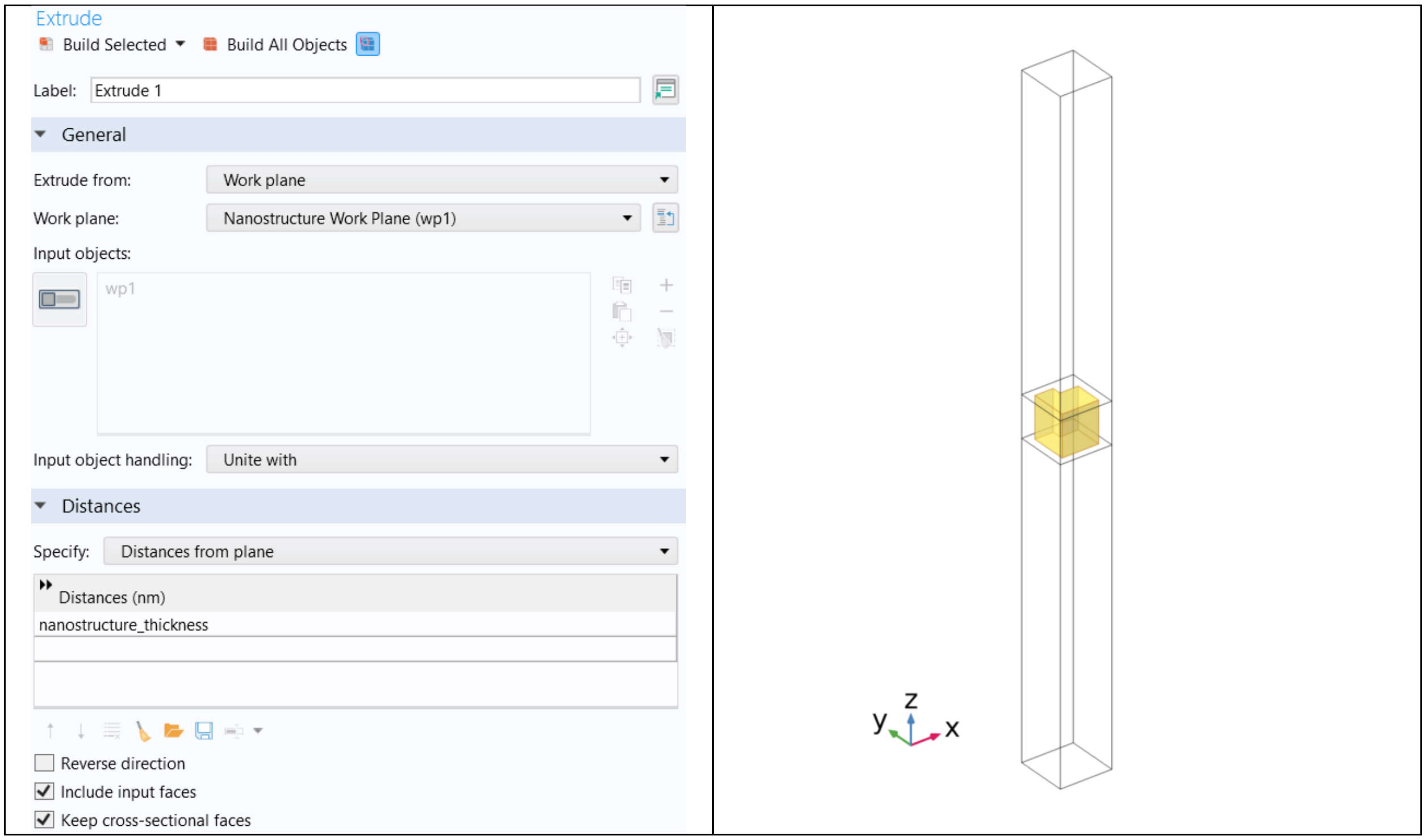

FIG S10. (Left) A screenshot of the arguments & parameters used in the Extrude layer. (Right) The resultant output is highlighted in yellow, showing a now-3D nanoparticle within the overall model.

### *B.3. Materials*

The Materials grouping lets us define material properties for each selection in our model (Fig. S11). The incident selection is defined as air [4], the nanostructure selection is defined as silicon nitride [5], and the transmitted selection is defined as silicon dioxide [6] (Fig. S12).

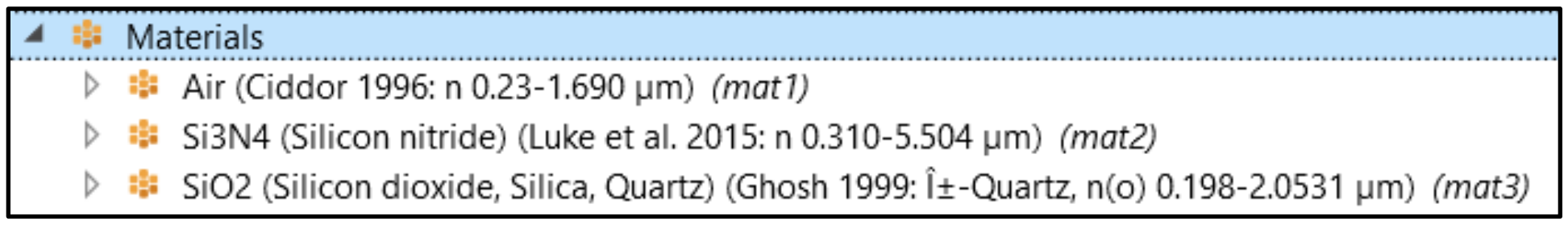

FIG S11. A screenshot of the Materials grouping and the three material properties layers created within it. Each layer defines a material property: air, silicon nitride, or silicon dioxide.

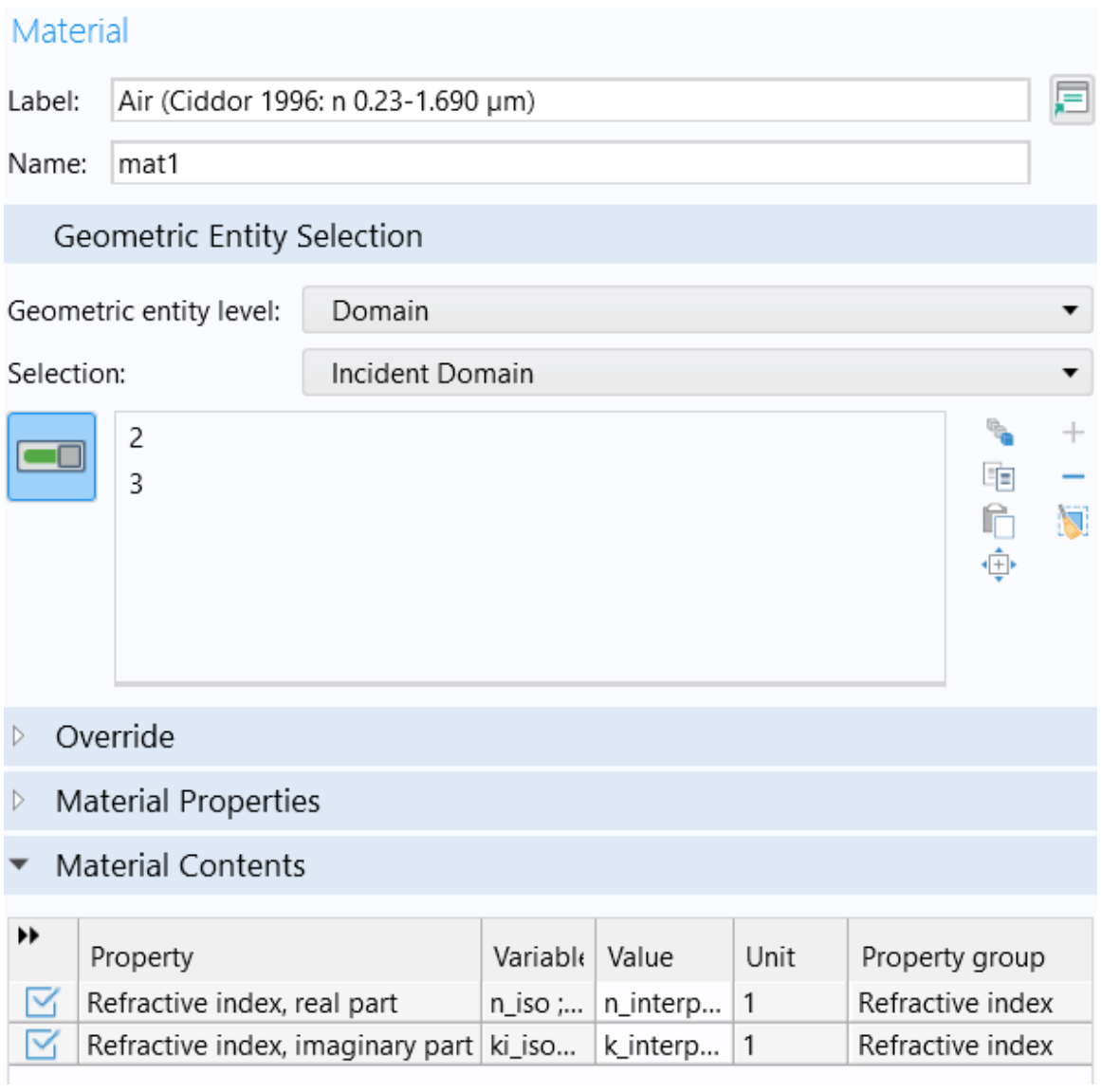
Material
Label: Air (Ciddor 1996: n 0.23-1.690 µm)
Name: mat1
Geometric Entity Selection
Geometric entity level: Domain
Selection: Incident Domain
2
3
Override
Material Properties
Material Contents
| Property | Variable | Value | Unit | Property group |
|---|---|---|---|---|
| Refractive index, real part | n_iso ;... | n_interp... | 1 | Refractive index |
| Refractive index, imaginary part | ki_iso... | k_interp... | 1 | Refractive index |

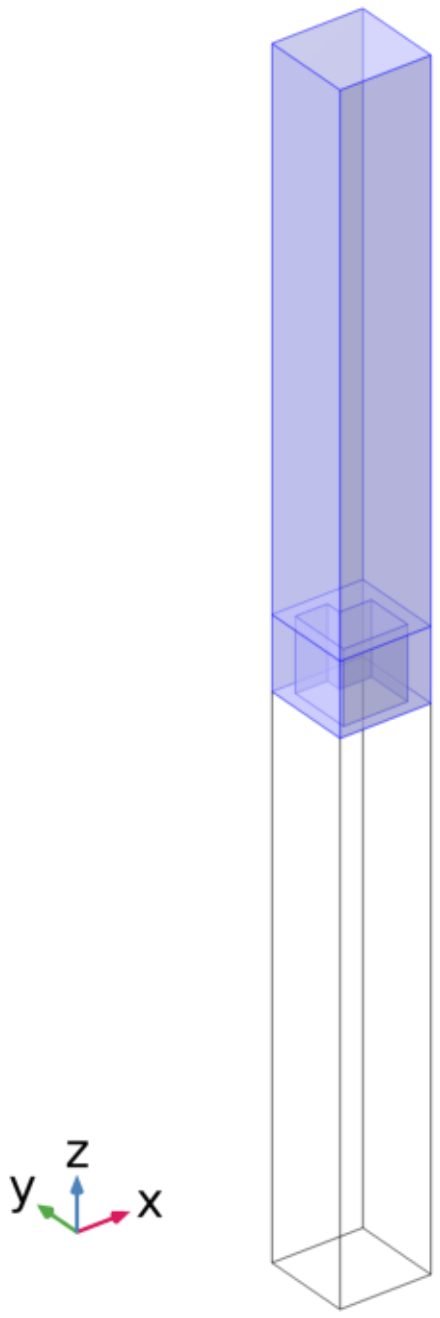
y
z
x

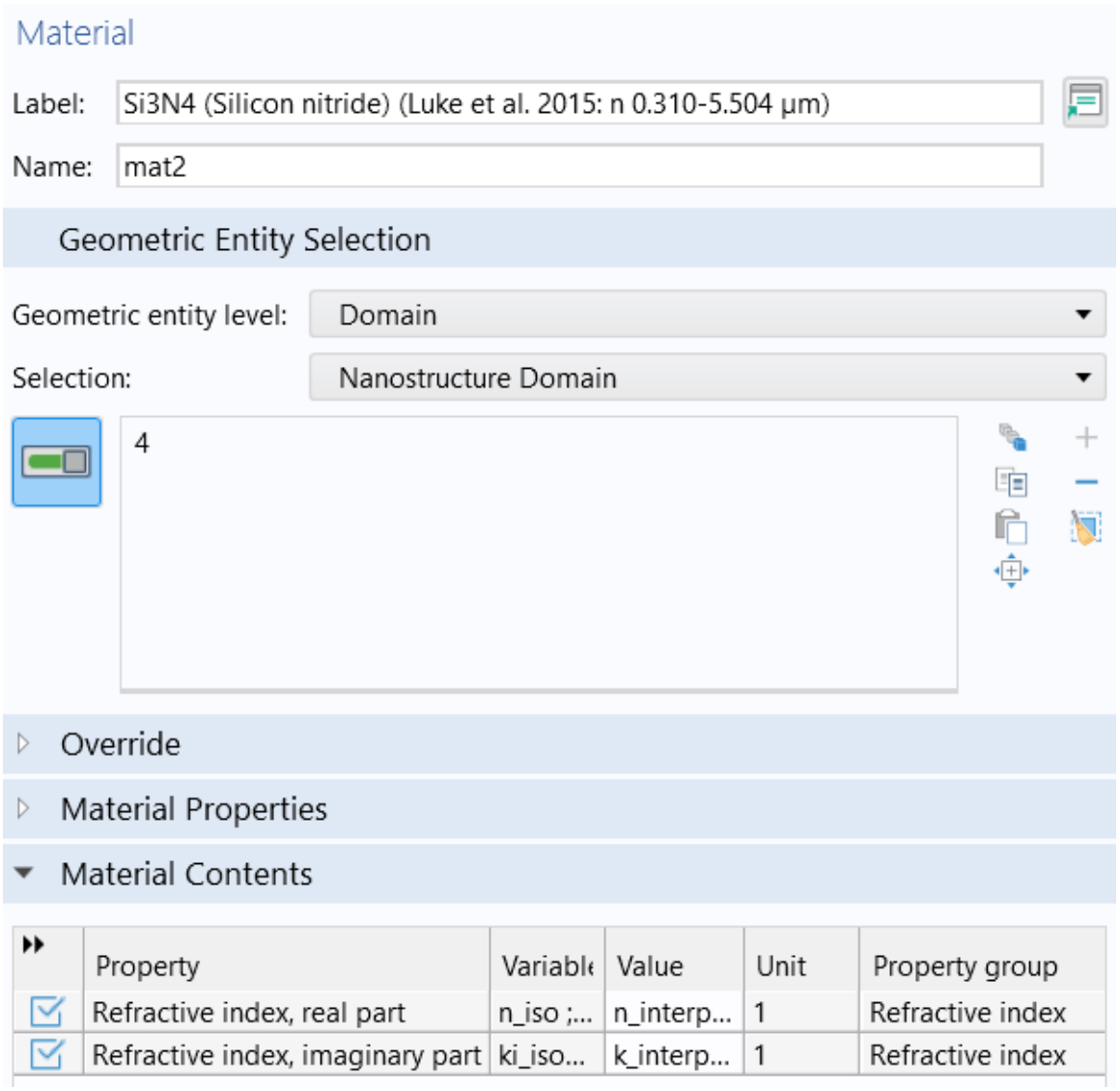
Material
Label: Si3N4 (Silicon nitride) (Luke et al. 2015: n 0.310-5.504 µm)
Name: mat2
Geometric Entity Selection
Geometric entity level: Domain
Selection: Nanostructure Domain
4
Override
Material Properties
Material Contents
| Property | Variable | Value | Unit | Property group |
|---|---|---|---|---|
| Refractive index, real part | n_iso ;... | n_interp... | 1 | Refractive index |
| Refractive index, imaginary part | ki_iso... | k_interp... | 1 | Refractive index |

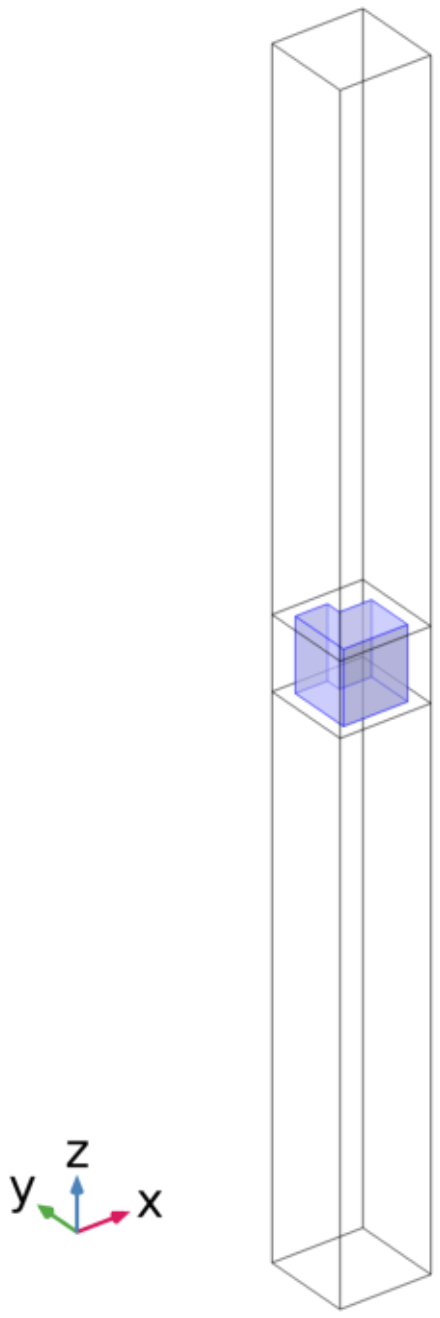
y
z
x

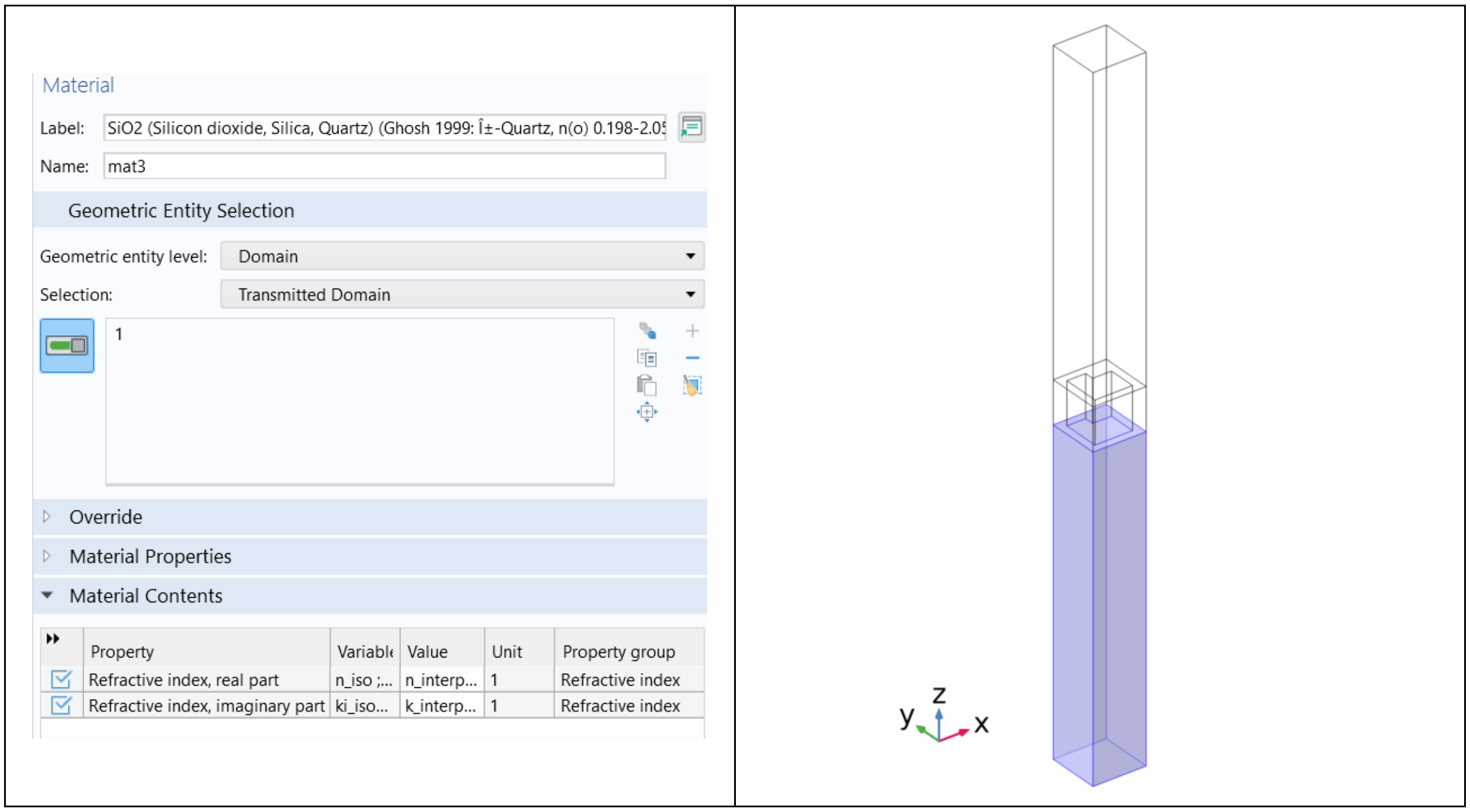

FIG S12. (Left) Screenshots of the (Top) air, (Middle) silicon nitride, and (Bottom) silicon dioxide assignments for the incident, nanostructure, and transmitted selections, respectively. (Right) Screenshots of the model with the blue highlighted portions corresponding to the material selections on the left.

### *B.4. Electromagnetic Waves, Frequency Domain*

The Electromagnetic Waves, Frequency Domain grouping defines the formulation of the incident electromagnetic wave and how the simulation is to approximate an infinite array solution. We do this by creating two Port layers and two Periodic Condition layers – these are in addition to the Wave Equation and Initial Values default layers (Fig. S13). We ignore the default-added Periodic Electric Conductor layer.

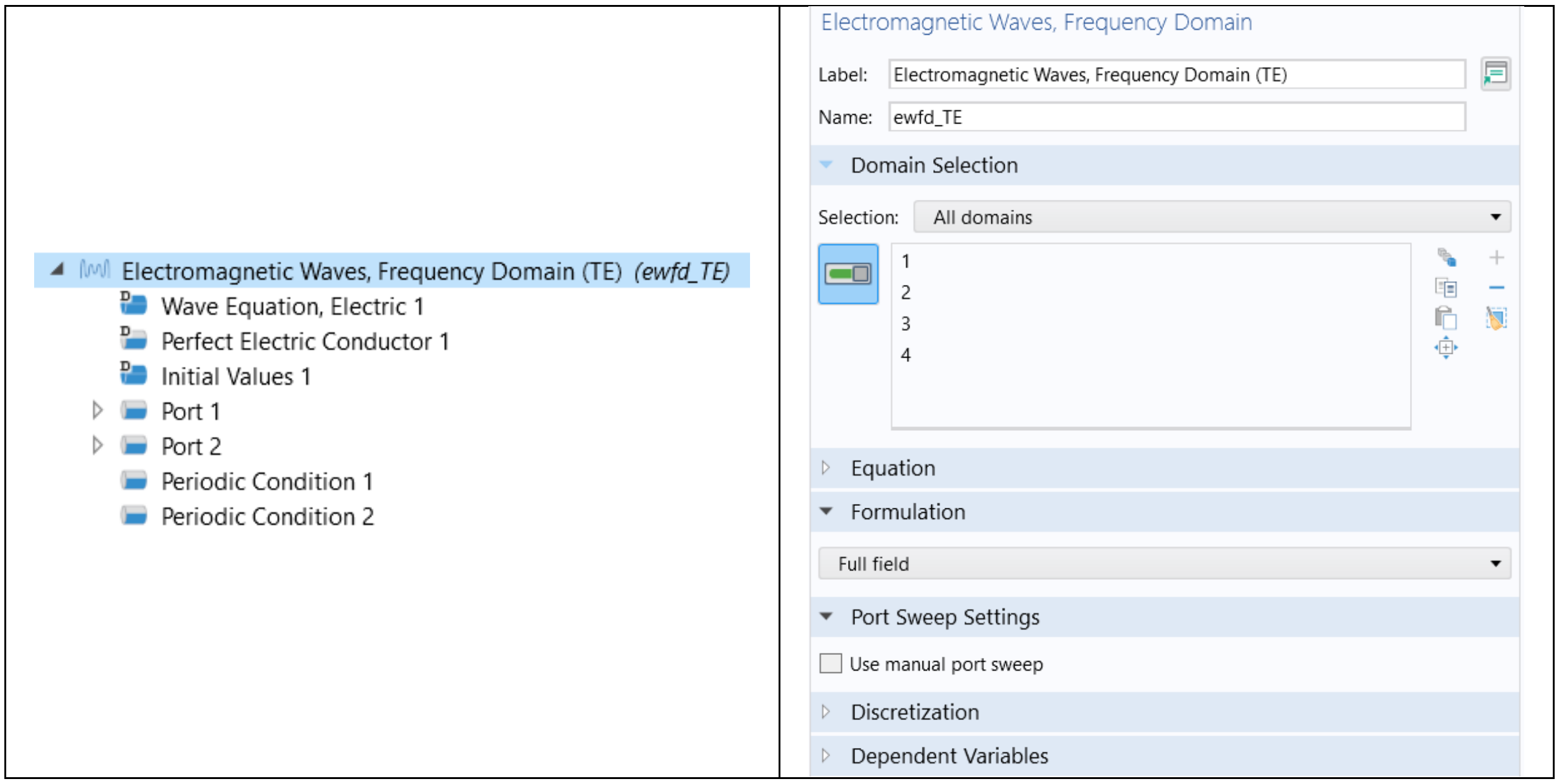

FIG S13. (Left) The layers created under the Electromagnetic Waves, Frequency Domain grouping. The Perfect Electric Conductor layer is ignored for this work. (Right) The parent layer is set to "Full field" formulation for all domains.

The Wave Equation and Initial Values default layers are kept to the default values, while ensuring that all domains in the model are selected, and that the initial values are set to 0 V/m for the x, y, and z components of the electric field. The two Port layers allow us to define the incoming electric field and act as probes for where far-field measurements are taken (Fig. S14). Importantly, the second Port layer is set to not create an electric field as we only need a single source; however, it allows us to take transmission measurements. The two Periodic Condition layers allow us to approximate an infinite array using Floquet periodicity boundary conditions (Fig. S15).

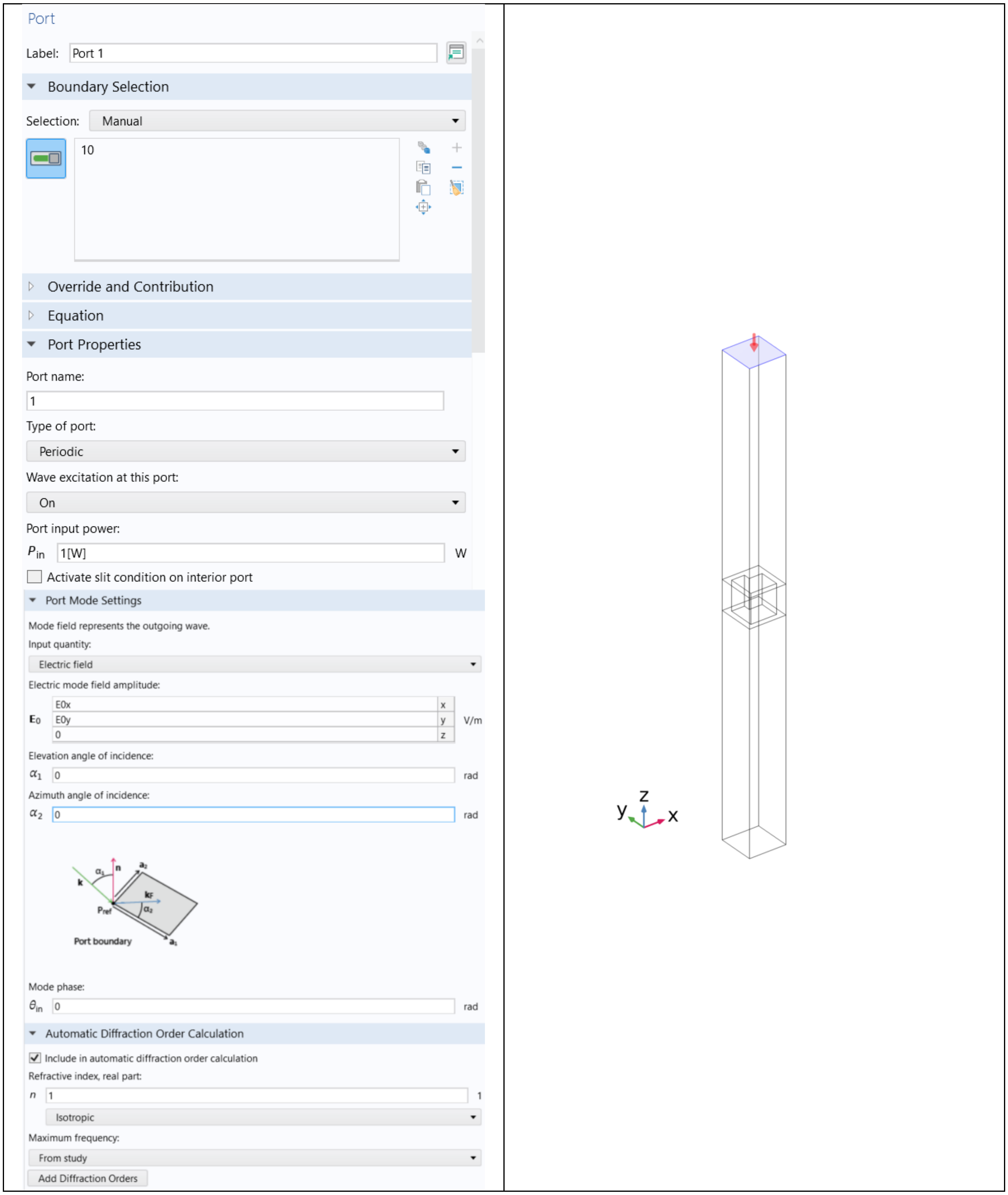

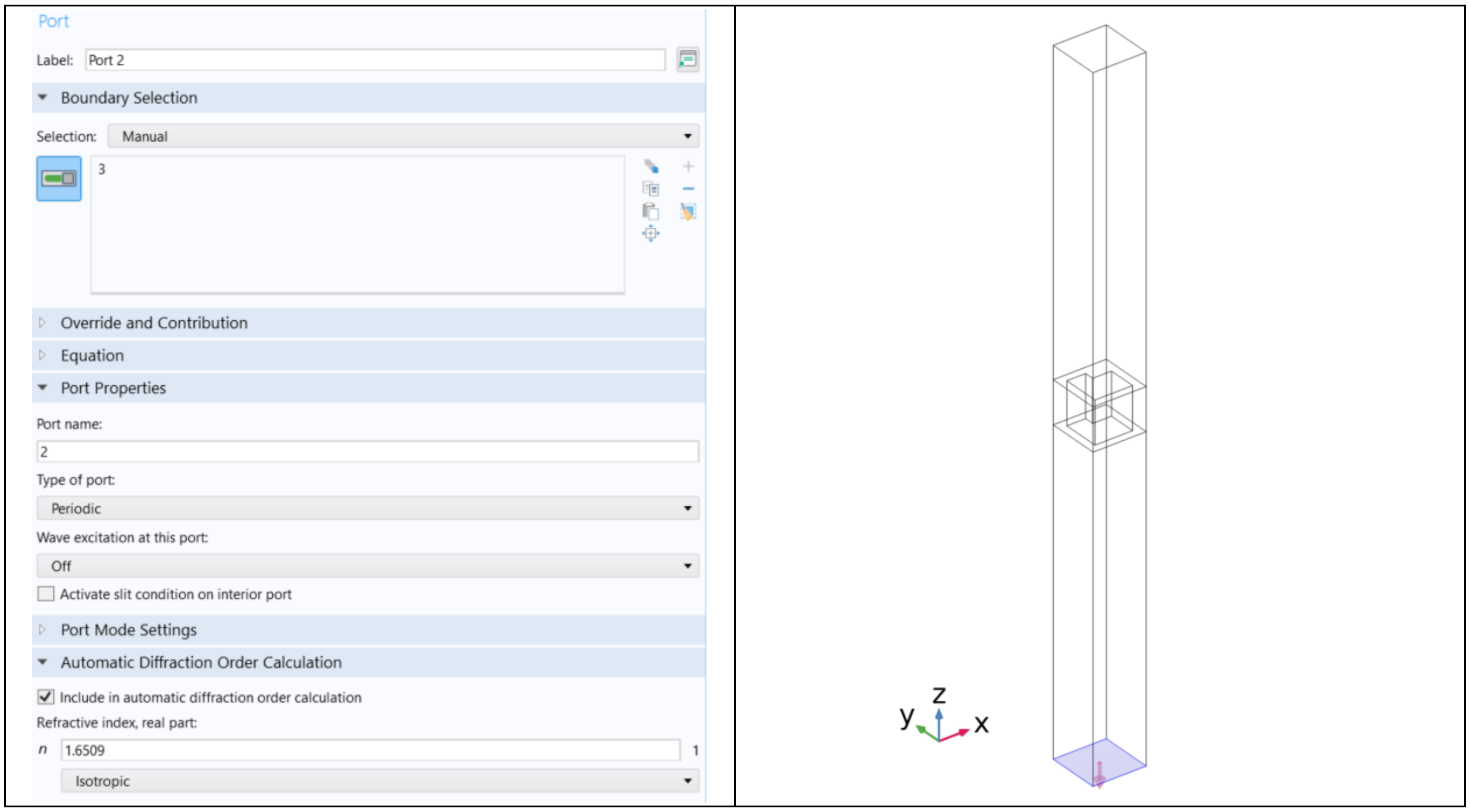

FIG S14. (Left) Screenshots of the (Top) Port 1 and (Bottom) Port 2 layers. Port 1 defines the incoming electric field wave. Port 2 is turned off and only used for data collection – however, it is also used to manually define the refractive index of the silicon dioxide substrate as 1.6509. (Right) The blue highlighted portions define the port locations, and the red arrows define the direction of the electric field propagation.

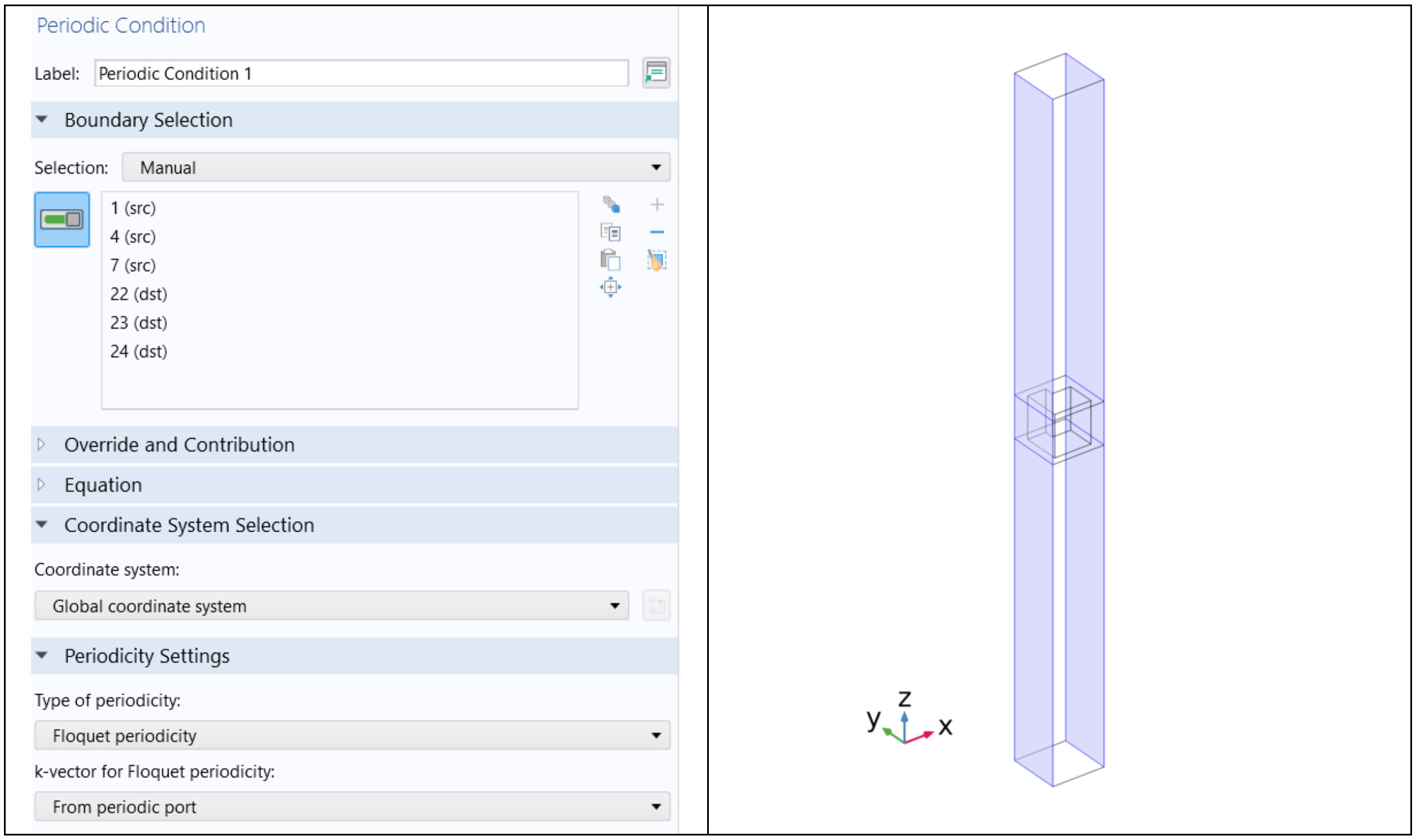

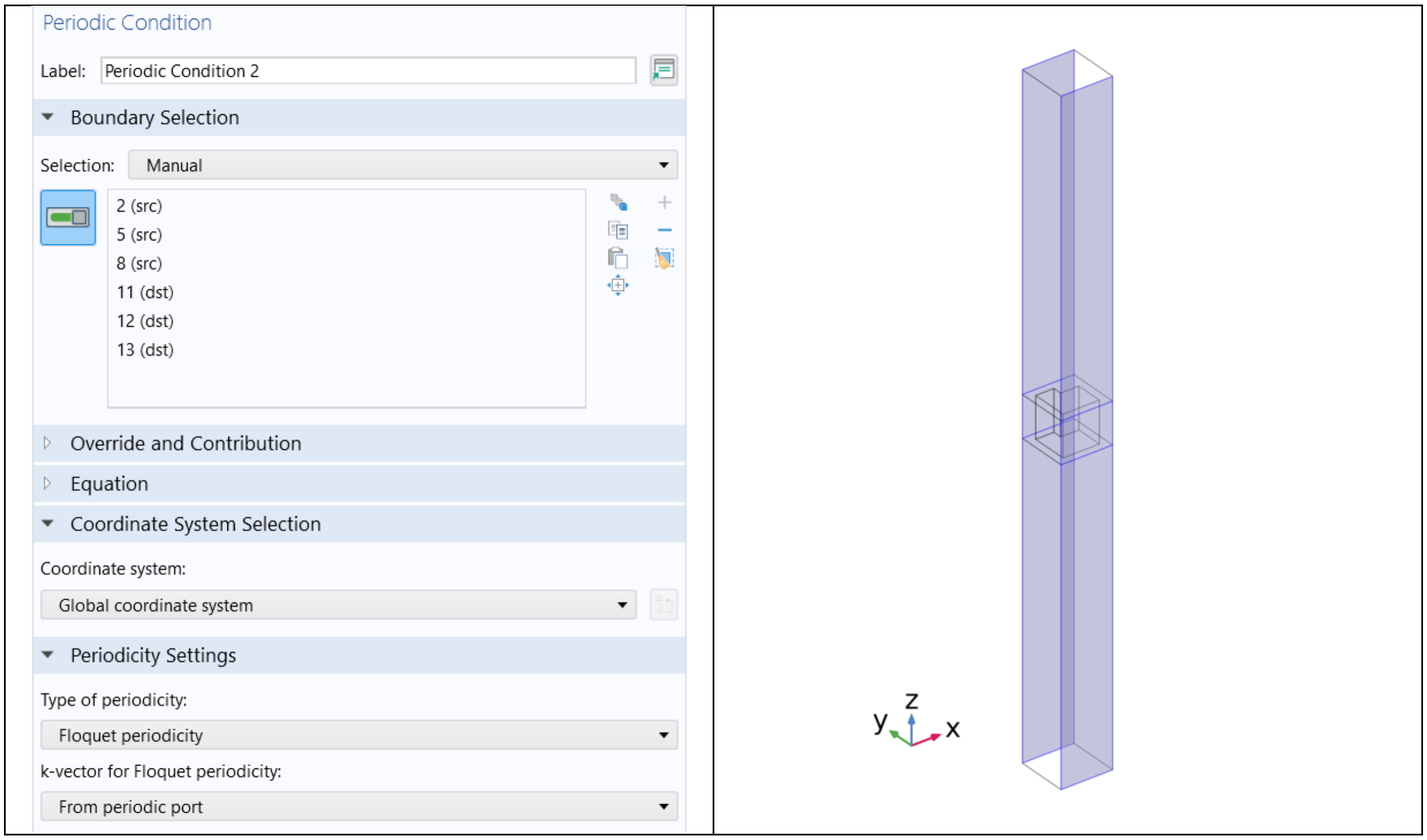

FIG S15. (Left) Screenshots of the parameters & arguments for both Periodic Condition layers. (Right) The portions of the model that are selected in each Periodic Condition layer are highlighted in blue.

### *B.5. Mesh*

The Mesh grouping divides the model's geometry into smaller and discrete elements to solve numerical calculations. For our work, we used a physics-controlled mesh with a normal element size, which are default options in COMSOL Multiphysics v6.0 (Fig. S16).

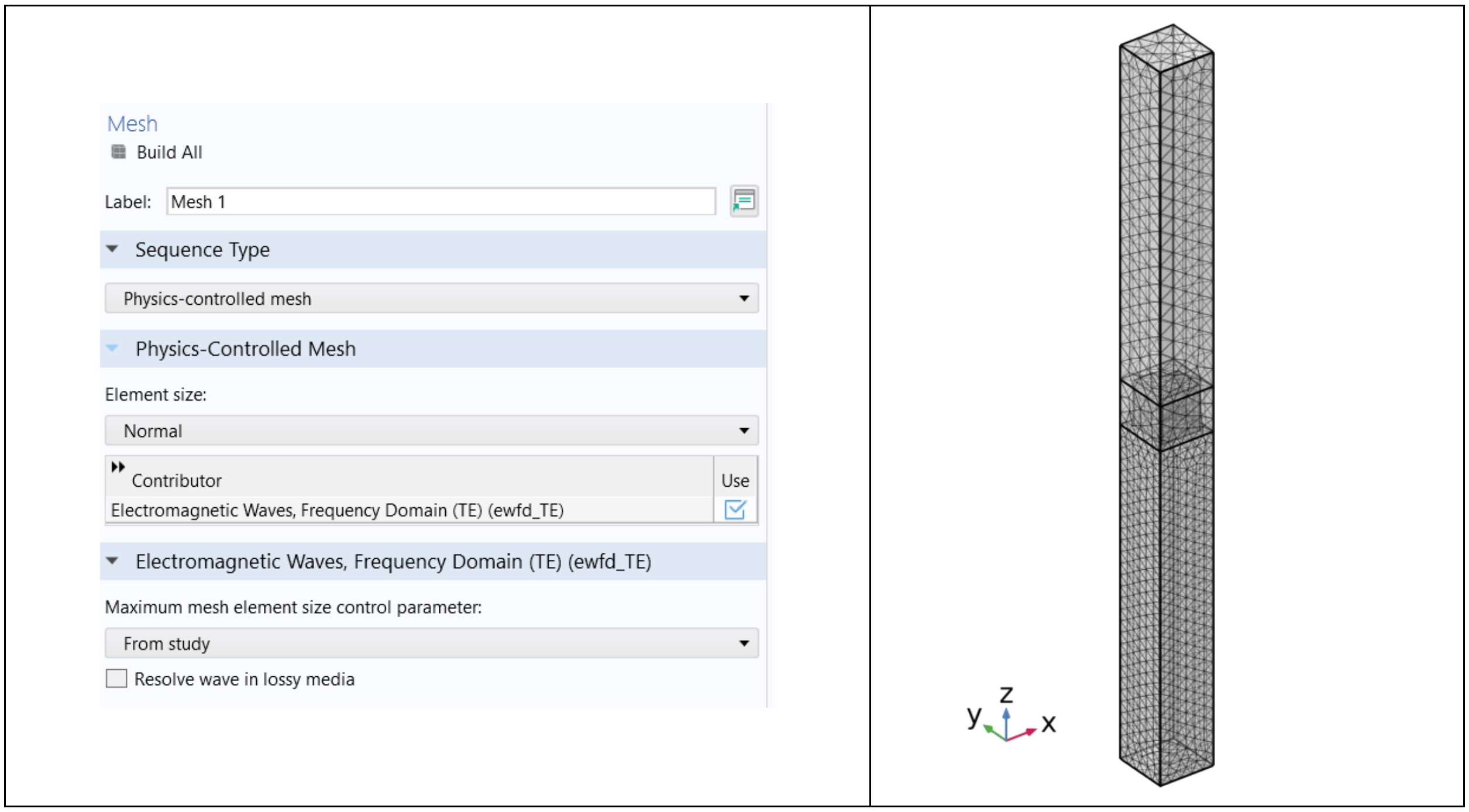

FIG S16. (Left) A screenshot of the Mesh parameters and arguments used. (Right) The rendered meshed model.

## C. Study

The Study grouping defines our simulation for us. We defined a Wavelength Domain layer that iterated through the wavelength_min and wavelength_max in steps of wavelength_step (Fig. S17), all of which were defined in the Group Definitions variable layers. This study also performed an auxiliary sweep between clockwise and counterclockwise circular light by switching between -90 deg and 90 deg values for the polarisation_rotation_angle parameter.

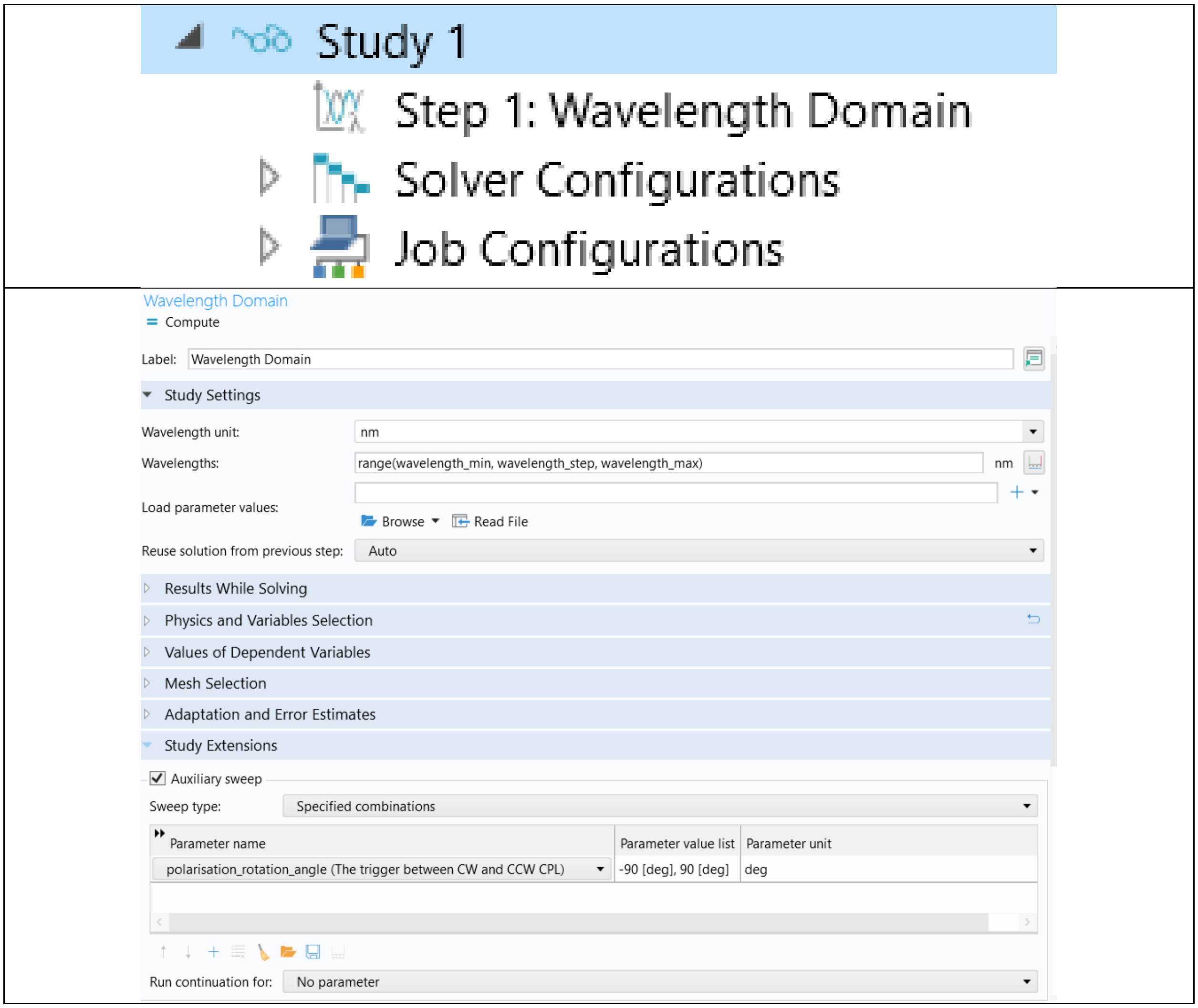

FIG S17. (Top) The Study groupings. The Solver Configurations & Job Configuration layers are untouched and created by the software while running the simulation. (Bottom) The arguments and parameters used for the Wavelength Domain layer.

## D. Results

Once the simulation is solved, we then extract the results we want through the Results grouping. In this grouping, we create four Global Evaluation and two Surface Integral layers under the "Derived Layers" sub-layer (Fig. S18). The expressions used for each of the Global Evaluation layers ("Reflectance, Transmittance, & Absoprtance"; "Stokes Vectors in Reflectance"; "Optical Chirality Density at Nanoparticle"; and "Optical Chirality Generation at Nanoparticle") are provided in Tables S08-11. The selected parameters and expressions for the two Surface Integral Layers ("Net Magnetic Flux" and "Net Electric Flux") can be seen in Fig. S19-S20.

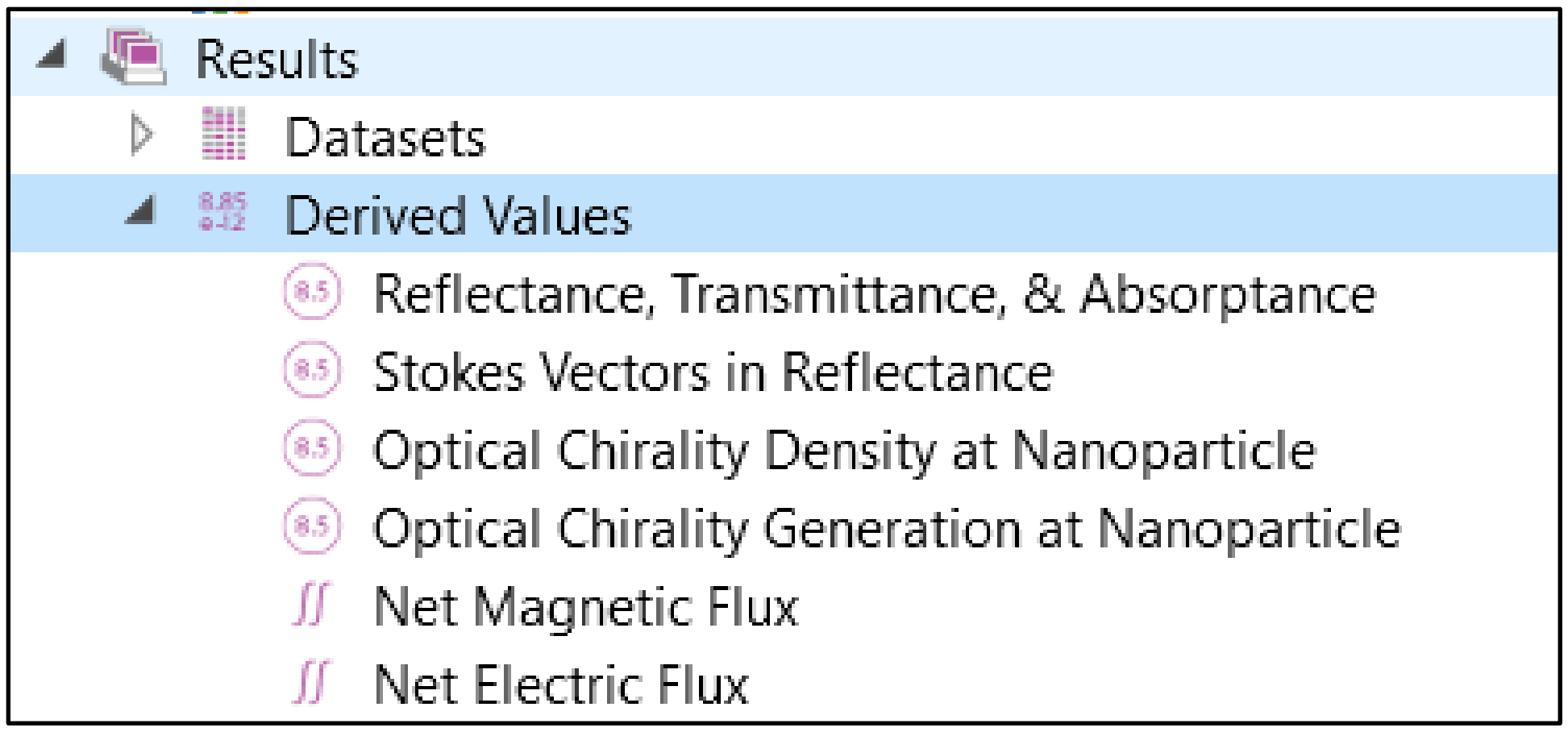

FIG S18. The Results grouping automatically creates a "Datasets" layer once a simulation is solved. The Derived Values layers were created to extract the measurements presented in the main text.

TABLE S08. Reflectance, Transmittance, & Absoprtance

| Expression | Unit | Description |
|---|---|---|
| **ewfd_TE.Rtotal** | 1 | Total reflectance |
| **ewfd_TE.Ttotal** | 1 | Total transmittance |
| **ewfd_TE.Atotal** | 1 | Absorptance |

TABLE S09. Stokes Vectors in Reflectance

| Expression* | Unit | Description |
|---|---|---|
| **abs(ewfd_TE.JROOP_0_0/ewfd_TE.normJR_0_0)^2 + abs(ewfd_TE.JRIP_0_0/ewfd_TE.normJR_0_0)^2** | 1 | S0 (Total Intensity) |
| **abs(ewfd_TE.JROOP_0_0/ewfd_TE.normJR_0_0)^2 - abs(ewfd_TE.JRIP_0_0/ewfd_TE.normJR_0_0)^2** | 1 | S1 (Horizontal - Vertical) |
| **2 * real( (ewfd_TE.JROOP_0_0/ewfd_TE.normJR_0_0) * conj(ewfd_TE.JRIP_0_0/ewfd_TE.normJR_0_0) )** | 1 | S2 (+45 - -45) |
| **-2 * imag( (ewfd_TE.JROOP_0_0/ewfd_TE.normJR_0_0) * conj(ewfd_TE.JRIP_0_0/ewfd_TE.normJR_0_0) )** | 1 | S3 (Right Circular - Left Circular) |

*The expressions used are the conversion from Jones Vectors to Stokes Vectors [7]. JROOP, JRIP, and normJR are in-built COMSOL commands for the output, input, and normalized Jones Vectors.

TABLE S10. Optical Chirality Density at Nanoparticle

| Expression | Unit | Description |
|---|---|---|
| **-ewfd_TE.omega / 2 * chiral_density_NP** | N/rad | Optical Chirality Density |

TABLE S11. Optical Chirality Generation at Nanoparticle

| Expression | Unit | Description |
|---|---|---|
| **ewfd_TE.lambda0 * ewfd_TE.omega * 0.5 * chiral_generation_NP** | N*m/s | Optical Chirality Generation |

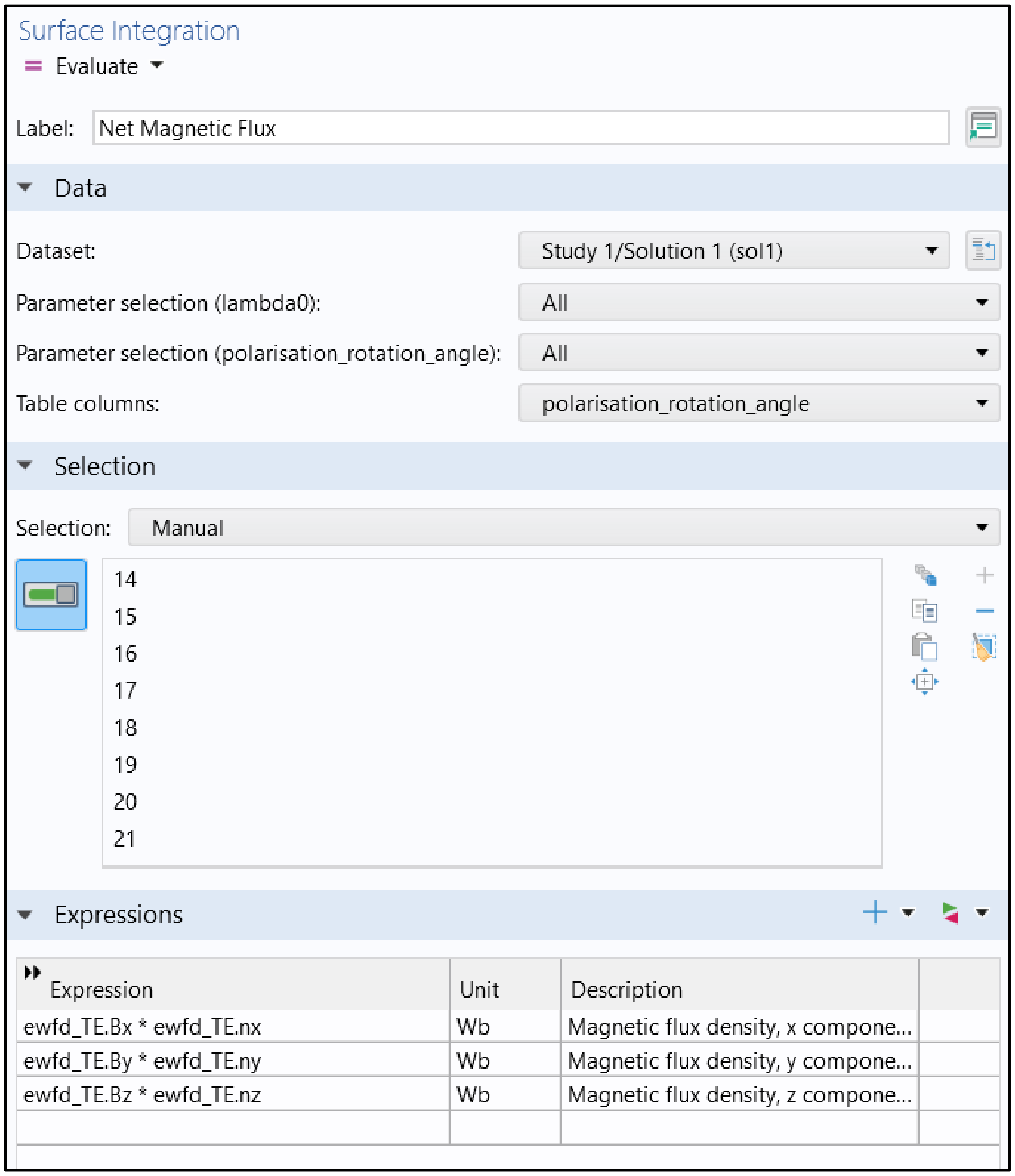

FIG S19. The “Net Magnetic Flux” layer, including the parameter selections and expressions.

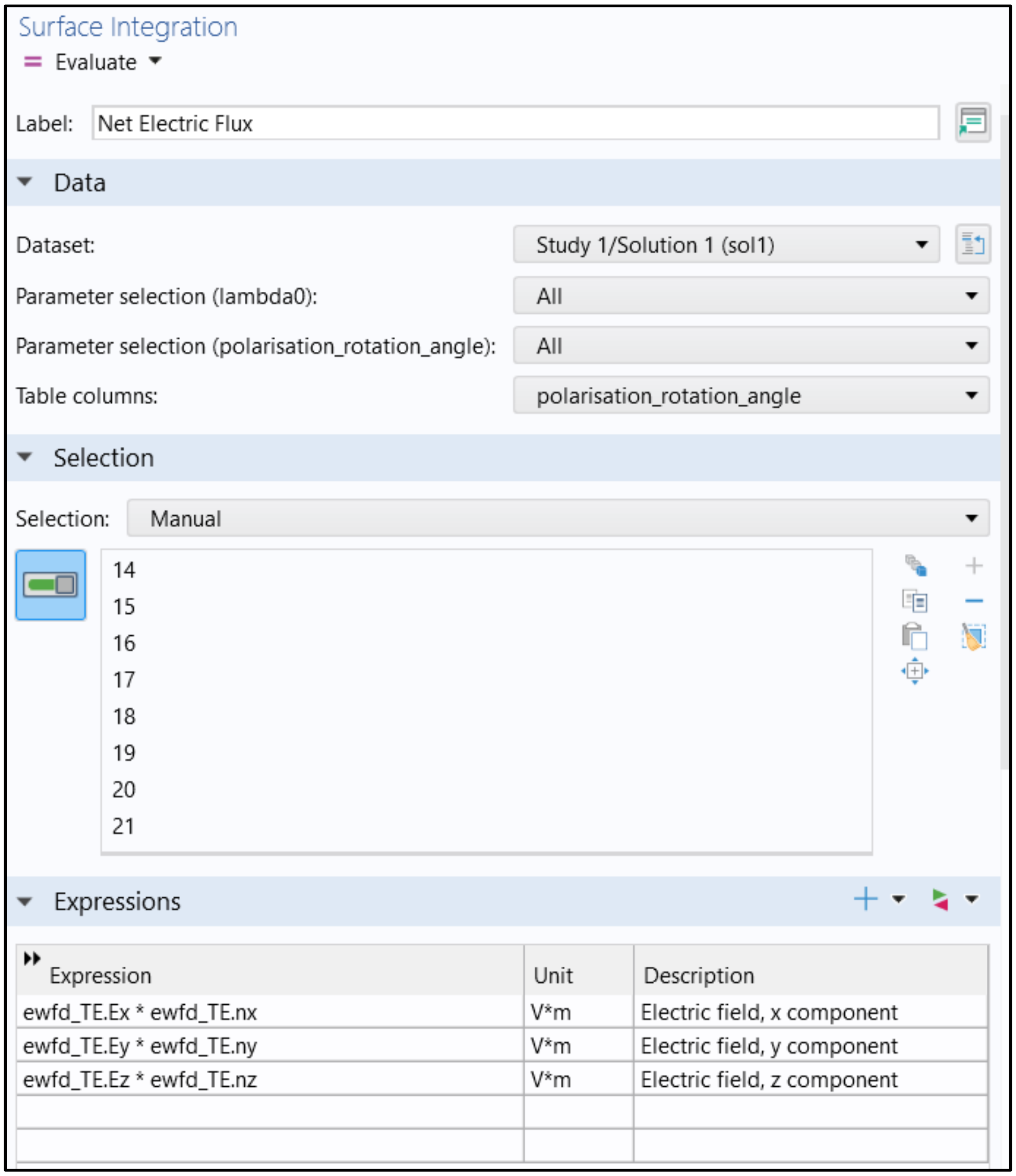

FIG S20. The “Net Electric Flux” layer, including the parameter selections and expressions.

## S2. CONSEQUENCE OF THE OPTICAL CHIRALITY CONTINUITY EQUATION

To investigate the physical origin of optical chirality generation, we begin by identifying the optical chirality continuity equation [8,9]:

$$\frac{1}{\mu_0}\nabla\cdot\boldsymbol{F}+\frac{\partial C}{\partial t}=-\frac{1}{2}[\boldsymbol{j_{tot}}\cdot(\nabla\times\boldsymbol{E})+\boldsymbol{E}\cdot(\nabla\times\boldsymbol{j_{tot}})] \tag{S1.1}$$

where $C\equiv\frac{\epsilon_0}{2}\boldsymbol{E}\cdot(\nabla\times\boldsymbol{E})+\frac{1}{2\mu_0}\boldsymbol{B}\cdot(\nabla\times\boldsymbol{B})$ is the optical chirality density, $\boldsymbol{F}\equiv\frac{1}{2}[\boldsymbol{E}\times(\nabla\times\boldsymbol{B})-\boldsymbol{B}\times(\nabla\times\boldsymbol{E})]$ is the optical chirality flux, and $\boldsymbol{j_{tot}}$ is the total current density containing all primary and secondary sources. $\boldsymbol{E}$ & $\boldsymbol{B}$ are the time-dependent electric and magnetic induction fields, respectively, and $\mu_0$ & $\epsilon_0$ are the vacuum magnetic permeability and electric permittivity, respectively. The right-hand side of Eqn. (S1.1) corresponds to the generation or dissipation of optical chirality.

Eqn. (S1.1) parallels Poynting's theorem, which governs energy conservation in electromagnetism [8,10]:

$$\frac{1}{\mu_0}\nabla\cdot\mathbf{S}+\frac{\delta u}{\delta t}=-\boldsymbol{j_{tot}}\cdot\boldsymbol{E} \tag{S1.2}$$

where $u=\frac{\epsilon_0}{2}\boldsymbol{E}\cdot\boldsymbol{E}+\frac{1}{2\mu_0}\boldsymbol{B}\cdot\boldsymbol{B}$ is the energy density, $\boldsymbol{S}=\boldsymbol{E}\times\boldsymbol{B}$ is the Poynting vector representing energy flow, and $\boldsymbol{j_{tot}}\cdot\boldsymbol{E}$ describes the power supplied by the total current density $\boldsymbol{j_{tot}}$.

The structural similarities between Eqns. (3.1) and (3.2) led previous studies to draw parallels between energy conservation and chiral conservation [11–13]. Building on this structural analogy, we show that a consequence of Eqn. (3.1) can elucidate the physical origin of optical chirality sources. The key steps are highlighted in the main text (see Main Text Section III.A). A full, step-by-step derivation is provided below.

### A. Rewriting the divergence of the optical chirality flux

We begin by taking the divergence of the optical chirality flux:

$$\nabla\cdot\boldsymbol{F}=\nabla\cdot\frac{1}{2}[\boldsymbol{E}\times(\nabla\times\boldsymbol{B})-\boldsymbol{B}\times(\nabla\times\boldsymbol{E})] \tag{S1.3}$$

Applying the vector identity $\nabla\cdot(\boldsymbol{a}\times\boldsymbol{b})=\boldsymbol{b}\cdot(\nabla\times\boldsymbol{a})-\boldsymbol{a}\cdot(\nabla\times\boldsymbol{b})$ [14], where $\boldsymbol{a}$ and $\boldsymbol{b}$ are arbitrary vector fields:

$$\begin{aligned}\nabla\cdot\boldsymbol{F}&=\frac{1}{2}\Big[\big((\nabla\times\boldsymbol{B})\cdot(\nabla\times\boldsymbol{E})-\boldsymbol{E}\cdot\big(\nabla\times(\nabla\times\boldsymbol{B})\big)\big)-\big((\nabla\times\boldsymbol{E})\cdot(\nabla\times\boldsymbol{B})-\boldsymbol{B}\cdot\big(\nabla\times(\nabla\times\boldsymbol{E})\big)\big)\Big]\\&=\frac{1}{2}\big[(\nabla\times\boldsymbol{B})\cdot(\nabla\times\boldsymbol{E})-\boldsymbol{E}\cdot\big(\nabla\times(\nabla\times\boldsymbol{B})\big)-(\nabla\times\boldsymbol{E})\cdot(\nabla\times\boldsymbol{B})+\boldsymbol{B}\cdot\big(\nabla\times(\nabla\times\boldsymbol{E})\big)\big]\\&=\frac{1}{2}\big[(\nabla\times\boldsymbol{B})\cdot(\nabla\times\boldsymbol{E})-(\nabla\times\boldsymbol{E})\cdot(\nabla\times\boldsymbol{B})-\boldsymbol{E}\cdot\big(\nabla\times(\nabla\times\boldsymbol{B})\big)+\boldsymbol{B}\cdot\big(\nabla\times(\nabla\times\boldsymbol{E})\big)\big]\\&=\frac{1}{2}\big[(\nabla\times\boldsymbol{B})\cdot(\nabla\times\boldsymbol{E})-(\nabla\times\boldsymbol{B})\cdot(\nabla\times\boldsymbol{E})+\boldsymbol{B}\cdot\big(\nabla\times(\nabla\times\boldsymbol{B})\big)-\boldsymbol{E}\cdot\big(\nabla\times(\nabla\times\boldsymbol{E})\big)\big]\\&=\frac{1}{2}\big[\boldsymbol{B}\cdot\big(\nabla\times(\nabla\times\boldsymbol{E})\big)-\boldsymbol{E}\cdot\big(\nabla\times(\nabla\times\boldsymbol{B})\big)\big]\end{aligned}$$

Thus,

$$\nabla\cdot\boldsymbol{F}=\frac{1}{2}\big[\boldsymbol{B}\cdot\big(\nabla\times(\nabla\times\boldsymbol{E})\big)-\boldsymbol{E}\cdot\big(\nabla\times(\nabla\times\boldsymbol{B})\big)\big] \tag{S1.4}$$

Next, we introduce the wave equations for $\boldsymbol{E}$ and $\boldsymbol{B}$ [15]:

$$\nabla\times(\nabla\times\boldsymbol{E})=-\mu_0\frac{\partial\boldsymbol{j_{tot}}}{\partial t}-\frac{1}{c^2}\frac{\partial^2\boldsymbol{E}}{\partial t^2}$$

$$\nabla \times (\nabla \times \boldsymbol{B}) = \mu_0 \nabla \times \boldsymbol{j_{tot}} - \frac{1}{c^2}\frac{\partial^2 \boldsymbol{B}}{\partial t^2}$$

where $c \cong 3 \times 10^8 \frac{m}{s}$ is the speed of light in a vacuum. Substituting the wave equations into Eqn. (S1.4):

$$\begin{aligned}
\nabla \cdot \boldsymbol{F} &= \frac{1}{2}\left[\boldsymbol{B} \cdot \left(-\mu_0 \frac{\partial \boldsymbol{j_{tot}}}{\partial t} - \frac{1}{c^2}\frac{\partial^2 \boldsymbol{E}}{\partial t^2}\right) - \boldsymbol{E} \cdot \left(\mu_0 \cdot (\nabla \times \boldsymbol{j_{tot}}) - \frac{1}{c^2}\frac{\partial^2 \boldsymbol{B}}{\partial t^2}\right)\right] \\
&= \frac{1}{2}\left[-\boldsymbol{B} \cdot \mu_0 \frac{\partial \boldsymbol{j_{tot}}}{\partial t} - \boldsymbol{B} \cdot \frac{1}{c^2}\frac{\partial^2 \boldsymbol{E}}{\partial t^2} - \boldsymbol{E} \cdot \mu_0 \cdot (\nabla \times \boldsymbol{j_{tot}}) + \boldsymbol{E} \cdot \frac{1}{c^2}\frac{\partial^2 \boldsymbol{B}}{\partial t^2}\right] \\
&= \frac{1}{2}\left[\left(-\boldsymbol{B} \cdot \mu_0 \frac{\partial \boldsymbol{j_{tot}}}{\partial t} - \boldsymbol{E} \cdot \mu_0 \cdot (\nabla \times \boldsymbol{j_{tot}})\right) + \left(-\boldsymbol{B} \cdot \frac{1}{c^2}\frac{\partial^2 \boldsymbol{E}}{\partial t^2} + \boldsymbol{E} \cdot \frac{1}{c^2}\frac{\partial^2 \boldsymbol{B}}{\partial t^2}\right)\right] \\
&= \frac{1}{2}\left[-\mu_0 \left(\boldsymbol{B} \cdot \frac{\partial \boldsymbol{j_{tot}}}{\partial t} + \boldsymbol{E} \cdot (\nabla \times \boldsymbol{j_{tot}})\right) + \frac{1}{c^2}\left(-\boldsymbol{B} \cdot \frac{\partial^2 \boldsymbol{E}}{\partial t^2} + \boldsymbol{E} \cdot \frac{\partial^2 \boldsymbol{B}}{\partial t^2}\right)\right] \\
&= \frac{1}{2}\left[-\mu_0 \left(\boldsymbol{B} \cdot \frac{\partial \boldsymbol{j_{tot}}}{\partial t} + \boldsymbol{E} \cdot (\nabla \times \boldsymbol{j_{tot}})\right) + \frac{1}{c^2}\frac{\partial}{\partial t}\left(-\boldsymbol{B} \cdot \frac{\partial \boldsymbol{E}}{\partial t} + \boldsymbol{E} \cdot \frac{\partial \boldsymbol{B}}{\partial t}\right)\right] \\
&= -\frac{\mu_0}{2}\left[\boldsymbol{B} \cdot \frac{\partial \boldsymbol{j_{tot}}}{\partial t} + \boldsymbol{E} \cdot (\nabla \times \boldsymbol{j_{tot}})\right] + \frac{1}{2c^2}\frac{\partial}{\partial t}\left(-\boldsymbol{B} \cdot \frac{\partial \boldsymbol{E}}{\partial t} + \boldsymbol{E} \cdot \frac{\partial \boldsymbol{B}}{\partial t}\right)
\end{aligned}$$

Subtracting $\frac{1}{2c^2}\frac{\partial}{\partial t}\left(-\boldsymbol{B} \cdot \frac{\partial \boldsymbol{E}}{\partial t} + \boldsymbol{E} \cdot \frac{\partial \boldsymbol{B}}{\partial t}\right)$ from both sides:

$$\begin{aligned}
\nabla \cdot \boldsymbol{F} - \frac{1}{2c^2}\frac{\partial}{\partial t}\left(-\boldsymbol{B} \cdot \frac{\partial \boldsymbol{E}}{\partial t} + \boldsymbol{E} \cdot \frac{\partial \boldsymbol{B}}{\partial t}\right) &= -\frac{\mu_0}{2}\left[\boldsymbol{B} \cdot \frac{\partial \boldsymbol{j_{tot}}}{\partial t} + \boldsymbol{E} \cdot (\nabla \times \boldsymbol{j_{tot}})\right] \\
\nabla \cdot \boldsymbol{F} - \frac{\partial}{\partial t}\left[\frac{1}{2c^2}\left(-\boldsymbol{B} \cdot \frac{\partial \boldsymbol{E}}{\partial t} + \boldsymbol{E} \cdot \frac{\partial \boldsymbol{B}}{\partial t}\right)\right] &= -\frac{\mu_0}{2}\left[\boldsymbol{B} \cdot \frac{\partial \boldsymbol{j_{tot}}}{\partial t} + \boldsymbol{E} \cdot (\nabla \times \boldsymbol{j_{tot}})\right]
\end{aligned}$$

Multiplying both sides with $\frac{1}{\mu_0}$:

$$\frac{1}{\mu_0}\nabla \cdot \boldsymbol{F} - \frac{1}{\mu_0}\frac{\partial}{\partial t}\left[\frac{1}{2c^2}\left(-\boldsymbol{B} \cdot \frac{\partial \boldsymbol{E}}{\partial t} + \boldsymbol{E} \cdot \frac{\partial \boldsymbol{B}}{\partial t}\right)\right] = -\frac{1}{\mu_0}\frac{\mu_0}{2}\left[\boldsymbol{B} \cdot \frac{\partial \boldsymbol{j_{tot}}}{\partial t} + \boldsymbol{E} \cdot (\nabla \times \boldsymbol{j_{tot}})\right]$$

Thus,

$$\frac{1}{\mu_0}\nabla \cdot \boldsymbol{F} - \frac{\partial}{\partial t}\left[\frac{1}{2\mu_0 c^2}\left(-\boldsymbol{B} \cdot \frac{\partial \boldsymbol{E}}{\partial t} + \boldsymbol{E} \cdot \frac{\partial \boldsymbol{B}}{\partial t}\right)\right] = -\frac{1}{2}\left[\boldsymbol{B} \cdot \frac{\partial \boldsymbol{j_{tot}}}{\partial t} + \boldsymbol{E} \cdot (\nabla \times \boldsymbol{j_{tot}})\right] \quad (S1.5)$$

Recognizing that the bracketed term on the left-hand side corresponds to the optical chiral density $C$ and the bracketed term on the right-hand side corresponds to the optical chirality source term, then Eqn. (S1.5) resembles the optical chirality continuity equation [8] as presented in Eqn. (S1.1).

### B. Rewriting the optical chirality source term

We now focus on simplifying the optical chirality source term $\boldsymbol{B} \cdot \frac{\partial \boldsymbol{j_{tot}}}{\partial t} - \boldsymbol{E} \cdot (\nabla \times \boldsymbol{j_{tot}})$ from Eqn. (S1.5). Applying the vector identity $\nabla \cdot (\boldsymbol{a} \times \boldsymbol{b}) = \boldsymbol{b} \cdot (\nabla \times \boldsymbol{a}) - \boldsymbol{a} \cdot (\nabla \times \boldsymbol{b})$ [14], where $\boldsymbol{a}$ and $\boldsymbol{b}$ are arbitrary vector fields:

$$\boldsymbol{B} \cdot \frac{\partial \boldsymbol{j_{tot}}}{\partial t} + \boldsymbol{E} \cdot (\nabla \times \boldsymbol{j_{tot}}) = \boldsymbol{B} \cdot \frac{\partial \boldsymbol{j_{tot}}}{\partial t} + \big(\nabla \cdot (\boldsymbol{j_{tot}} \times \boldsymbol{E}) + \boldsymbol{j_{tot}} \cdot (\nabla \times \boldsymbol{E})\big)$$

Applying Faraday's Law $\left(\nabla \times \boldsymbol{E} = -\frac{\partial \boldsymbol{B}}{\partial t}\right)$ [16], we obtain:

$$\boldsymbol{B} \cdot \frac{\partial \boldsymbol{j_{tot}}}{\partial t} + \boldsymbol{E} \cdot (\nabla \times \boldsymbol{j_{tot}}) = \boldsymbol{B} \cdot \frac{\partial \boldsymbol{j_{tot}}}{\partial t} + \left(\nabla \cdot (\boldsymbol{j_{tot}} \times \boldsymbol{E}) + \boldsymbol{j_{tot}} \cdot \left(-\frac{\partial \boldsymbol{B}}{\partial t}\right)\right)$$

$$
\begin{aligned}
&= \boldsymbol{B} \cdot \frac{\partial \boldsymbol{j_{tot}}}{\partial t} + \nabla \cdot (\boldsymbol{j_{tot}} \times \boldsymbol{E}) + \boldsymbol{j_{tot}} \cdot \left(-\frac{\partial \boldsymbol{B}}{\partial t}\right) \\
&= \boldsymbol{B} \cdot \frac{\partial \boldsymbol{j_{tot}}}{\partial t} + \boldsymbol{j_{tot}} \cdot \left(-\frac{\partial \boldsymbol{B}}{\partial t}\right) + \nabla \cdot (\boldsymbol{j_{tot}} \times \boldsymbol{E}) \\
&= \boldsymbol{B} \cdot \frac{\partial \boldsymbol{j_{tot}}}{\partial t} - \boldsymbol{j_{tot}} \cdot \frac{\partial \boldsymbol{B}}{\partial t} + \nabla \cdot (\boldsymbol{j_{tot}} \times \boldsymbol{E}) \\
&= \boldsymbol{B} \cdot \frac{\partial \boldsymbol{j_{tot}}}{\partial t} + (\boldsymbol{B} \cdot \frac{\partial \boldsymbol{j_{tot}}}{\partial t} - \boldsymbol{B} \cdot \frac{\partial \boldsymbol{j_{tot}}}{\partial t}) - \boldsymbol{j_{tot}} \cdot \frac{\partial \boldsymbol{B}}{\partial t} + \nabla \cdot (\boldsymbol{j_{tot}} \times \boldsymbol{E}) \\
&= \boldsymbol{B} \cdot \frac{\partial \boldsymbol{j_{tot}}}{\partial t} + \boldsymbol{B} \cdot \frac{\partial \boldsymbol{j_{tot}}}{\partial t} - \boldsymbol{B} \cdot \frac{\partial \boldsymbol{j_{tot}}}{\partial t} - \boldsymbol{j_{tot}} \cdot \frac{\partial \boldsymbol{B}}{\partial t} + \nabla \cdot (\boldsymbol{j_{tot}} \times \boldsymbol{E}) \\
&= 2\boldsymbol{B} \cdot \frac{\partial \boldsymbol{j_{tot}}}{\partial t} - \boldsymbol{B} \cdot \frac{\partial \boldsymbol{j_{tot}}}{\partial t} - \boldsymbol{j_{tot}} \cdot \frac{\partial \boldsymbol{B}}{\partial t} + \nabla \cdot (\boldsymbol{j_{tot}} \times \boldsymbol{E})
\end{aligned}
$$

Recognizing that $-\boldsymbol{B} \cdot \frac{\partial \boldsymbol{j_{tot}}}{\partial t} - \boldsymbol{j_{tot}} \cdot \frac{\partial \boldsymbol{B}}{\partial t}$ can be acquired via application of the product rule on $-\frac{\partial}{\partial t}(\boldsymbol{j_{tot}} \cdot \boldsymbol{B})$, we can make the substitution:

$$
\boldsymbol{B} \cdot \frac{\partial \boldsymbol{j_{tot}}}{\partial t} + \boldsymbol{E} \cdot (\nabla \times \boldsymbol{j_{tot}}) = 2\boldsymbol{B} \cdot \frac{\partial \boldsymbol{j_{tot}}}{\partial t} - \frac{\partial}{\partial t}(\boldsymbol{j_{tot}} \cdot \boldsymbol{B}) + \nabla \cdot (\boldsymbol{j_{tot}} \times \boldsymbol{E}) \quad (S1.6)
$$

Substituting Eqn. (S1.6) into Eqn. (S1.5):

$$
\frac{1}{\mu_0}\nabla \cdot \boldsymbol{F} - \frac{\partial}{\partial t}\left[\frac{1}{2\mu_0 c^2}\left(-\boldsymbol{B} \cdot \frac{\partial \boldsymbol{E}}{\partial t} + \boldsymbol{E} \cdot \frac{\partial \boldsymbol{B}}{\partial t}\right)\right] = -\frac{1}{2}\left[2\boldsymbol{B} \cdot \frac{\partial \boldsymbol{j_{tot}}}{\partial t} - \frac{\partial}{\partial t}(\boldsymbol{j_{tot}} \cdot \boldsymbol{B}) + \nabla \cdot (\boldsymbol{j_{tot}} \times \boldsymbol{E})\right] \quad (S1.7)
$$

### C. Rewriting the optical chirality density term

We now focus on simplifying the optical chirality density term $\frac{1}{2\mu_0 c^2}\left(-\boldsymbol{B} \cdot \frac{\partial \boldsymbol{E}}{\partial t} + \boldsymbol{E} \cdot \frac{\partial \boldsymbol{B}}{\partial t}\right)$ from Eqn. (S1.7). Applying the Ampère-Maxwell Law [16] $\left(\nabla \times \boldsymbol{B} = \mu_0\left(\boldsymbol{j_{tot}} + \epsilon_0 \frac{\partial \boldsymbol{E}}{\partial t}\right) \rightarrow \frac{\partial \boldsymbol{E}}{\partial t} = \frac{\nabla \times \boldsymbol{B}}{\mu_0 \epsilon_0} - \frac{\boldsymbol{j_{tot}}}{\epsilon_0}\right)$:

$$
\frac{1}{2\mu_0 c^2}\left(-\boldsymbol{B} \cdot \frac{\partial \boldsymbol{E}}{\partial t} + \boldsymbol{E} \cdot \frac{\partial \boldsymbol{B}}{\partial t}\right) = \frac{1}{2\mu_0 c^2}\left[-\boldsymbol{B} \cdot \left(\frac{\nabla \times \boldsymbol{B}}{\mu_0 \epsilon_0} - \frac{\boldsymbol{j_{tot}}}{\epsilon_0}\right) + \boldsymbol{E} \cdot \frac{\partial \boldsymbol{B}}{\partial t}\right]
$$

Applying Faraday's Law $\left(\nabla \times \boldsymbol{E} = -\frac{\partial \boldsymbol{B}}{\partial t} \rightarrow \frac{\partial \boldsymbol{B}}{\partial t} = -\nabla \times \boldsymbol{E}\right)$ [16], we obtain:

$$
\begin{aligned}
\frac{1}{2\mu_0 c^2}\left(-\boldsymbol{B} \cdot \frac{\partial \boldsymbol{E}}{\partial t} + \boldsymbol{E} \cdot \frac{\partial \boldsymbol{B}}{\partial t}\right) &= \frac{1}{2\mu_0 c^2}\left[-\boldsymbol{B} \cdot \left(\frac{\nabla \times \boldsymbol{B}}{\mu_0 \epsilon_0} - \frac{\boldsymbol{j_{tot}}}{\epsilon_0}\right) + \boldsymbol{E} \cdot (-\nabla \times \boldsymbol{E})\right] \\
&= \frac{1}{2\mu_0 c^2}\left[\frac{\boldsymbol{B} \cdot (\nabla \times \boldsymbol{B})}{\mu_0 \epsilon_0} - \frac{\boldsymbol{B} \cdot \boldsymbol{j_{tot}}}{\epsilon_0} - \boldsymbol{E} \cdot (\nabla \times \boldsymbol{E})\right]
\end{aligned}
$$

The curl of the magnetic induction field $(\nabla \times \boldsymbol{B})$ will always be orthogonal to the magnetic induction field $\boldsymbol{B}$; thus, $\boldsymbol{B} \cdot (\nabla \times \boldsymbol{B}) = \boldsymbol{0}$. Similarly, $\boldsymbol{E} \cdot (\nabla \times \mathrm{E}) = \boldsymbol{0}$. Applying this, we obtain:

$$
\begin{aligned}
\frac{1}{2\mu_0 c^2}\left(-\boldsymbol{B} \cdot \frac{\partial \boldsymbol{E}}{\partial t} + \boldsymbol{E} \cdot \frac{\partial \boldsymbol{B}}{\partial t}\right) &= \frac{1}{2\mu_0 c^2}\left[\frac{0}{\mu_0 \epsilon_0} - \frac{\boldsymbol{B} \cdot \boldsymbol{j_{tot}}}{\epsilon_0} - 0\right] \\
&= \frac{1}{2\mu_0 c^2}\left[-\frac{\boldsymbol{B} \cdot \boldsymbol{j_{tot}}}{\epsilon_0}\right]
\end{aligned}
$$

Electromagnetic signals in a vacuum travel at $c = \frac{1}{\sqrt{\mu_o \epsilon_0}}$. Substituting this into the equation:

$$
\frac{1}{2\mu_0 c^2}\left(-\boldsymbol{B} \cdot \frac{\partial \boldsymbol{E}}{\partial t} + \boldsymbol{E} \cdot \frac{\partial \boldsymbol{B}}{\partial t}\right) = \frac{1}{2\mu_0\left(\frac{1}{\sqrt{\mu_o \epsilon_0}}\right)^2}\left[-\frac{\boldsymbol{B} \cdot \boldsymbol{j_{tot}}}{\epsilon_0}\right]
$$

$$= \frac{1}{2\mu_0\left(\frac{1}{\mu_o\epsilon_0}\right)}\left[-\frac{\boldsymbol{B}\cdot\boldsymbol{j_{tot}}}{\epsilon_0}\right]$$
$$= \frac{\mu_0\epsilon_0}{2\mu_0}\left[-\frac{\boldsymbol{B}\cdot\boldsymbol{j_{tot}}}{\epsilon_0}\right]$$
$$= \frac{1}{2}[-\boldsymbol{B}\cdot\boldsymbol{j_{tot}}]$$

Thus,

$$\frac{1}{2\mu_0 c^2}\left(-\boldsymbol{B}\cdot\frac{\partial\boldsymbol{E}}{\partial t}+\boldsymbol{E}\cdot\frac{\partial\boldsymbol{B}}{\partial t}\right) = -\frac{1}{2}\,\boldsymbol{B}\cdot\boldsymbol{j_{tot}} \qquad (S1.8)$$

Substituting Eqn. (S1.8) into Eqn. (S1.7):

$$\frac{1}{\mu_0}\nabla\cdot\boldsymbol{F}-\frac{\partial}{\partial t}\left(-\frac{1}{2}\,\boldsymbol{B}\cdot\boldsymbol{j_{tot}}\right) = -\frac{1}{2}\left[2\boldsymbol{B}\cdot\frac{\partial\boldsymbol{j_{tot}}}{\partial t}-\frac{\partial}{\partial t}(\boldsymbol{j_{tot}}\cdot\boldsymbol{B})+\nabla\cdot(\boldsymbol{j_{tot}}\times\boldsymbol{E})\right] \qquad (S1.9)$$

**D. Simplifying Eqn. (S1.9)**

Next, we focus on simplifying the expression obtained in Eqn. (S1.9):

$$\frac{1}{\mu_0}\nabla\cdot\boldsymbol{F}+\frac{1}{2}\frac{\partial}{\partial t}(\boldsymbol{B}\cdot\boldsymbol{j_{tot}}) = -\frac{1}{2}\left[2\boldsymbol{B}\cdot\frac{\partial\boldsymbol{j_{tot}}}{\partial t}-\frac{\partial}{\partial t}(\boldsymbol{j_{tot}}\cdot\boldsymbol{B})+\nabla\cdot(\boldsymbol{j_{tot}}\times\boldsymbol{E})\right]$$
$$\frac{1}{\mu_0}\nabla\cdot\boldsymbol{F}+\frac{1}{2}\frac{\partial}{\partial t}(\boldsymbol{B}\cdot\boldsymbol{j_{tot}}) = -\left[\boldsymbol{B}\cdot\frac{\partial\boldsymbol{j_{tot}}}{\partial t}-\frac{1}{2}\frac{\partial}{\partial t}(\boldsymbol{j_{tot}}\cdot\boldsymbol{B})+\frac{1}{2}\nabla\cdot(\boldsymbol{j_{tot}}\times\boldsymbol{E})\right]$$
$$\frac{1}{\mu_0}\nabla\cdot\boldsymbol{F}+\frac{1}{2}\frac{\partial}{\partial t}(\boldsymbol{B}\cdot\boldsymbol{j_{tot}}) = -\boldsymbol{B}\cdot\frac{\partial\boldsymbol{j_{tot}}}{\partial t}+\frac{1}{2}\frac{\partial}{\partial t}(\boldsymbol{j_{tot}}\cdot\boldsymbol{B})-\frac{1}{2}\nabla\cdot(\boldsymbol{j_{tot}}\times\boldsymbol{E})$$
$$\frac{1}{\mu_0}\nabla\cdot\boldsymbol{F}+\frac{1}{2}\frac{\partial}{\partial t}(\boldsymbol{B}\cdot\boldsymbol{j_{tot}})-\frac{1}{2}\frac{\partial}{\partial t}(\boldsymbol{j_{tot}}\cdot\boldsymbol{B}) = -\boldsymbol{B}\cdot\frac{\partial\boldsymbol{j_{tot}}}{\partial t}-\frac{1}{2}\nabla\cdot(\boldsymbol{j_{tot}}\times\boldsymbol{E})$$
$$\frac{1}{\mu_0}\nabla\cdot\boldsymbol{F}+\frac{1}{2}\left[\frac{\partial}{\partial t}(\boldsymbol{B}\cdot\boldsymbol{j_{tot}})-\frac{\partial}{\partial t}(\boldsymbol{j_{tot}}\cdot\boldsymbol{B})\right] = -\boldsymbol{B}\cdot\frac{\partial\boldsymbol{j_{tot}}}{\partial t}-\frac{1}{2}\nabla\cdot(\boldsymbol{j_{tot}}\times\boldsymbol{E})$$

Recognizing that the difference rule can be applied on $\frac{\partial}{\partial t}(\boldsymbol{B}\cdot\boldsymbol{j_{tot}})-\frac{\partial}{\partial t}(\boldsymbol{j_{tot}}\cdot\boldsymbol{B})$:

$$\frac{1}{\mu_0}\nabla\cdot\boldsymbol{F}+\frac{1}{2}\frac{\partial}{\partial t}[(\boldsymbol{B}\cdot\boldsymbol{j_{tot}})-(\boldsymbol{j_{tot}}\cdot\boldsymbol{B})] = -\boldsymbol{B}\cdot\frac{\partial\boldsymbol{j_{tot}}}{\partial t}-\frac{1}{2}\nabla\cdot(\boldsymbol{j_{tot}}\times\boldsymbol{E})$$
$$\frac{1}{\mu_0}\nabla\cdot\boldsymbol{F}+\frac{1}{2}\frac{\partial}{\partial t}[(\boldsymbol{B}\cdot\boldsymbol{j_{tot}})-(\boldsymbol{B}\cdot\boldsymbol{j_{tot}})] = -\boldsymbol{B}\cdot\frac{\partial\boldsymbol{j_{tot}}}{\partial t}-\frac{1}{2}\nabla\cdot(\boldsymbol{j_{tot}}\times\boldsymbol{E})$$
$$\frac{1}{\mu_0}\nabla\cdot\boldsymbol{F}+\frac{1}{2}\frac{\partial}{\partial t}[0] = -\boldsymbol{B}\cdot\frac{\partial\boldsymbol{j_{tot}}}{\partial t}-\frac{1}{2}\nabla\cdot(\boldsymbol{j_{tot}}\times\boldsymbol{E})$$

Thus,

$$\frac{1}{\mu_0}\nabla\cdot\boldsymbol{F} = -\boldsymbol{B}\cdot\frac{\partial\boldsymbol{j_{tot}}}{\partial t}-\frac{1}{2}\nabla\cdot(\boldsymbol{j_{tot}}\times\boldsymbol{E}) \qquad (S1.10)$$

**E. Time-averaging under a steady-state assumption**

To obtain a physically measurable form, we consider the time-averaged version of Eqn. (S1.10) over a period $T$ (denoted by $<\cdot>_T$) [18,19]:

$$\frac{1}{\mu_0}<\nabla\cdot\boldsymbol{F}>_T = -<\boldsymbol{B}\cdot\frac{\partial\boldsymbol{j_{tot}}}{\partial t}>_T-\frac{1}{2}<\nabla\cdot(\boldsymbol{j_{tot}}\times\boldsymbol{E})>_T \qquad (S1.11)$$

We now assume the steady-state solution where $\boldsymbol{\mathcal{F}}$, $\boldsymbol{\mathcal{B}}$, $\boldsymbol{\mathcal{J}_{tot}}$, and $\boldsymbol{\mathcal{E}}$ are the time-harmonic optical chirality flux, magnetic induction field, total current density, and electric field, respectively, such that for some real-valued, time-dependent field $\boldsymbol{X}$, it relates to its time-harmonic field amplitude $\boldsymbol{\mathcal{X}}$ by $\boldsymbol{X} = Re[\boldsymbol{\mathcal{X}}e^{-i\omega t}]$, where $\omega$ is the angular frequency, and that each field varies harmonically with a single frequency [18,20]. Writing Eqn. (S1.11) with the time-harmonic notation:

$$\frac{1}{\mu_0} < \nabla \cdot \boldsymbol{\mathcal{F}} >_T = -< \boldsymbol{\mathcal{B}} \cdot \frac{\partial \boldsymbol{\mathcal{J}_{tot}}}{\partial t} >_T - \frac{1}{2} < \nabla \cdot (\boldsymbol{\mathcal{J}_{tot}} \times \boldsymbol{\mathcal{E}}) >_T \qquad (S1.12)$$

**F. Integrating over a volume of interest**

Next, we integrate the time-averaged form shown in Eqn. (S1.12) over a volume $V$, which encompasses all primary and secondary sources:

$$\frac{1}{\mu_0} \iiint_V < \nabla \cdot \boldsymbol{\mathcal{F}} >_T dV = - \iiint_V < \boldsymbol{\mathcal{B}} \cdot \frac{\partial \boldsymbol{\mathcal{J}_{tot}}}{\partial t} >_T dV - \frac{1}{2} \iiint_V < \nabla \cdot (\boldsymbol{\mathcal{J}_{tot}} \times \boldsymbol{\mathcal{E}}) >_T dV \qquad (S1.13)$$

We turn our attention to the right-most term in Eqn. (S1.13): $\iiint_V < \nabla \cdot (\boldsymbol{\mathcal{J}_{tot}} \times \boldsymbol{\mathcal{E}}) >_T$. Applying Gauss' theorem to this term $\left(\iiint_V < \nabla \cdot (\boldsymbol{\mathcal{J}_{tot}} \times \boldsymbol{\mathcal{E}}) >_T dV = \iint_{\partial V} < \boldsymbol{\mathcal{J}_{tot}} \times \boldsymbol{\mathcal{E}} >_T \cdot \hat{\boldsymbol{n}}\, da\right.$, where $\hat{\boldsymbol{n}}$ is the unit vector$\left.\right)$ [21] can show that the divergence of the current density for any closed surface defining a volume will equal to zero under steady-state conditions. $\iint_{\partial V} < \boldsymbol{\mathcal{J}_{tot}} \times \boldsymbol{\mathcal{E}} >_T \cdot \hat{n}\, da$ represents the net flux of the time-averaged vector field $\boldsymbol{\mathcal{J}_{tot}} \times \boldsymbol{\mathcal{E}}$ through the closed surface $\partial V$ [21]. This flux will be zero if all current sources – both primary and secondary – are accounted for by $\boldsymbol{\mathcal{J}_{tot}}$ and contained in the volume $V$ [22,23]. $\boldsymbol{\mathcal{J}_{tot}}$ is the total current density that contains all primary and secondary sources, as established at the start of this derivation. Thus, in the absence of additional net sources or sinks of current on the boundary $\partial V$, the volume integral $\iiint_V < \nabla \cdot (\boldsymbol{\mathcal{J}_{tot}} \times \boldsymbol{\mathcal{E}}) >_T$ will also equal to zero. Therefore,

$$\frac{1}{\mu_0} \iiint_V < \nabla \cdot \boldsymbol{\mathcal{F}} >_T dV = - \iiint_V < \boldsymbol{\mathcal{B}} \cdot \frac{\partial \boldsymbol{\mathcal{J}_{tot}}}{\partial t} >_T dV \qquad (S1.14)$$

The partial time-derivative for some harmonic field $\boldsymbol{\mathcal{X}}$ can be written as $\frac{\partial \boldsymbol{\mathcal{X}}}{\partial t} = i\omega \boldsymbol{\mathcal{X}}$ [18]. Applying this to $\frac{\partial \boldsymbol{\mathcal{J}_{tot}}}{\partial t}$:

$$\begin{aligned} \frac{1}{\mu_0} \iiint_V < \nabla \cdot \boldsymbol{\mathcal{F}} >_T dV &= - \iiint_V i\omega_0 < \boldsymbol{\mathcal{B}} \cdot \boldsymbol{\mathcal{J}}_{\mathbf{tot}} >_T dV \\ &= -\omega_0 \iiint_V i < \boldsymbol{\mathcal{B}} \cdot \boldsymbol{\mathcal{J}}_{\mathbf{tot}} >_T dV \end{aligned}$$

To evaluate the time-averaged $\boldsymbol{\mathcal{B}} \cdot \boldsymbol{\mathcal{J}_{tot}}$ expression, we can use the following rule: $< \boldsymbol{\mathcal{X}} \cdot \boldsymbol{\mathcal{Y}} >_T = \frac{1}{2} Re(\boldsymbol{\mathcal{X}} \cdot \boldsymbol{\mathcal{Y}}^*)$ for time-harmonic fields $\boldsymbol{\mathcal{X}}$ and $\boldsymbol{\mathcal{Y}}$ [18]:

$$\frac{1}{\mu_0} \iiint_V \nabla \cdot \frac{1}{2} \mathrm{Re}(\boldsymbol{\mathcal{F}}) dV = -\omega_0 \iiint_V i \cdot \frac{1}{2} Re(\boldsymbol{\mathcal{B}} \cdot \boldsymbol{\mathcal{J}}^*_{\mathbf{tot}})\, dV$$

Simplifying and rearranging:

$$\begin{aligned} \frac{1}{2\mu_0} \iiint_V \nabla \cdot \mathrm{Re}(\boldsymbol{\mathcal{F}}) dV &= -\frac{\omega_0}{2} \iiint_V i \cdot Re(\boldsymbol{\mathcal{B}} \cdot \boldsymbol{\mathcal{J}}^*_{\mathbf{tot}})\, dV \\ \frac{1}{2\mu_0} \iiint_V \nabla \cdot \mathrm{Re}(\boldsymbol{\mathcal{F}}) dV &= -\frac{\omega_0}{2} \iiint_V Im(\boldsymbol{\mathcal{B}} \cdot \boldsymbol{\mathcal{J}}^*_{\mathbf{tot}})\, dV \\ \iiint_V \nabla \cdot \mathrm{Re}(\boldsymbol{\mathcal{F}}) dV &= -2\mu_0 \cdot \frac{\omega_0}{2} \iiint_V Im(\boldsymbol{\mathcal{B}} \cdot \boldsymbol{\mathcal{J}}^*_{\mathbf{tot}})\, dV \end{aligned}$$

Thus,

$$\iiint_V \nabla \cdot \mathrm{Re}(\boldsymbol{\mathcal{F}})dV = -\mu_0\omega_0 \iiint_V Im(\boldsymbol{\mathcal{B}} \cdot \boldsymbol{\mathcal{J}}^*_{\mathbf{tot}})dV \tag{S1.15}$$

Applying Gauss's theorem [21] to the term on the left:

$$\iint_{\partial V} \mathrm{Re}(\boldsymbol{\mathcal{F}}) \cdot \hat{\boldsymbol{n}}\, da = -\mu_0\omega_0 \iiint_V Im(\boldsymbol{\mathcal{B}} \cdot \boldsymbol{\mathcal{J}}^*_{\mathbf{tot}})dV \tag{S1.16}$$

Thus, to generate or dissipate optical chirality in a closed volume, $Im(\boldsymbol{\mathcal{B}} \cdot \boldsymbol{\mathcal{J}}^*_{\mathbf{tot}})$ must be non-zero. This demonstrates that the interaction of the magnetic induction field with the total current density leads to a source or sink of optical chirality flux. Notably, the interaction of the electric field with the total current density does not contribute to the source term in the optical chirality continuity equation.

## S3. NET MAGNETIC & ELECTRIC FLUXES

### A. Net Magnetic Flux

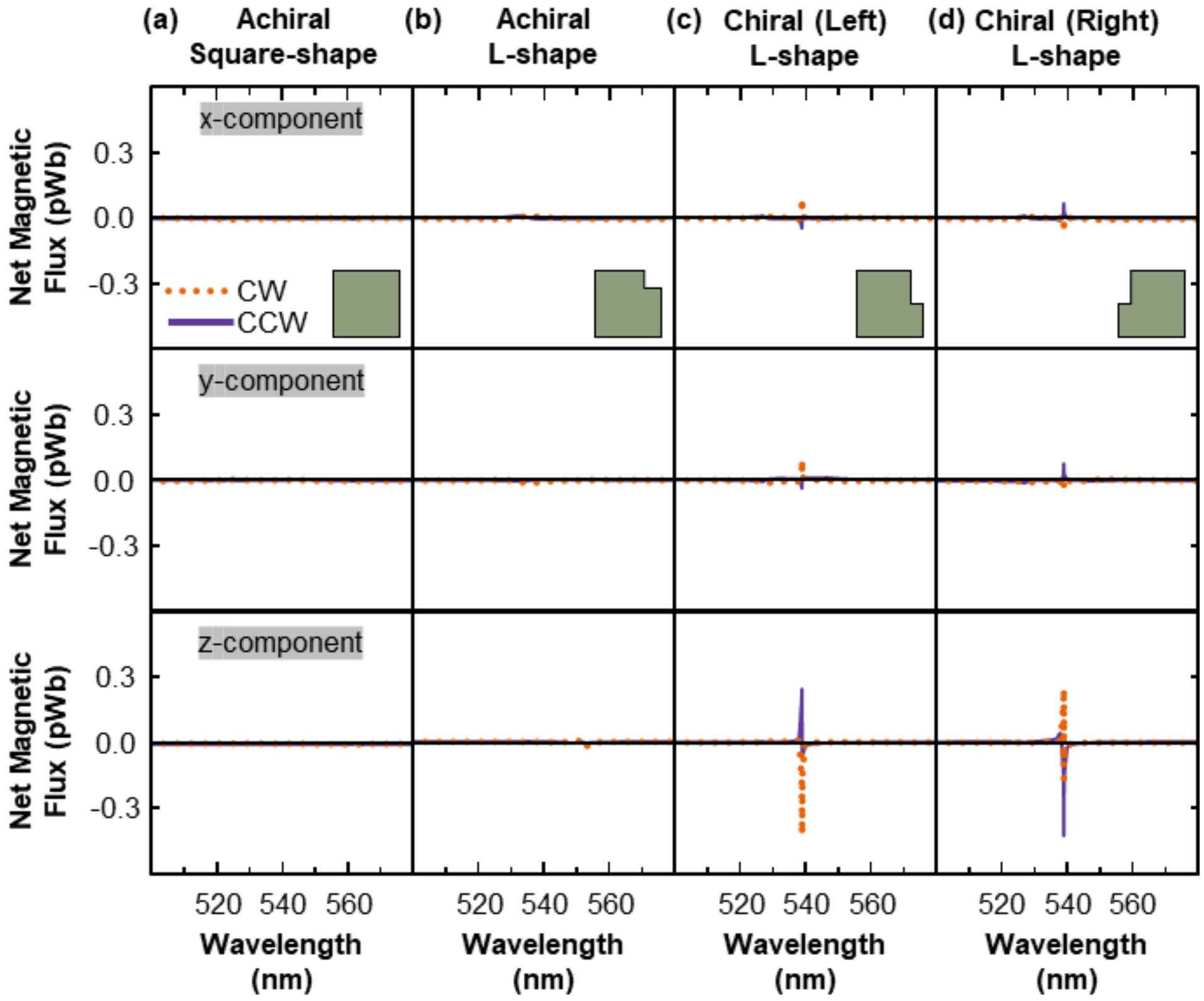

FIG S21. Net magnetic flux under clockwise (CW; orange, dotted) and counterclockwise (CCW; purple, solid) CPL illumination for the (a) achiral square-shaped, (b) achiral L-shaped, (c) chiral (left) L-shaped, and (d) chiral (right) L-shaped structures for the (top) x-, (middle) y-, and (bottom) z-components of the net magnetic flux. (a-b) The achiral structures showed a zero or nearly-zero flux for all components of the net magnetic flux. (c-d) The chiral L-shaped structures showed mirror behavior between the (c) left- and (d) right-handed configurations of the L-shape; a sharp peak in the net magnetic flux was seen at the ~539 nm resonance of all components, with the highest-amplitude peak in the z-component.

### B. Net Electric Flux

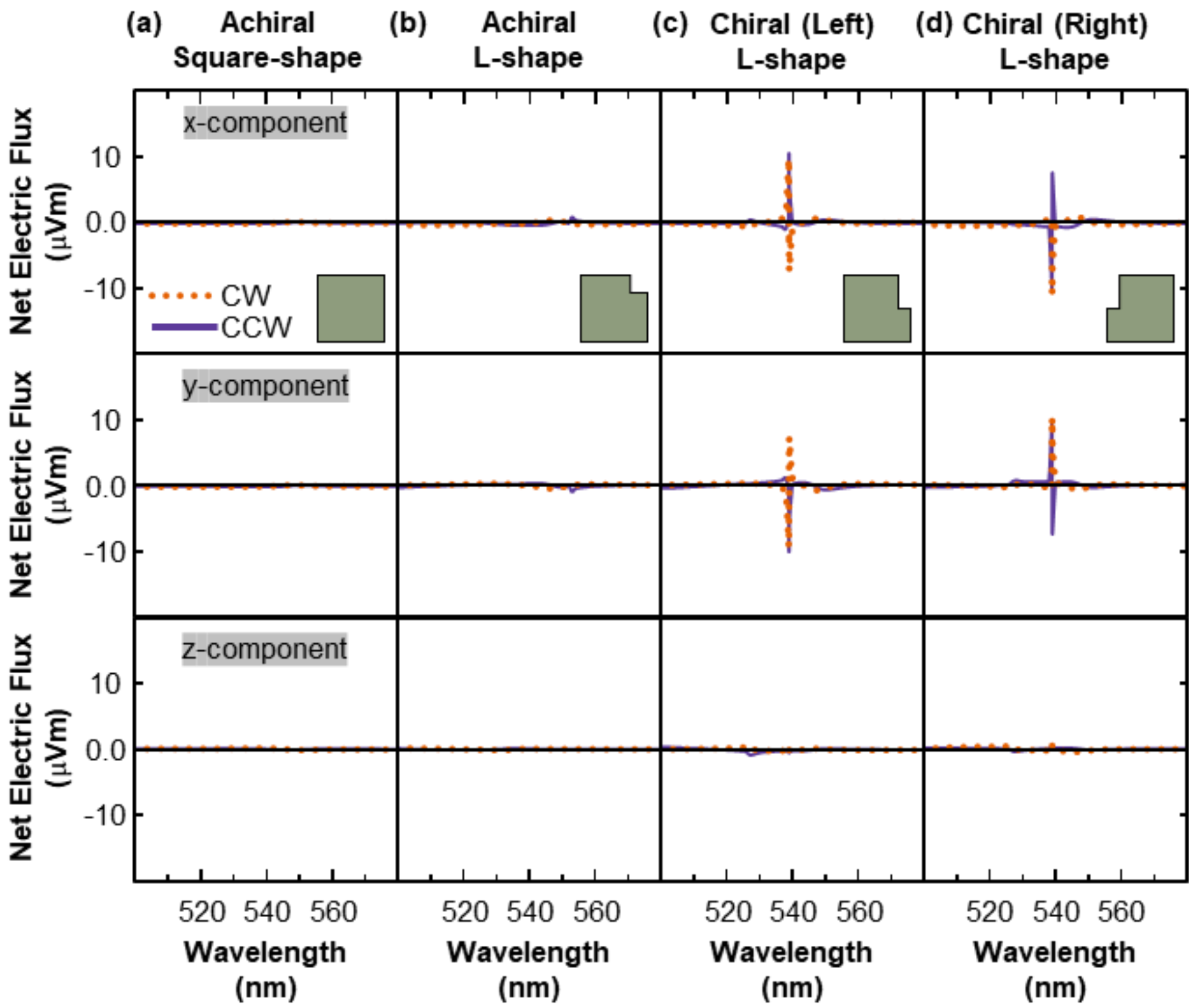

FIG S22. Net electric flux under clockwise (CW; orange, dotted) and counterclockwise (CCW; purple, solid) CPL illumination for the (a) achiral square-shaped, (b) achiral L-shaped, (c) chiral (left) L-shaped, and (d) chiral (right) L-shaped structures for the (top) x-, (middle) y-, and (bottom) z-components of the net electric flux. (a-b) The achiral structures showed a zero or nearly-zero values for all components of the net electric flux. (c-d) The chiral L-shaped structures showed mirror behavior between the (c) left- and (d) right-handed configurations of the L-shape; a sharp peak in the net electric flux was seen at the ~539 nm resonance of the x- and y-components.